\documentclass[a4paper,11pt]{article}
\usepackage{jheppub} 
\makeatletter
\gdef\@fpheader{\textcolor{white}{preprint}}
\makeatother
\usepackage{lineno}
\usepackage{comment}
\usepackage{ulem}
\usepackage{amsmath,amssymb,mathtools}
\usepackage{subcaption}
\usepackage{graphicx}
\usepackage{pdflscape}
\usepackage{placeins}

\title{\boldmath Regularized Master-Field Approximation for Mass-Deformed Supersymmetric Yang-Mills Matrix Models}

\author[b]{Yue Lei}
\author[a,b,c]{and Reishi Maeta}

\affiliation[a]{
Graduate School of Advanced Science and Engineering,\\
Hiroshima University,\\
Higashi-Hiroshima, Hiroshima 739-8526, Japan
}

\affiliation[b]{
Department of Physics,\\
McGill University,\\
Montreal, QC H3A 2T8, Canada
}

\affiliation[c]{
Theory Center, Institute of Particle and Nuclear Studies,\\
High Energy Accelerator Research Organization (KEK),\\
1-1 Oho, Tsukuba, Ibaraki 305-0801, Japan
}

\emailAdd{maeta-reishi@hiroshima-u.ac.jp}
\emailAdd{yue.lei@mail.mcgill.ca}

\abstract{
The regularized master-field approximation (RMA) is an efficient computational method for large-$N$ reduced matrix models, including those with multiple matrices or complex integration weights.
In this work, we apply the RMA to mass-deformed supersymmetric Yang--Mills-type matrix models with three, four, and ten matrices.
All three models admit fuzzy-sphere configurations as classical solutions.
We show that the RMA quantitatively reproduces these fuzzy-sphere solutions and their breakdown, and that zero modes of the Dirac operator can destabilize fuzzy-sphere configurations in certain cases.
}

\begin{document}
\maketitle
\flushbottom

\section{Introduction}\label{sec:intro
}
The large-\(N\) limit of gauge theories and matrix models has played an important role in the development of modern theoretical physics.
The observation that the rank of the gauge group can be used as an expansion parameter revealed a close connection between gauge theories and string theories \cite{tHooft:1973alw}.
Independently, large-N reduction \cite{Eguchi:1982nm, GROSS1982440, Parisi:1982gp} showed that gauge theories in the large N limit become volume-independent and can therefore be represented as a reduced matrix model. Matrix models later acquired a more direct role in nonperturbative formulations of superstring theories and M-theory through the BFSS and IKKT proposals \cite{deWit:1988wri, Banks:1996vh, Ishibashi:1996xs}. 
Large-\(N\) matrix models have also provided useful theoretical laboratories for studying strongly coupled quantum systems \cite{Wadia:1980cp, Gross:1980he, Gonzalez-Arroyo:1982hwr, Gonzalez-Arroyo:2010omx}, noncommutative geometry \cite{Connes:1997cr, Douglas:2001ba, Szabo:2001kg}, emergent spacetime \cite{Jevicki:1979mb, DiFrancesco:1993cyw, Steinacker:2010rh}, and possible microscopic descriptions of gravity \cite{Itzhaki:1998dd, Steinacker:2008ri, Steinacker:2024unq, Ho:2025htr}.

One of the most important simplifications in the planar limit is large-\(N\) factorization. 
Correlations between gauge-invariant normalized single-trace operators become suppressed, suggesting that the matrix integral becomes effectively deterministic at the level of such observables.
This observation motivates the master-field conjecture: expectation values of arbitrary single-trace operators may be reproduced by evaluating the corresponding operators on a distinguished infinite-dimensional matrix configuration \cite{Witten:1979pi}.
In this sense, the master field provides a deterministic description of the planar theory while retaining the complete information contained in the noncommutative matrix observables.

Despite this simplification, it remains difficult to obtain analytic or nonperturbative results for general matrix models.
Exact analytic solutions are available only in special cases, and numerical or approximate methods are usually required.
Monte Carlo simulations have been successfully applied to many bosonic matrix models and certain supersymmetric matrix models, including \(D=4\) supersymmetric Yang-Mills model and finite temperature BFSS and BMN matrix quantum mechanics \cite{Ambjorn:2000bf, Anagnostopoulos:2007fw, Hanada:2008gy, Asano:2018nol}.
More recently, matrix-bootstrap methods have made substantial progress by combining the loop equations with positivity conditions, thereby constraining physical moments within rigorous allowed regions \cite{Anderson:2018xuq, Lin:2020mme, Kazakov:2021lel}.

These approaches nevertheless have important limitations.
Conventional Monte Carlo methods rely on interpreting the integration weight as a probability distribution and are therefore most effective when the weight is real and non-negative.
The most direct example is the real-time or Minkowski-type theory whose matrix integral weight is oscillatory, typically of the form \(e^{iS}\), and therefore cannot be interpreted as a probability measure.
A related difficulty arises in Euclidean supersymmetric matrix models: after integrating out the fermions, the resulting Pfaffian may be negative or complex. 
In that case, the fermion-induced phase again prevents the full integration weight from being treated as a positive probability measure. 
These difficulties are collectively referred to as the sign or phase problem.
Positivity-based bootstrap methods are also most naturally formulated when a positive measure is available and cannot be applied directly to general reduced matrix models with complex weights.\footnote{
In this paper, a (large-$N$) reduced matrix model refers to a matrix model without explicit coordinate dependence such as $t$ or $x$.
}
Considerable effort has therefore been devoted to alternative methods, including complex Langevin methods, generalized Lefschetz-thimble techniques, and bootstrap formulations that do not rely directly on positivity \cite{Aarts:2009uq, Nishimura:2019qal, Cristoforetti:2012su, Alexandru:2015sua,
Li:2024ggr, Maeta:2026oku, Berenstein:2026wky}.

The regularized master-field approximation, or RMA, provides a complementary approach to the numerical study of large-\(N\) reduced matrix models \cite{Maeta:2026miu}.
The RMA builds on the procedure introduced in \cite{Jevicki:1982jj,Jevicki:1983hb,Koch:2021yeb,Mathaba:2023non}, reinterpreting it as a finite-dimensional regularization of the planar master field and further developing the associated numerical optimization procedure. 
The regularized master field is represented by a finite set of $M\times M$ matrices whose components are determined numerically by requiring a finite set of planar loop equations to be satisfied as accurately as possible.
Here $M$ denotes the dimension of the regularized master field and is not identified with the rank $N$ of the original matrix integral.

A particular advantage of the RMA is that it works directly with the planar loop equations and is therefore free from the sign problem associated with complex or oscillatory integration weights.
The method was recently tested in Hermitian one- and two-matrix models \cite{Maeta:2026miu}. 
In Euclidean-type examples, the resulting regularized moments were found to agree well with known analytic results, numerical simulations, and positivity-based bootstrap calculations.
The method was also applied to Minkowski-type matrix integrals with oscillatory weights, where it reproduced perturbative and formal solutions with good accuracy. 
The physically relevant solutions were distinguished by their stability under changes in the regularization scheme, including variations of the matrix dimension, the number of loop equations, and the initial condition. 
These results provide evidence that a useful finite-dimensional regularization of the master field exists, at least for the bosonic one- and two-matrix models considered there.

The next natural step is to extend the RMA to matrix models containing fermionic matrices. 
Such an extension is essential for supersymmetric Yang-Mills matrix models and, ultimately, for the IKKT matrix model. Unlike bosonic matrices, however, fermionic matrices have Grassmann-odd entries and cannot be represented by ordinary numerical configurations. 
There is therefore no obvious fermionic analogue of the bosonic master field.

The problem of how to treat the fermionic matrices has already been addressed in Sec.~5 of \cite{Maeta:2026miu}: one first integrates out all the fermionic degrees of freedom and treats the resulting theory as an effective bosonic theory.
However, integrating out the fermions introduces a new weight for the bosonic matrices, namely the Pfaffian.
In addition, for observables involving fermions, Wick's theorem must be applied to eliminate the fermionic variables, and the resulting fermion propagator is a complicated function of the bosonic matrices.
These are new ingredients that are absent in purely bosonic matrix models, and it is therefore not immediately obvious whether the same master-field picture as before continues to hold for the effective bosonic theory obtained after integrating out the fermions.
We therefore first clarify this issue and show analytically that the effective bosonic theory can indeed be expected to possess a master field.

For theories with a quadratic fermionic action, we integrate out the fermions before applying the RMA.
This produces a Pfaffian contribution to the effective bosonic action, while observables containing fermions are reduced by Wick contraction using the inverse fermion kernel.
The resulting Wick-reduced quantities are then evaluated on the regularized bosonic master field.
The presence of fermions also modifies the loop equations through their graded cyclicity and Grassmann signs, so bosonic and fermionic Schwinger--Dyson equations must be treated separately.
We develop these ingredients explicitly in Sec.~2.

In the actual numerical calculations, we use mass-deformed supersymmetric Yang--Mills-type matrix models.
These models were introduced in \cite{Martina:2025kwc} and contain three, four, and ten matrices, respectively, and all the models considered in this work admit the fuzzy sphere as a solution of the equations of motion.
We then explicitly show that the RMA reproduces the fuzzy-sphere structure well.
From a theoretical point of view, the mass-deformed model containing ten matrices is also known as the polarised IKKT matrix model and has been intensively studied in recent years, mainly in the context of holographic duality \cite{Bonelli:2002mb,Hartnoll:2024csr, Komatsu:2024bop, Komatsu:2024ydh, Hartnoll:2025ecj}.
Through numerical calculations of this model, we find that not only the three matrices constituting the fuzzy sphere but also the remaining seven matrices exhibit characteristic patterns.
Another issue that deserves attention is the massless limit of the mass-deformed models, for which some subtle issues have been pointed out \cite{Komatsu:2024ydh, Chou:2025rwy}, and we discuss these issues as well.

The organization of this paper is as follows.
In Section \ref{sec:method}, we first review the basic concepts of the RMA and discuss how it can be applied to systems involving fermions.
In particular, we show that the theory obtained after integrating out the fermions can indeed possess a master field, and that the expectation values of physical quantities involving fermions can indeed be described in terms of the master field.
Section \ref{sec:three_matrix}, \ref{sec:four_matrix} and \ref{sec:ten_matrix} are devoted to the actual numerical calculations, where we apply the RMA to the three-matrix model, the four-matrix model, and the ten-matrix model, also known as the polarised IKKT matrix model.
We show that the RMA reproduces the fuzzy-sphere phase and captures transitions away from it, and that zero modes of the Dirac operator can sometimes play a decisive role in destabilizing the fuzzy-sphere configuration.
Finally, in Section \ref{sec:discussion}, we present our discussion and summarize the results.
The calculations and discussion of the structure of the saddle points, i.e., solutions to the equations of motion, in the three-matrix and four-matrix models are presented in Appendix \ref{app:classical-saddles}.

\section{Method}\label{sec:method}

\subsection{Brief review of the regularized master-field approximation}
\label{sec:rma-review}
We briefly review the regularized master-field approximation and refer to \cite{Maeta:2026miu} for a more detailed discussion. 
Consider a matrix model of \(N\times N\) matrices \(A_\mu\), and denote the normalized trace by \(\operatorname{tr} \equiv N^{-1}\operatorname{Tr}\).
For two gauge-invariant single-trace observables, large-\(N\) factorization implies
\begin{equation}
\left\langle \operatorname{tr} {\cal O}_1[A]\,\operatorname{tr} {\cal O}_2[A]\right\rangle
=
\left\langle \operatorname{tr} {\cal O}_1[A]\right\rangle
\left\langle \operatorname{tr} {\cal O}_2[A]\right\rangle
+
{\cal O}\left(N^{-2}\right).
\end{equation}
In particular, the variance of any normalized single-trace observable is suppressed by \(1/N^2\)
\begin{equation}
\left\langle
\left|
\operatorname{tr} {\cal O}[A]
-
\left\langle\operatorname{tr} {\cal O}[A]\right\rangle
\right|^2
\right\rangle
=
{\cal O}\left(N^{-2}\right).
\end{equation}
Thus, any gauge-invariant normalized trace observables become sharply concentrated around definite values as \(N\to\infty\). It is in this sense that the large-\(N\) matrix integral becomes effectively deterministic.

The master-field conjecture postulates a collection of generally noncommuting, formally infinite-dimensional matrices $\hat A_\mu$ such that, for a gauge-invariant single-trace observable,
\begin{equation}
\lim_{N\to\infty}
\left\langle\operatorname{tr}{\cal O}[A]\right\rangle
=
\operatorname{tr}{\cal O}[\hat A].
\label{eq:general_master_field}
\end{equation}
Here $\hat A_\mu$ is the master field associated with $A_\mu$.
For example, we consider operators of the form
\[
{\cal O}[\hat A]
=
\hat A_{\mu_1}\cdots\hat A_{\mu_n},
\qquad n=0,1,\ldots.
\]

The master field \(\hat A_\mu\) is formally infinite dimensional and therefore cannot be treated directly in a numerical calculation. The basic assumption of the RMA is that its planar trace data can be approximated by a collection of finite \(M\times M\) matrices \(\hat A_\mu^{(M)}\). The parameter \(M\) is a regulator for the master field and is not identified with the rank \(N\) of the original matrix integral.
The finite-dimensional regularization is defined by replacing each moment, represented as a normalized trace of the master field, by the corresponding normalized trace of finite \(M\times M\) matrices:
\begin{equation}
\operatorname{tr}_N\! {\cal O}[\hat A]
\quad\xrightarrow{\;\text{regularization}\;}\quad
\operatorname{tr}_M\!{\cal O}[\hat A^{(M)}],
\label{eq:regularized-moment}
\end{equation}
where $\operatorname{tr}_M$ is the normalized trace of dimensions $M$.
This is the finite-dimensional replacement used in every planar loop equation. 
This replacement also requires an appropriate normalization so that the quantities on the two sides remain of order unity in the large-$N$ and large-$M$ limits respectively.
For a product of $n$ matrices, this normalization can be implemented either by rescaling each matrix or by including the equivalent overall factor in front of the trace.

After this replacement, each planar loop equation becomes a function of
the finite-dimensional matrices $\hat A_\mu^{(M)}$.
We denote the $\alpha$-th regularized loop equation by
\begin{equation}
\mathcal{L}^{(M)}_\alpha
\!\left[\hat A^{(M)}\right]=0.
\label{eq:rma_residual}
\end{equation}
An exact finite-dimensional solution would satisfy $\mathcal{L}^{(M)}_\alpha=0$ for every retained loop equation.
The matrices $\hat A_\mu^{(M)}$ are determined by minimizing the residuals of a finite set of loop equations.

The components of \(\hat A_\mu^{(M)}\) are numerically determined by minimizing the objective function
\begin{equation}
F^{(M)}\!\left[\hat A^{(M)}\right]
=
\sum_{\alpha\in S_X}
r_\alpha
\left|
\mathcal L_\alpha^{(M)}
\!\left[\hat A^{(M)}\right]
\right|^2,
\qquad r_\alpha>0 ,
\label{eq:objectivefunction}
\end{equation}
where $S_X$ denotes the finite set of $X$ loop equations retained in the optimization, and $r_\alpha$ is a positive numerical weight assigned to the $\alpha$-th equation.
In practice, the independent components of the regularized matrices \(\hat A_\mu^{(M)}\) are determined by minimizing the sum of the squared loop-equation residuals using a nonlinear least-squares method.
Since the loop equations are nonlinear, a square system need not have a unique solution even when the number of independent constraints equals the number of variables.
If such an ambiguity arises, additional independent loop equations can be included so that the system becomes overdetermined, thereby further constraining the regularized master field.
The reliability of the resulting solution is then examined by varying the set of loop equations, the matrix size \(M\), and the initial conditions.

\subsection{Integrating out the fermionic matrices}
\label{sec:fermionic-integration}
We now explain how the RMA is extended to matrix models containing fermions.
The components of a fermionic matrix are Grassmann odd, so notions such as an ordinary numerical configuration, a saddle point, or a dominant configuration do not apply to the fermionic variables themselves.
There is therefore no natural fermionic analogue of the bosonic master field used in the RMA.
Instead, we first integrate out the fermionic matrices and formulate the theory in terms of an effective action for the bosonic matrices. The RMA is then applied to the bosonic master field of this effective theory.

The general procedure is as follows.
First, the fermionic action is expressed as a quadratic form whose kernel depends on the bosonic matrices.
Next, integration over the fermionic matrices produces the Pfaffian of the quadratic kernel and we obtain an effective action for the bosonic matrices.
At fixed bosonic matrices, the fermion propagator is given by the inverse of the quadratic kernel, and Wick's theorem is used to rewrite all observables containing fermionic matrices as bosonic functionals.
Finally, these Wick-reduced functionals are evaluated using the regularized bosonic master field.

More explicitly, we expand the fermionic matrices as
\(\Psi_\alpha=\Psi_\alpha^a t^a\), where \(t^a\) are Hermitian generators of \(u(N)\) satisfying
\(\operatorname{Tr}(t^a t^b)=\delta^{ab}\). The adjoint index runs over
\(a=1,\ldots,N^2\). 
Since the fermionic action is quadratic in the Grassmann-odd coefficients $\Psi_\alpha^a$, we may combine the spinor and adjoint labels into a single composite index $I=(\alpha,a)$ and write the fermionic action exactly as
\begin{equation}
S_{\mathrm F}[A,\Psi]
=
\frac{N}{2}\,
\Psi_I\,{\cal M}_{IJ}[A]\,\Psi_J,
\qquad
I=(\alpha,a),
\label{eq:general_fermion_kernel}
\end{equation}
where \({\cal M}[A]\) is an antisymmetric kernel acting on spinor--adjoint space.
The fermionic integral produces the Pfaffian
\begin{equation}
\int d\Psi\,
e^{-S_{\mathrm F}[A,\Psi]}
=
\operatorname{Pf}(N{\cal M}[A])
=
N^{D_{\mathrm F}/2}\operatorname{Pf}{\cal M}[A],
\label{eq:fermion-pfaffian}
\end{equation}
where $D_{\mathrm F}$ is the dimension of ${\cal M}[A]$.
Since the factor $N^{D_{\mathrm F}/2}$ is independent of the bosonic matrices, we omit it from the effective action.
The partition function can therefore be expressed entirely as a bosonic matrix integral,
\begin{equation}
Z
=
\int dA\,d\Psi\,
e^{-S_{\mathrm B}[A]-S_{\mathrm F}[A,\Psi]}
=
\int dA\,e^{-S_{\mathrm{eff}}[A]},
\qquad
S_{\mathrm{eff}}[A]
\equiv
S_{\mathrm B}[A]
-
\log\operatorname{Pf}{\cal M}[A].
\label{eq:effective-bosonic-action}
\end{equation}
Since the dimension of \({\cal M}[A]\) is proportional to the dimension of the adjoint representation and is therefore of order \(N^2\), the Pfaffian term is expected to contribute to the effective action at order \(N^2\).
It is thus natural to expect the large-\(N\) factorization underlying the bosonic master-field description to remain applicable after the fermions have been integrated out.

The same procedure applies to an arbitrary Grassmann-even observable
\({\cal O}[A,\Psi]\).
We define its fermion-integrated form
\({\cal O}_{\mathrm B}[A]\) and the corresponding bosonic expectation value by
\begin{align}
\operatorname{Pf}{\cal M}[A]\,
{\cal O}_{\mathrm B}[A]
&\equiv
\int d\Psi\,
{\cal O}[A,\Psi]\,
e^{-S_{\mathrm F}[A,\Psi]},
\nonumber\\
\left\langle{\cal O}[A,\Psi]\right\rangle
&=
\frac{
\displaystyle
\int dA\,
{\cal O}_{\mathrm B}[A]\,
e^{-S_{\mathrm{eff}}[A]}
}{
\displaystyle
\int dA\,
e^{-S_{\mathrm{eff}}[A]}
}.
\label{eq:fermion-integrated-observable}
\end{align}

Since the fermionic action is quadratic, \({\cal O}_{\mathrm B}[A]\) is obtained exactly by Wick contraction.
With the bosonic matrices held fixed, the fermionic propagator is
\begin{equation}
\left\langle
\Psi_{I}\Psi_{J}
\right\rangle_{\mathrm F}
=
\frac{1}{N}
\left({\cal M}^{-1}[A]\right)_{IJ}
\equiv
\frac{1}{N}\,
{\cal G}_{IJ}[A].
\label{eq:fermion-propagator}
\end{equation}
The factor $1/N$ follows from writing the fermionic action as
$S_{\mathrm F}=\frac{N}{2}\Psi_I{\cal M}_{IJ}[A]\Psi_J$, thus the matrix entering the quadratic form is $N{\cal M}[A]$.
One could instead absorb $N$ into the definition of the kernel; we leave it explicit to display the overall normalization of the action.

Correlation functions containing an odd number of fermionic coefficients vanish, whereas those containing an even number are reduced to sums over pairwise contractions, with the signs determined by fermionic statistics.

A point that requires some care is that Wick's theorem is applied to the Grassmann-odd coefficients \(\Psi_\alpha^a\), rather than directly to the fermionic matrices \(\Psi_\alpha\).
One must therefore first expand each fermionic matrix in the generator basis, \(\Psi_\alpha=\Psi_\alpha^a t^a\), and then contract the adjoint coefficients using the inverse quadratic kernel.
For example, consider a moment containing two adjacent fermionic insertions.
Expanding the fermionic matrices in adjoint components and applying the contraction above gives
\begin{equation}
\begin{aligned}
\left\langle
\operatorname{tr}\!\left(
A_i\Psi_\alpha\Psi_\beta
\right)
\right\rangle
&=
\left\langle
\Psi_\alpha^a\Psi_\beta^b\,
\operatorname{tr}\!\left(
A_i t^a t^b
\right)
\right\rangle
=
\left\langle
\frac{1}{N^2}
\operatorname{Tr}\!\left(
A_i\,{\cal G}_{\alpha\beta}[A]
\right)
\right\rangle,
\end{aligned}
\label{eq:two-fermion-example}
\end{equation}
where \({\cal G}_{\alpha\beta}[A]\equiv
{\cal G}_{\alpha\beta}^{ab}[A]\, t^a t^b\), and repeated adjoint indices are summed.

Applying the master-field prescription to the complete Wick-reduced observable and subsequently applying the regularization, we obtain
\begin{equation}
\begin{aligned}
\left\langle
\operatorname{tr}\!\left(
A_i\Psi_\alpha\Psi_\beta
\right)
\right\rangle
&=
\frac{1}{N^2}
\operatorname{Tr}\!\left(
\hat A_i\,
{\cal G}_{\alpha\beta}[\hat A]
\right)
\\
&\xrightarrow{\;\mathrm{regularization}\;}
\frac{1}{M^2}
\operatorname{Tr}\!\left(
\hat A_i^{(M)}
{\cal G}_{\alpha\beta}^{(M)}
[\hat A^{(M)}]
\right).
\end{aligned}
\label{eq:master-field-two-fermion-example}
\end{equation}
This is the general procedure for integrating-out the fermionic matrices and applying the master field prescription to the effective bosonic theory.
We illustrate this procedure explicitly for the three-matrix model in Sec.~\ref{subsec:three_matrix_setup}.  

\subsection{Worked example: the three-matrix model}
\label{subsec:three_matrix_setup}

We illustrate the method explicitly with the mass-deformed supersymmetric
Yang--Mills matrix model used in \cite{Martina:2025kwc}.
We derive its loop equations below before illustrating the integration over the fermions. Its numerical results are presented in Sec.~\ref{sec:three_matrix}.
The $D=3$ model has the lowest dimensionality and hence the simplest structure.
Its dynamical variables are three Hermitian
bosonic matrices \(A_i\), \(i=1,2,3\), and two Hermitian fermionic matrices \(\Psi_\alpha\), \(\alpha=1,2\), whose components are Grassmann odd.
In the conventions used throughout this work, the
action is
\begin{align}
 S_{3\mathrm{M}}[A,\Psi]
 =N\operatorname{Tr}\bigg\{&
 -\frac14[A_i,A_j][A_i,A_j]
 -\frac{i}{2}\bar\Psi\Gamma^i[A_i,\Psi]
 +m_1m_2 A_iA_i \nonumber\\
 &+\frac{2i}{3}(m_1+m_2)\epsilon_{ijk}A_iA_jA_k
 +i m_1\bar\Psi\Psi
 \bigg\}.
 \label{eq:three_matrix_action}
\end{align}
We use the Euclidean Majorana representation
\begin{equation}
 \Gamma^i=\sigma^i,
 \qquad
 C=\Gamma^2,
 \qquad
 \bar\Psi=\Psi^T C,
 \label{eq:d3_spinor_conventions}
\end{equation}
where \(\sigma^i\), \(i=1,2,3\), are the Pauli matrices, and the transpose
in \(\Psi^T\) acts only on the two-component spinor index.
We denote the normalized trace by \(\operatorname{tr}=N^{-1}\operatorname{Tr}\).

To write the loop equations compactly, we introduce a collective index
\(z\in\{1,2,3,4,5\}\) and define
\begin{equation}
\begin{aligned}
X_1 &\equiv A_1,
&
X_2 &\equiv A_2,
&
X_3 &\equiv A_3,
\\
X_4 &\equiv \Psi_1,
&
X_5 &\equiv \Psi_2.
\end{aligned}
\label{eq:collective_matrix_letters}
\end{equation}
The bosonic labels \(i\) and fermionic labels \(\alpha\) are understood as belonging to disjoint index sets. A general single-trace moment is
\begin{equation}
 w\{z_1\cdots z_n\}
 \equiv
 \left\langle
 \operatorname{tr}(X_{z_1}\cdots X_{z_n})
 \right\rangle,
 \qquad
 w\{\emptyset\}\equiv 1,
 \label{eq:graded_moments}
\end{equation}
Here $(z_1,\ldots,z_n)$ specifies the order of the matrices in the product $X_{z_1}\cdots X_{z_n}$.
Each $z_k$ labels a bosonic or a fermionic matrix.
The quantity $w\{z_1\cdots z_n\}$ is the corresponding single-trace moment.
A moment can be nonzero only if the product contains an even number of fermionic matrices.
The cyclicity of the trace gives the graded relation
\begin{equation}
 w\{z_1\cdots z_n\}
 =
 (-1)^{|X_{z_n}|
 |X_{z_1}\cdots X_{z_{n-1}}|}
 w\{z_n z_1\cdots z_{n-1}\},
 \label{eq:graded_cyclicity}
\end{equation}
where \(|{\cal O}|=0\) or \(1\) according to whether \({\cal O}\) is
Grassmann even or odd.

\subsubsection{Quadratic fermion kernel and propagators}

To perform the Grassmann integral, we expand the matrices in an orthonormal
basis \(t^a\in u(N)\),
\begin{equation}
 A_i=A_i^a t^a,
 \qquad
 \Psi_\alpha=\Psi_\alpha^a t^a,
 \qquad
 \operatorname{Tr}(t^a t^b)=\delta^{ab},
 \qquad
 [t^a,t^b]=i f^{abc}t^c.
 \label{eq:generator_conventions}
\end{equation}
The adjoint-representation matrix associated with \(A_i\) is
\begin{equation}
 A_i^{ab}
 \equiv
 \operatorname{Tr}\!\left(t^a[A_i,t^b]\right)
 =-i f^{abc}A_i^c.
 \label{eq:adjoint_A_definition}
\end{equation}
The fermionic part of the action is then the antisymmetric quadratic form
\begin{equation}
 S_{\mathrm F}[A,\Psi]
 =
 \frac{N}{2}
 \Psi_\alpha^a {\cal M}_{\alpha\beta}^{ab}[A]\Psi_\beta^b,
 \qquad
 {\cal M}_{\alpha\beta}^{ab}
 =
 -i(C\Gamma^i)_{\alpha\beta}A_i^{ab}
 +2i m_1 C_{\alpha\beta}\delta^{ab}.
 \label{eq:fermion_kernel}
\end{equation}
Thus \({\cal M}\) acts on the composite index \((\alpha,a)\) and has dimension
\(2N^2\) for \(U(N)\).
Defining the adjoint matrices
\((\mathbf A_i)^{ab}\equiv A_i^{ab}\), its spinor block form is
\begin{equation}
 {\cal M}[A]=
 \begin{pmatrix}
 -\mathbf A_1-i\mathbf A_2
     & \mathbf A_3+2m_1\mathbf{1}_{N^2}\\
 \mathbf A_3-2m_1\mathbf{1}_{N^2}
     & \mathbf A_1-i\mathbf A_2
 \end{pmatrix}.
 \label{eq:fermion_kernel_block}
\end{equation}

At fixed bosonic matrices, we define
\begin{equation}
 {\cal G}[A]\equiv {\cal M}^{-1}[A].
 \label{eq:fermion_G_definition}
\end{equation}
The fermionic two-point function is
\begin{equation}
 \left\langle
 \Psi_\alpha^a\Psi_\beta^b
 \right\rangle_{\mathrm F}
 =
 \frac{1}{N}{\cal G}_{\alpha\beta}^{ab}[A].
 \label{eq:fermion_propagator}
\end{equation}
where the normalized fermionic expectation value is defined by
\begin{equation}
\left\langle
{\cal O}[A,\Psi]
\right\rangle_{\mathrm F}
\equiv
\frac{
\displaystyle
\int d\Psi\,
{\cal O}[A,\Psi]\,
e^{-S_{\mathrm F}[A,\Psi]}
}{
\displaystyle
\int d\Psi\,
e^{-S_{\mathrm F}[A,\Psi]}
}.
\label{eq:fixed_A_fermionic_expectation}
\end{equation}

All higher fermionic correlators follow from Wick's theorem. For example,
\begin{align}
 \left\langle
 \Psi_\alpha^a\Psi_\beta^b
 \Psi_\gamma^c\Psi_\delta^d
 \right\rangle_{\mathrm F}
 =\frac{1}{N^2}\big(
 &{\cal G}_{\alpha\beta}^{ab}{\cal G}_{\gamma\delta}^{cd}
 -{\cal G}_{\alpha\gamma}^{ac}{\cal G}_{\beta\delta}^{bd}
 +{\cal G}_{\alpha\delta}^{ad}{\cal G}_{\beta\gamma}^{bc}
 \big).
 \label{eq:fermion_wick_four_point}
\end{align}

\subsubsection{Large-\(N\) loop equations}
\label{subsec:three_matrix_loops}

The loop equations follow from infinitesimal changes of variables in the matrix integral. At large \(N\), products of single-trace observables factorize, while the joining terms are suppressed by \(1/N^2\). 
We formulate the loop equations for the single-trace moments
\begin{equation}
w\{z_1\cdots z_n\}
=
\left\langle
\operatorname{tr}
\left(
X_{z_1}\cdots X_{z_n}
\right)
\right\rangle, 
\end{equation}
and take as our starting point the trivial identity that the matrix integral under an
infinitesimal change of integration variables is invariant
\begin{equation}
\int dX\,
\operatorname{tr}
\left(
t^aX_{z_1}\cdots X_{z_n}
\right)
e^{-S[X]}
=
\int dX'\,
\operatorname{tr}
\left(
t^aX'_{z_1}\cdots X'_{z_n}
\right)
e^{-S[X']},
\label{eq:loop_equation_starting_identity}
\end{equation}
where \(dX\equiv dA\,d\Psi\) collectively denotes integration over the bosonic and fermionic matrix components.
The primed matrices are obtained by varying one of the five matrices while keeping the remaining four fixed:
\begin{equation}
\begin{aligned}
A_i
&\longrightarrow
A_i'
=
A_i+\delta A_i,
&
\delta A_i
&\equiv
\epsilon_B t^a,
\\
\Psi_\alpha
&\longrightarrow
\Psi_\alpha'
=
\Psi_\alpha+\delta\Psi_\alpha,
&
\delta\Psi_\alpha
&\equiv
\epsilon_F t^a.
\end{aligned}
\label{eq:bosonic_fermionic_variations}
\end{equation}
Here \(\epsilon_B\) is a Grassmann-even infinitesimal parameter, whereas
\(\epsilon_F\) is Grassmann odd. Expanding
Eq.~\eqref{eq:loop_equation_starting_identity} to first order in the corresponding infinitesimal parameter gives the Schwinger--Dyson identity.
Equivalently, the same identity is obtained by integrating a total derivative with respect to the varied matrix component.

For the bosonic variation of \(A_i\), a nontrivial equation is obtained when
\(X_{z_1}\cdots X_{z_n}\) contains an even number of fermionic matrices.
After taking the large-\(N\) limit, applying large-\(N\) factorization, and
dropping the joining terms suppressed by \(1/N^2\), the resulting bosonic loop equation is
\begin{align}
0={}&
 \sum_{k=1}^{n}
 \delta_{z_k,i}\,
 w\{z_1\cdots z_{k-1}\}\,
 w\{z_{k+1}\cdots z_n\}
 -w\{z_1\cdots z_n[j,[j,i]]\}
 \nonumber\\
&-\frac{i}{2}(C\Gamma^i)_{\alpha\beta}
 w\{z_1\cdots z_n\{\alpha,\beta\}\}
 -2m_1m_2\,w\{z_1\cdots z_n i\}
 \nonumber\\
&-i(m_1+m_2)\epsilon_{ijk}
 w\{z_1\cdots z_n[j,k]\} .
\label{eq:bosonic_loop_equation}
\end{align}
The notation for the inserted operators is, for example,
\begin{align}
 w\{z_1\cdots z_n[j,[j,i]]\}
 &\equiv
 \left\langle
 \operatorname{tr}\!\left(
 X_{z_1}\cdots X_{z_n}[A_j,[A_j,A_i]]
 \right)
 \right\rangle,
 \nonumber\\
 w\{z_1\cdots z_n\{\alpha,\beta\}\}
 &\equiv
 \left\langle
 \operatorname{tr}\!\left(
 X_{z_1}\cdots X_{z_n}
 \{\Psi_\alpha,\Psi_\beta\}
 \right)
 \right\rangle,
 \label{eq:inserted_moment_notation_bosonic}
\end{align}
with analogous definitions for
\(w\{z_1\cdots z_n i\}\) and \(w\{z_1\cdots z_n[j,k]\}\). If one of the two factors in the splitting term contains an odd number of fermions, that factor vanishes.

For the fermionic variation of \(\Psi_\alpha\), a nontrivial equation is instead obtained when \(X_{z_1}\cdots X_{z_n}\) contains an odd number of fermionic matrices.
Since both $\epsilon_F$ and the product $X_{z_1}\cdots X_{z_n}$ are then Grassmann odd, the coefficient of $\epsilon_F$ in the varied integrand is Grassmann even.
The corresponding fermionic loop equation is
\begin{align}
0={}&
 \sum_{k=1}^{n}
 (-1)^{|X_{z_1}\cdots X_{z_{k-1}}|}
 \delta_{z_k,\alpha}\,
 w\{z_1\cdots z_{k-1}\}\,
 w\{z_{k+1}\cdots z_n\}
 \nonumber\\
&-i(C\Gamma^i)_{\alpha\beta}
 w\{z_1\cdots z_n[i,\beta]\}
 +2i m_1 C_{\alpha\beta}
 w\{z_1\cdots z_n\beta\} .
\label{eq:fermionic_loop_equation}
\end{align}
Here
\begin{align}
 w\{z_1\cdots z_n[i,\beta]\}
 &\equiv
 \left\langle
 \operatorname{tr}\!\left(
 X_{z_1}\cdots X_{z_n}[A_i,\Psi_\beta]
 \right)
 \right\rangle,
 \nonumber\\
 w\{z_1\cdots z_n\beta\}
 &\equiv
 \left\langle
 \operatorname{tr}\!\left(
 X_{z_1}\cdots X_{z_n}\Psi_\beta
 \right)
 \right\rangle.
 \label{eq:inserted_moment_notation_fermionic}
\end{align}
The sign in the splitting term is determined by the number of fermionic matrices preceding the varied matrix $\Psi_\alpha$ in the ordered product.

We have now obtained all the loop equations that are needed.
The type of loop equation is specified by the varied matrix.
Varying $A_i$ gives a bosonic loop equation, while varying $\Psi_\alpha$ gives a fermionic loop equation.
A nontrivial bosonic loop equation requires the inserted product
$X_{z_1}\cdots X_{z_n}$ to contain an even number of fermionic
matrices.
A nontrivial fermionic loop equation instead requires an odd number of fermionic matrices.

More precisely, the loop equations are organized by the number $n=0,1,\ldots$ of matrices in the inserted product $X_{z_1}\cdots X_{z_n}$.
For each $n$, the $5^n$ index sequences $(z_1,\ldots,z_n)$,
with $z_k\in\{1,\ldots,5\}$, are enumerated in lexicographic order.
Products containing an even number of fermionic matrices give the three bosonic loop equations, while products containing an odd number give the two fermionic loop equations.
The equations are ordered first by $n$, then by the index
sequence, and finally by the variation index $i$ or $\alpha$.

For the numerical implementation, we select equations from this
ordered list according to the restrictions described in
Sec.~\ref{sec:three_matrix}.
We denote the number of retained equations by $X$ and their
set of labels by $S_X$, consistently with the objective function
in Eq.~\eqref{eq:objectivefunction}.

\subsubsection{Integrating out the fermions}

We now apply the fermion integration prescription of Sec.~\ref{sec:fermionic-integration} to the loop equations derived above.
For each observable ${\cal O}[A,\Psi]$ containing fermionic matrices,
we construct the bosonic functional ${\cal O}_{\mathrm B}[A]$ defined
in Eq.~\eqref{eq:fermion-integrated-observable} by Wick contraction at
fixed bosonic matrices.

Consider the bosonic loop equation obtained by varying $A_i$ in
the matrix integral with the insertion $\operatorname{tr}(t^a A_{i_1}\cdots A_{i_n})$.
The variation acts both on this insertion and on the action.
Although this insertion contains only bosonic matrices, varying the fermionic part of the action produces the bilinear contribution in Eq.~\eqref{eq:bosonic_loop_equation}, whose fermion-integrated form we now evaluate explicitly.

Since the action contains fermionic matrices, the observable appearing in this term is
\begin{equation}
{\cal O}[A,\Psi]
=
\operatorname{tr}\!\left(
A_{i_1}\cdots A_{i_n}
\{\Psi_\alpha,\Psi_\beta\}
\right)
\label{eq:three_matrix_fermionic_observable}
\end{equation}
arising from the variation of the action.

Expanding the fermionic matrices as $\Psi_\alpha=\Psi_\alpha^a t^a$ and using the propagator in Eq.~\eqref{eq:fermion_propagator}, we obtain
\begin{equation}
\begin{aligned}
\left\langle{\cal O}[A,\Psi]\right\rangle_{\mathrm F}
&=
\frac{1}{N}
\operatorname{Tr}\!\left(
A_{i_1}\cdots A_{i_n}t^a t^b
\right)
\left(
\langle\Psi_\alpha^a\Psi_\beta^b\rangle_{\mathrm F}
+
\langle\Psi_\beta^a\Psi_\alpha^b\rangle_{\mathrm F}
\right)
\\
&=
\frac{1}{N^2}
\left(
{\cal G}_{\alpha\beta}^{ab}[A]
+
{\cal G}_{\beta\alpha}^{ab}[A]
\right)
\operatorname{Tr}\!\left(
A_{i_1}\cdots A_{i_n}t^a t^b
\right).
\end{aligned}
\label{eq:three_matrix_OB_example}
\end{equation}
Here repeated adjoint indices are summed. The spinor indices $\alpha,\beta$ are held fixed in this expression and are contracted with $(C\Gamma^i)_{\alpha\beta}$ when the result is inserted into Eq.~\eqref{eq:bosonic_loop_equation}.
One factor of $1/N$ comes from the normalized trace and the other from the fermionic propagator.

Equation~\eqref{eq:three_matrix_OB_example} explicitly evaluates the normalized fermionic expectation $\langle{\cal O}[A,\Psi]\rangle_{\mathrm F}$ defined in Eq.~\eqref{eq:fixed_A_fermionic_expectation}.
The bosonic matrices are held fixed, and Wick contraction replaces the pair of Grassmann coefficients by the fermionic propagator.
The result therefore depends only on the bosonic matrices through the explicit matrix product and the inverse fermion kernel.
The full expectation value is subsequently obtained by averaging this result with the effective bosonic weight as in Eq.~\eqref{eq:fermion-integrated-observable}.

For each trial configuration $\hat A_i^{(M)}$, we construct the finite-dimensional fermion kernel and its inverse ${\cal G}^{(M)}[\hat A^{(M)}]$.
The corresponding finite-dimensional functional is
\begin{equation}
\begin{aligned}
\left\langle
{\cal O}[\hat A^{(M)},\Psi]
\right\rangle_{\mathrm F}^{(M)}
={}&
\frac{1}{M^2}
\left(
{\cal G}_{\alpha\beta}^{(M)\,ab}[\hat A^{(M)}]
+
{\cal G}_{\beta\alpha}^{(M)\,ab}[\hat A^{(M)}]
\right)
\\
&\times
\operatorname{Tr}\!\left(
\hat A_{i_1}^{(M)}\cdots
\hat A_{i_n}^{(M)}t^a t^b
\right),
\end{aligned}
\label{eq:three_matrix_OB_regularized}
\end{equation}
The resulting bosonic functional is evaluated together with the
other terms in the loop equation. 
This example shows explicitly how the fermionic contribution is retained after the fermionic matrices have been integrated out.
Its evaluation therefore requires only the bosonic matrices and the inverse fermion kernel constructed from them.

\section{Three-Matrix Model}\label{sec:three_matrix}

We now present numerical results for the three-matrix model, using the setup and loop equations developed in Sec.~\ref{subsec:three_matrix_setup}.

Under the setup introduced in Sec.~\ref{subsec:three_matrix_setup}, we performed numerical calculations using the regularized master-field approximation. The regularized master field is represented by three complex $M\times M$ matrices. We fix the gauge by taking $\widehat A_1^{(M)}$ to be diagonal and complex, while $\widehat A_2^{(M)}$ and $\widehat A_3^{(M)}$ are taken to be generic complex matrices. The number of independent real optimization variables is \(4M^2+2\) after taking into account the residual gauge redundancy after diagonalizing $\widehat A_1^{(M)}$.

Diagonalizing $\widehat A_1^{(M)}$ leaves $M$ complex diagonal entries and $2M^2$ complex entries in the other two matrices, giving $4M^2+2M$ real parameters. For distinct eigenvalues of $\widehat A_1^{(M)}$, the remaining continuous redundancy consists of simultaneous diagonal complex similarity transformations, $\widehat A_i^{(M)}\mapsto D\widehat A_i^{(M)}D^{-1}$ with $D=\operatorname{diag}(d_1,\ldots,d_M)$ and $d_a\neq0$. An overall rescaling of all $d_a$ acts trivially, so this residual symmetry has $M-1$ independent complex parameters, or $2(M-1)$ real parameters. Removing this redundancy therefore leaves $4M^2+2M-2(M-1)=4M^2+2$ independent real parameters for a generic configuration.

The minimization of the objective function defined in \eqref{eq:objectivefunction} is performed using the \texttt{least\_squares} function in \texttt{scipy.optimize} with the \texttt{trf} algorithm. We use $\texttt{ftol}=\texttt{xtol}=\texttt{gtol}=10^{-14}$ and $\texttt{max\_nfev}=1000$.

The loop equations are generated in increasing word length and, within each word length, in lexicographic order. For a retained loop equation $\mathcal L^{(M)}_\alpha$ generated from a word of length $n_\alpha$, we assign the weight
\begin{equation}
r_\alpha
=
\frac{1}{N_{n_\alpha}},
\end{equation}
where $N_n$ denotes the number of retained loop equations generated from words of length $n$. This normalization prevents a sector containing a larger number of equations from dominating the objective function solely because of its multiplicity.

In the numerical implementation, we retain loop equations obtained by varying the bosonic matrices $A_i$ with insertions $\operatorname{tr}(t^a A_{i_1}\cdots A_{i_n})$, as in Eq.~\eqref{eq:loop_equation_starting_identity}. The matrix products in these insertions contain only bosonic matrices. However, the variation also acts on the fermionic part of the action and therefore produces fermion bilinears. Their contributions are retained by evaluating the fixed-background fermionic expectations through Wick contraction, as illustrated in Eq.~\eqref{eq:three_matrix_OB_example}.

We also examined loop equations obtained from fermionic variations with inserted products containing one fermionic matrix. Numerically, these equations were found to be satisfied almost automatically once the fermionic expectations were evaluated using the inverse fermion kernel. This is consistent with the fact that, for a quadratic fermionic action, the fermionic Schwinger--Dyson equations express the Gaussian Wick-contraction identities. They therefore do not provide additional independent constraints in this implementation and are not included in the RMA objective function.

To assess the number of locally independent constraints, we compute the rank of the real Jacobian formed by differentiating the real and imaginary parts of the retained loop-equation residuals with respect to the independent real parameters of the trial matrices. This rank is compared with the number of optimization variables in the numerical results below.

We use both fuzzy sphere and random initial configurations for the three matrices.
Fuzzy-sphere configurations are constructed from the generators of the $M$-dimensional irreducible representation of $SU(2)$, and the radius is set by $\textrm{max}(\widetilde m_1,\widetilde m_2)$ with appropriate normalization of $M$ discussed below.
For the random configuration, the real and imaginary parts of the matrix entries are sampled independently from a Gaussian distribution with zero mean and standard deviation $10^{-3}$.
Our results below were obtained using the fuzzy sphere initial condition.
We also investigated random initial configurations, but these did not lead to fuzzy-sphere configurations in the runs studied here.

To compare configurations obtained at different regulator sizes \(M\), it is useful first to make explicit the normalization of the matrices and moments used in the numerical analysis. The classical bosonic equations of motion of this model admit fuzzy-sphere configurations, which provide natural reference saddles for the numerical solutions. We therefore use their \(M\)-dependence to motivate the normalization of the regularized matrices and moments, without yet relying on the detailed structure of these solutions.

For an \(M\)-dimensional irreducible representation of \(\mathfrak{su}(2)\), the generators satisfy
\begin{equation}
    \sum_{i=1}^{3} L_i^2
    =
    \frac{M^2-1}{4}\,\mathbf{1}_M ,
\end{equation}
so that the generators scale as \(L_i=\mathcal{O}(M)\) at large \(M\). A fuzzy-sphere configuration has the form
\begin{equation}
    \widehat A_i^{(M)} \sim m L_i .
\end{equation}

The mass scaling is chosen so that the bosonic action evaluated on a generic fuzzy-sphere saddle is of order $N^2$, as derived in Appendix  \ref{app:classical-saddles}. We use the corresponding scaling for the $M$-dimensional regularized configurations, 
\begin{equation}
    m_a=\frac{\widetilde m_a}{\sqrt{M}},
\label{eq:mass_rescale}
\end{equation}
where $a=1,2$ and the rescaled masses $\widetilde m_a$ are of order $\mathcal O(M^0)$ and therefore are held fixed as $M$ is varied. With this choice of mass rescaling, such configurations therefore satisfy
\begin{equation}
    \widehat A_i^{(M)}
    \sim
    \frac{1}{\sqrt{M}}\,\mathcal{O}(M)
    =
    \mathcal{O}(\sqrt{M}) .
\end{equation}

In terms of these variables, we define the rescaled single-trace moments by
\begin{equation}
    \bar w_{\{i_1\cdots i_n\}}^{(M)}
    =
    M^{-n/2}\frac{1}{M}
    \mathrm{Tr}\!\left[
        \widehat A_{i_1}^{(M)}
        \cdots
        \widehat A_{i_n}^{(M)}
    \right].
\end{equation}
This normalization removes the explicit growth with \(M\) expected from the fuzzy-sphere scaling and gives moments that can be compared directly across different regulator sizes.

A particularly useful quadratic observable is the sum of the three diagonal second moments,
\begin{equation}
    C_M
    \equiv
    \bar w_{\{11\}}^{(M)}
    +
    \bar w_{\{22\}}^{(M)}
    +
    \bar w_{\{33\}}^{(M)} .
\end{equation}
Using the definition above, this can be written as
\begin{equation}
    C_M
    =
    \frac{1}{M^2}
    \mathrm{Tr}\!\left[
        \sum_{i=1}^{3}
        \left(\widehat A_i^{(M)}\right)^2
    \right].
\label{eq:CM}
\end{equation}
Thus \(C_M\) measures the overall quadratic size of the three-matrix configuration in the same large-\(M\) normalization used for the moments. It is also invariant under rotations among the three bosonic matrices, making it a natural quantity for comparing the regularized configurations as the mass parameters are varied.

For the fuzzy-sphere configurations introduced above, \(C_M\) takes a simple analytic form, which provides a useful reference for the numerical RMA results. Substituting the fuzzy-sphere ansatz into the bosonic equations of motion gives
\begin{equation}
   m(m-m_1)(m-m_2)=0 .
\end{equation}
Thus, in addition to the trivial configuration \(m=0\), the classical equations admit fuzzy-sphere saddles at
\begin{equation}
    m=m_1,
    \qquad
    m=m_2 .
\end{equation}
There are therefore two fuzzy-sphere branches: for positive masses with $m_1>m_2$, the branch with radius parameter $m=m_1$ is radially stable, whereas for $m_2>m_1$, the branch with $m=m_2$ is radially stable. The special points $m_1=m_2$ and $m_2=2m_1$ correspond to coalescence of the two branches and equality of the fuzzy-sphere and trivial saddles respectively, as detailed in Appendix~\ref{app:classical-saddles}. Guided by the classical saddle structure, we examine the two mass regions $\widetilde m_2>\widetilde m_1$ and $\widetilde m_1>\widetilde m_2$. In each region, we compare the RMA results with the corresponding classical fuzzy-sphere.

We define the classical fuzzy-sphere Casimir in the large-$N$ theory before introducing the RMA regulator. On either nontrivial branch, $A_i^{\mathrm{FS}}=m_aL_i$, where $a=1,2$ and $L_i$ are the generators of the $N$-dimensional irreducible representation of $SU(2)$. Taking $m_a=\widetilde m_a/\sqrt{N}$ with $\widetilde m_a$ held fixed, we define the normalized large-$N$ Casimir by
\begin{equation}
C^{\mathrm{FS}}
\equiv
\lim_{N\to\infty}
\frac{1}{N^2}
\operatorname{Tr}\!\left[
\sum_{i=1}^{3}(A_i^{\mathrm{FS}})^2
\right]
=
\lim_{N\to\infty}
\frac{\widetilde m_a^{\,2}}{4}
\left(1-\frac{1}{N^2}\right)
=
\frac{\widetilde m_a^{\,2}}{4},
\qquad a=1,2.
\label{eq:classical_C}
\end{equation}
Here $C^{\mathrm{FS}}$ denotes the value on the branch specified by $a$. It is independent of the regulator dimension $M$ and provides the classical large-$N$ reference for the numerical observable $C_M$.

\begin{figure}[htbp]
\centering
\includegraphics[width=0.85\textwidth]{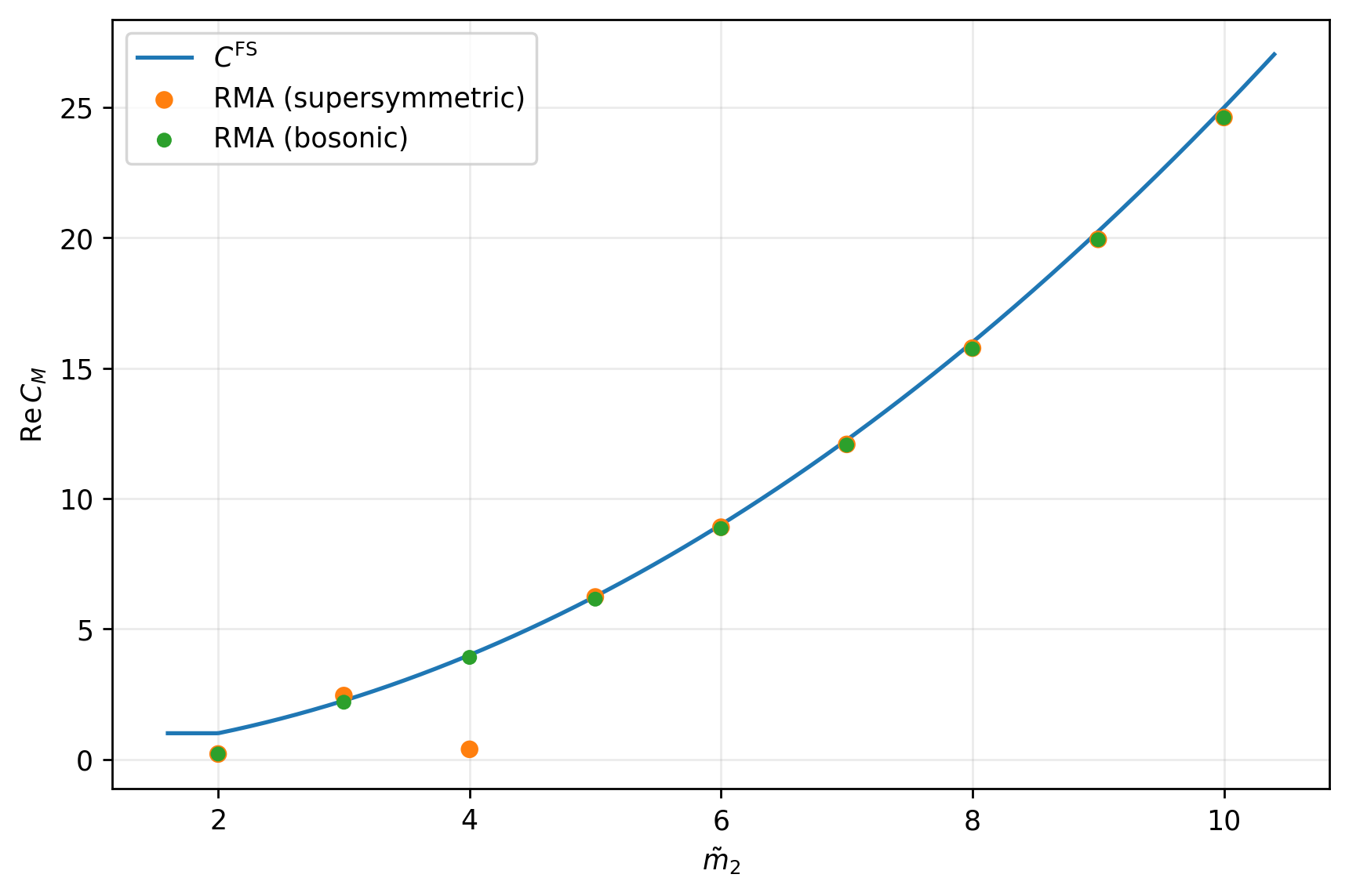}
\caption{Comparison of $C_M$ with $C^{\mathrm{FS}}$ for $M=8$ and $X=145$. $\widetilde m_2=2,3,\ldots,10$ at fixed $\widetilde m_1=2$. Orange and green dots represent $\operatorname{Re}C_M$ for the supersymmetric and bosonic models, respectively, while the solid blue line represents $C^{\mathrm{FS}}$.}
\label{fig:CM_m2_scan}
\end{figure}

Figure~\ref{fig:CM_m2_scan} compares the supersymmetric and bosonic RMA results at $M=8$ and $X=145$, comparing to $C^{\mathrm{FS}}=\max(\widetilde m_1,\widetilde m_2)^2/4$. Here, $\widetilde m_1=2$ is fixed and $\widetilde m_2>\tilde m_1$. At the sampled masses $\widetilde m_2>4$, both sets of results approximately follow the classical curve. The supersymmetric result is instead strongly suppressed at $\widetilde m_2=4$, while the solution at $\widetilde m_2=3$ remains close to the reference and approximately isotropic.

\begin{table}[t]
\centering
\begin{tabular}{c|cccc}
\hline
$\widetilde m_2$
& $\bar w^{(8)}_{\{11\}}$
& $\bar w^{(8)}_{\{22\}}$
& $\bar w^{(8)}_{\{33\}}$
& $\Delta_{SO(3)}$ \\
\hline
18 & $26.5791$ & $26.5794$ & $26.5792$
   & $6.08\times10^{-6}$ \\
16 & $21.0012$ & $21.0013+0.0001i$ & $21.0014$
   & $6.66\times10^{-6}$ \\
14 & $16.0796+0.0001i$ & $16.0795+0.0001i$ & $16.0795$
   & $1.95\times10^{-5}$ \\
12 & $11.8140$ & $11.8142$ & $11.8136+0.0001i$
   & $2.81\times10^{-5}$ \\
10 & $8.2059$ & $8.2065+0.0002i$ & $8.2065-0.0002i$
   & $7.27\times10^{-5}$ \\
9 & $6.6480$ & $6.6483+0.0001i$ & $6.6482+0.0001i$
  & $4.10\times10^{-5}$ \\
8 & $5.2563+0.0001i$ & $5.2579-0.0002i$
  & $5.2569-0.0002i$ & $1.45\times10^{-4}$ \\
7 & $4.0262$ & $4.0272$ & $4.0264$
  & $1.13\times10^{-4}$ \\
6 & $2.9670-0.0008i$ & $2.9673-0.0003i$
  & $2.9683-0.0004i$ & $4.06\times10^{-4}$ \\
5 & $2.0755-0.0006i$ & $2.0760-0.0004i$
  & $2.0762-0.0012i$ & $4.55\times10^{-4}$ \\
4 & $0.4549-0.0058i$ & $-0.0914-0.0192i$
  & $0.0205+0.0083i$ & $0.89731$ \\
3 & $0.8130+0.0002i$ & $0.8176+0.0003i$
  & $0.8145-0.0006i$ & $0.00287$ \\
2 & $0.0714+0.0051i$ & $0.0635-0.0005i$ & $0.0732+0.0152i$
  & $0.35244$ \\
\hline
\end{tabular}
\caption{Quadratic moments and their deviation from $SO(3)$ isotropy for the three-matrix RMA solutions at fixed $\widetilde m_1=2$, $M=8$, and $X=145$, as $\widetilde m_2$ is varied. The quantity $\Delta_{SO(3)}$ includes both diagonal and off-diagonal quadratic moments and is evaluated using the unrounded data. The displayed moments are rounded to four decimal places, with imaginary parts that round to zero omitted. }
\label{tab:three_matrix_mass_scan_M8}
\end{table}

To quantify deviations from $SO(3)$ symmetry in the quadratic moments, we define $\bar T_{ij}\equiv\bar w_{\{ij\}}^{(M)}$ for $i,j=1,2,3$ and compare it with the isotropic form $(C_M/3)\delta_{ij}$, where $C_M=\sum_i\bar T_{ii}$. We introduce the following measure
\begin{equation}
\Delta_{SO(3)}
\equiv
\sqrt{
\frac{
\displaystyle\sum_{i,j=1}^{3}
\left|\bar T_{ij}-\frac{C_M}{3}\delta_{ij}\right|^2
}{
\displaystyle\sum_{i,j=1}^{3}|\bar T_{ij}|^2
}
}.
\label{eq:SO3_deviation}
\end{equation}
For nonzero $\bar T$, this quantity vanishes when the quadratic moments are isotropic and measures their relative deviation from isotropy, including both diagonal and off-diagonal components. Note that this quantity itself is \(\textrm{SO}(3)\) invariant.

Table~\ref{tab:three_matrix_mass_scan_M8} presents the quadratic moments at fixed $\widetilde m_1=2$, $M=8$, and $X=145$, with $\widetilde m_2$ decreasing from $18$ to $2$. For $\widetilde m_2\geq6$, the three diagonal moments are nearly equal, and $\Delta_{SO(3)}$ remains below $5\times10^{-4}$, indicating approximate $SO(3)$ isotropy at the level of the quadratic moments.

At $\widetilde m_2=4$ and $2$, the quadratic moments become anisotropic, with $\Delta_{SO(3)}$ increasing to approximately $0.90$ and $0.35$, respectively. These departures occur at $\widetilde m_1=\widetilde m_2=2$ and $\widetilde m_2=2\widetilde m_1=4$, where the exact fuzzy-sphere backgrounds are unstable due to the zero modes of the Dirac operator, as discussed below.

A stronger characterization of the fuzzy-sphere configuration is obtained at the matrix level. Defining the Casimir matrix of the rescaled configuration by
\begin{equation}
    \mathcal{C}_M
    =\frac{1}{M}\sum_i\left(\widehat A_i^{(M)}\right)^2.
\end{equation}
Here $\mathcal C_M$ is an $M\times M$ matrix, whereas $C_M=M^{-1}\operatorname{Tr}\mathcal C_M$ is its normalized trace and is a scalar. For an irreducible fuzzy-sphere configuration, $\mathcal C_M$ is proportional to the identity. Checking this proportionality therefore provides an additional diagnostic of the fuzzy-sphere structure beyond comparing $C_M$ with $C^{\textrm{FS}}$.

\begin{figure}[t]
    \centering
    \includegraphics[width=\textwidth]{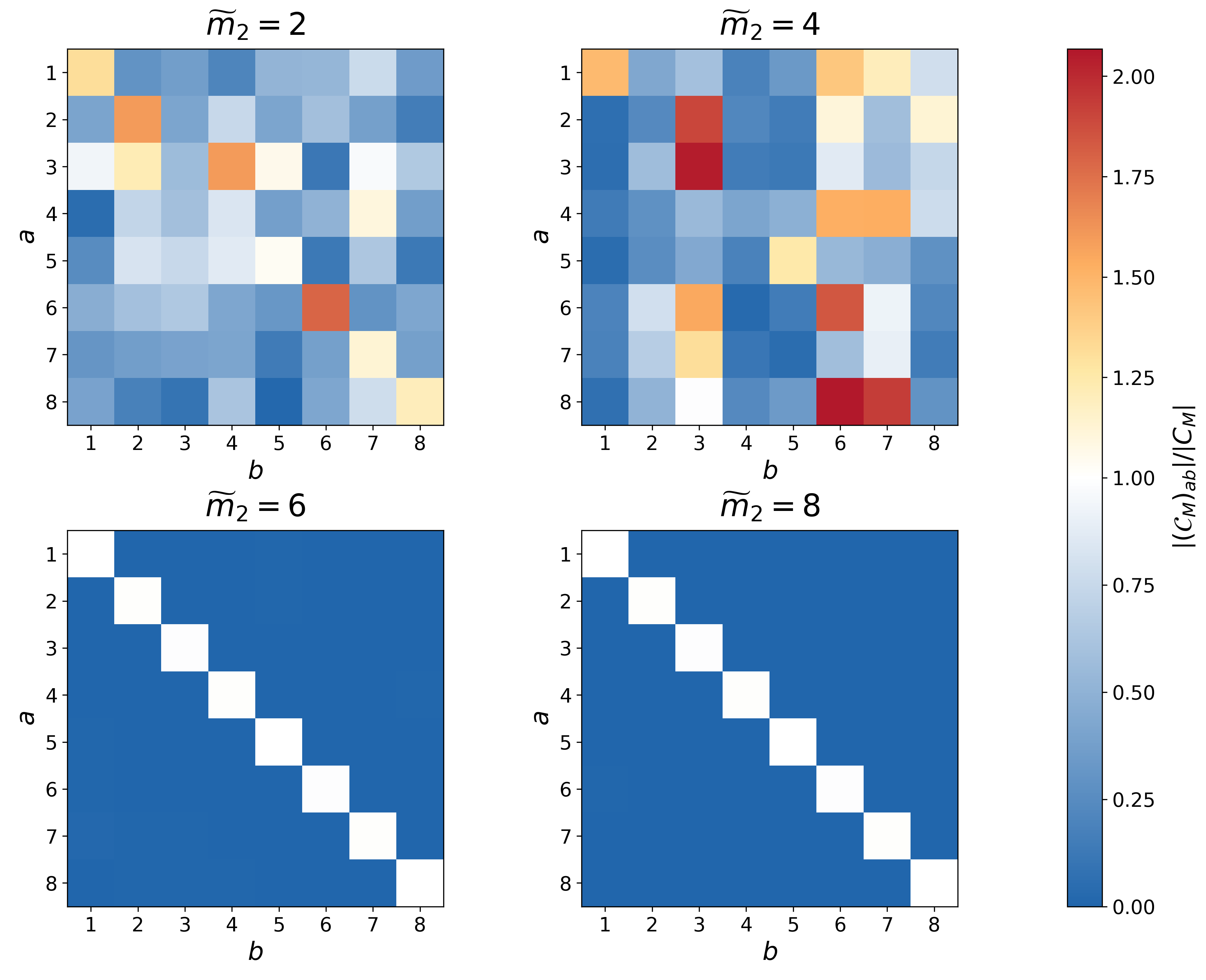}
    \caption{Heatmaps of the normalized Casimir matrix $|\mathcal{C}_{M,ab}/C_M|$ for $M=8$, $X=145$, and fixed $\tilde m_1=2$. The upper left, upper right, lower left, and lower right panels correspond to $\tilde m_2=2,4,6,8$, respectively. }
    \label{fig:casimir_3matrix}
\end{figure}

Figure~\ref{fig:casimir_3matrix} shows $|({\cal C}_{M})_{ab}|/|C_{M}|$ for $M=8$ and $X=145$, with $\tilde m_1=2$ fixed and $\tilde m_2=2,4,6,8$. At $\tilde m_2=6,8$, the normalized matrix is approximately equal to the identity matrix: the diagonal entries are close to unity, while the off-diagonal entries are strongly suppressed. This is consistent with an irreducible fuzzy-sphere configuration.

At $\tilde m_2=2,4$, the configuration appears to be random-like and the fuzzy-sphere structure is lost. This change suggests phase-transition-like behavior associated with the loss of the fuzzy-sphere structure as $\tilde m_2$ decreases.

\begin{figure}[htbp]
\centering
\includegraphics[width=0.85\textwidth]{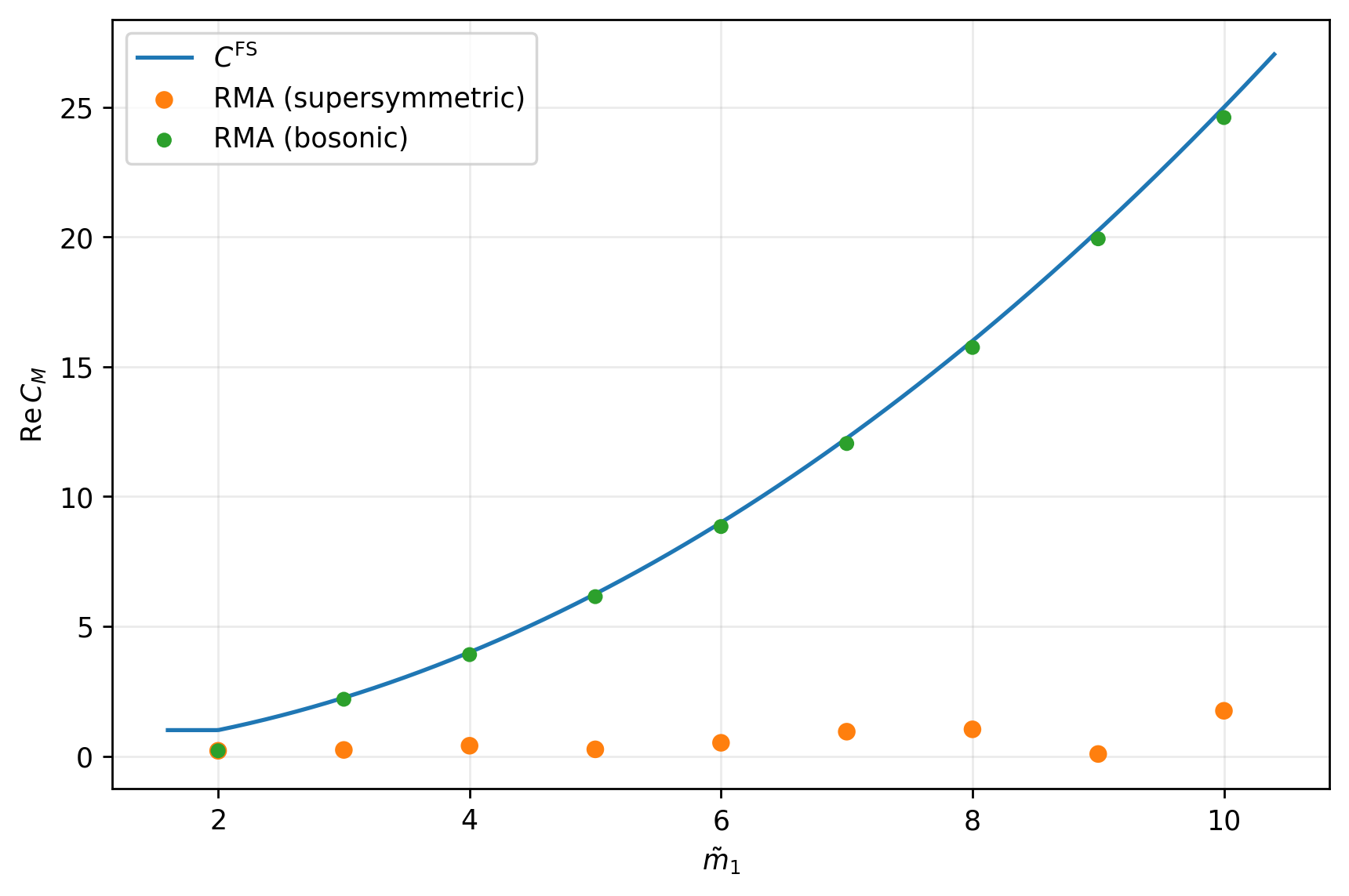}
\caption{Comparison of $C_M$ with $C^{\mathrm{FS}}$ for $M=8$ and $X=145$. $\widetilde m_1=2,3,\ldots,10$ at fixed $\widetilde m_2=2$. Orange and green dots represent $\operatorname{Re}C_M$ for the supersymmetric and bosonic models, respectively, while the solid blue line represents $C^{\mathrm{FS}}$.}
\label{fig:CM_m1_scan}
\end{figure}

For comparison, Figure~\ref{fig:CM_m1_scan} shows the scan in which $\widetilde m_2=2$ is fixed and $\widetilde m_1> \tilde m_2$. For $\widetilde m_1>\widetilde m_2$, the supersymmetric results depart strongly from the reference curve, whereas the bosonic results closely follow it. Both scans also show a departure at $\widetilde m_1=\widetilde m_2$.

The departures in both mass scans can be understood in terms of the zero modes of the Dirac operator (or, equivalently, of the fermion kernel) on the exact fuzzy-sphere background.\footnote{
In general, the Dirac operator ${\cal D}$ is the operator acting on a fermion field $\psi$ in a bilinear of the form $\bar{\psi}{\cal D}\psi$.
In the matrix models considered here, the fermionic action takes the form $\bar{\Psi}{\cal D}\Psi$, so the fermion kernel ${\cal M}$ is related to the Dirac operator by ${\cal M}=C{\cal D}$, where $C$ is the charge-conjugation matrix.
Since $C$ is invertible, ${\cal D}$ and ${\cal M}$ have identical zero modes.
In the following, we use the term Dirac operator when discussing zero modes.
}

From Eq.~\eqref{eq:fermion_kernel}, the Dirac operator acting on a matrix-valued spinor $\Psi$ is
\begin{equation}
\mathcal D\Psi
=-i\sum_{i=1}^{3}\sigma^i[\widehat A_i^{(M)},\Psi]
+2im_1\Psi,
\end{equation}
On the exact fuzzy-sphere background $\widehat A_i^{(M)}=aL_i$, with $a=\max(m_1,m_2)$, it becomes
\begin{equation}
\mathcal D\Psi
=-i\left(a\sum_{i=1}^{3}\sigma^i[L_i,\Psi]-2m_1\Psi\right).
\label{eq:three_matrix_Dirac_FS}
\end{equation}
The adjoint action of $L_i$ on matrices decomposes into angular-momentum sectors $l=0,\ldots,M-1$. Coupling each sector to the spin $1/2$ carried by the fermion gives total angular momentum $l+\tfrac12$ and, for $l\geq1$, $l-\tfrac12$. The operator $\sum_i\sigma^i[L_i,\cdot]$ has eigenvalues $l$ and $-(l+1)$ in these sectors, respectively. The corresponding Dirac eigenvalues are therefore $-i(al-2m_1)$, with multiplicity $2l+2$, and $i[a(l+1)+2m_1]$, with multiplicity $2l$.

For positive masses, only the first eigenvalue can vanish, giving
\begin{equation}
al=2m_1,\qquad l=1,\ldots,M-1.
\end{equation}
At $m_1=m_2>0$, we have $a=m_1$, so $l=2$ and there are six zero modes for $M\geq3$. These modes persist throughout $m_1>m_2$, where $a=m_1$ as well. For $m_2>m_1$, we instead have $a=m_2$, and the only possible integer solution is $l=1$ at $m_2=2m_1$, with four zero modes. These conditions depend only on mass ratios and therefore also hold for the rescaled masses.

In Figure~\ref{fig:CM_m1_scan}, the exact fuzzy sphere therefore has fermion zero modes throughout the region $\widetilde m_1\geq2$. In Figure~\ref{fig:CM_m2_scan}, they occur at $\widetilde m_2=2$ and $4$, corresponding to $l=2$ and $l=1$, respectively. At these configurations, the inverse fermion kernel used in the Wick contractions is singular. Even near the exact sphere, the propagator can become large, making the numerical optimization sensitive to small changes in the matrices. This suggests that the departures of the supersymmetric results from the classical curve are related to these zero modes. The bosonic calculation does not involve the fermion propagator and has no such singularity.

\begin{figure}[p]
\centering
\includegraphics[width=\textwidth,height=0.48\textheight,keepaspectratio]{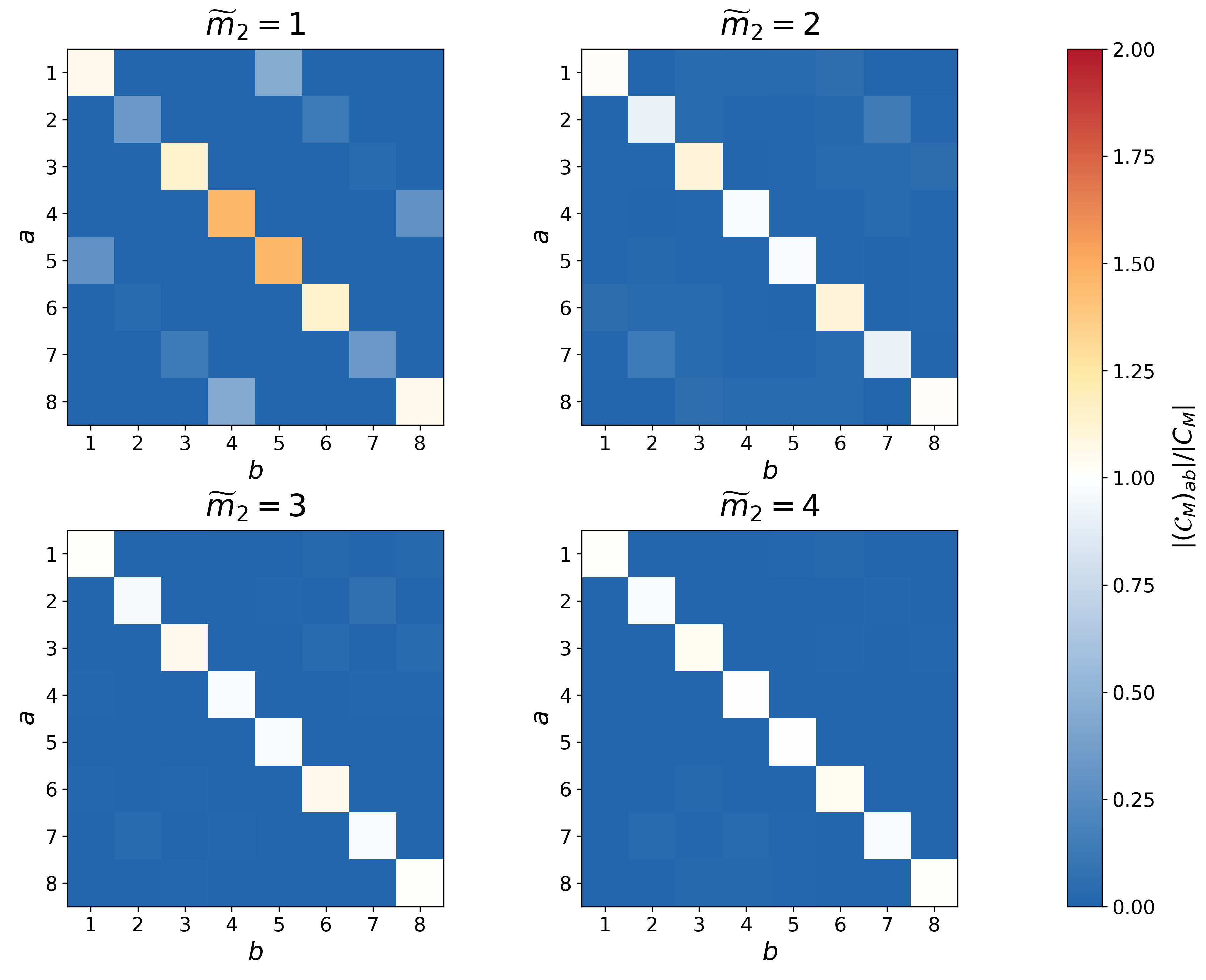}

\medskip
\includegraphics[width=\textwidth,height=0.48\textheight,keepaspectratio]{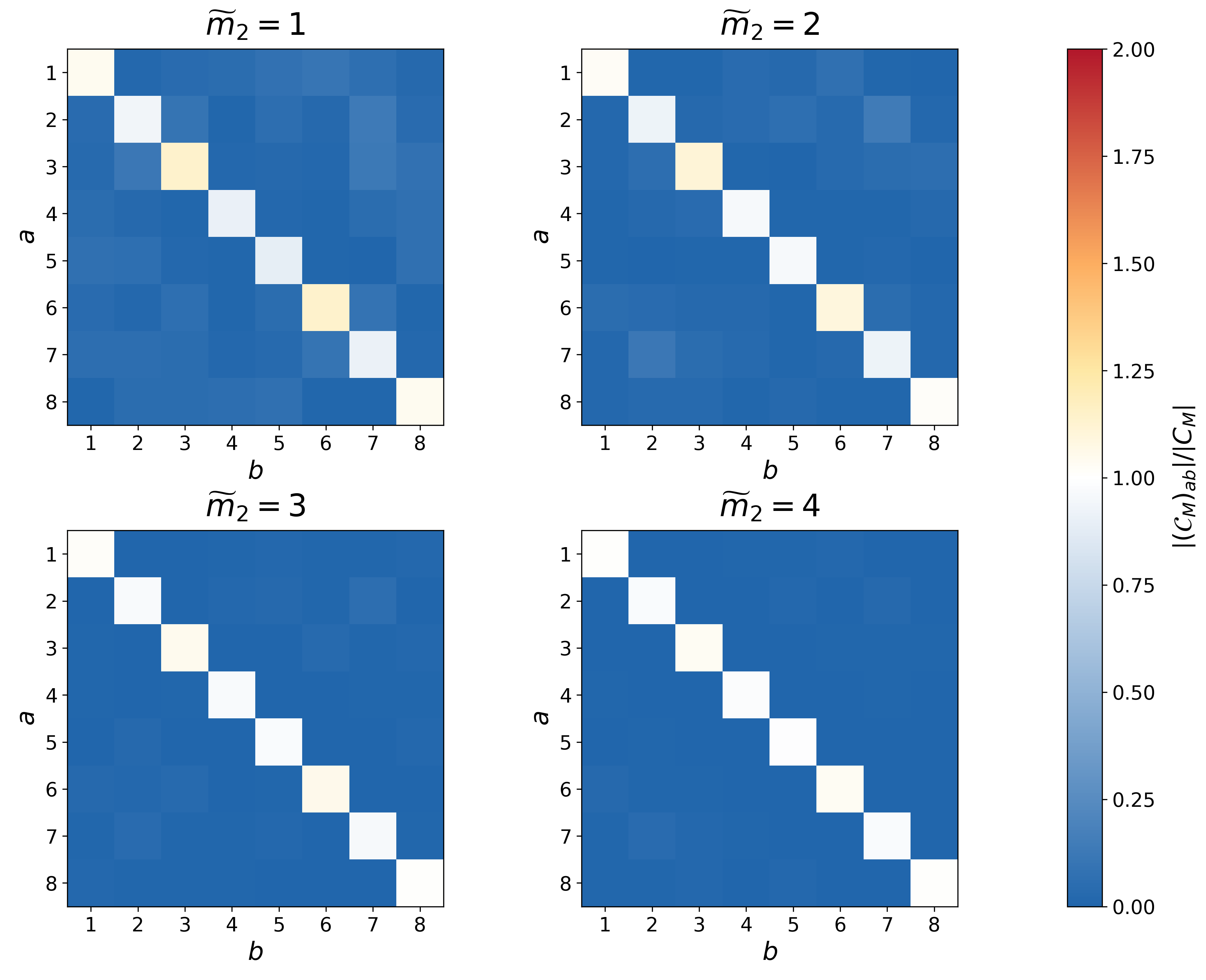}
\caption{Heatmaps of the normalized Casimir matrix $|({\cal C}_{M})_{ab}|/|C_{M}|$ for the three-matrix model at $M=8$ and $X=145$. The upper and lower grids have fixed $\widetilde m_1=0.2$ and $0.8$, respectively. Within each grid, the upper left, upper right, lower left, and lower right panels correspond to $\widetilde m_2=1,2,3,4$, respectively.}
\label{fig:small_mass_heatmaps}
\end{figure}

To examine small-mass behavior away from these zero-mode mass ratios for \(\tilde m_2 > \tilde m_1\), Figure~\ref{fig:small_mass_heatmaps} compares $|({\cal C}_{M})_{ab}|/|C_{M}|$ at fixed $\widetilde m_1=0.2$ and $0.8$, with $\widetilde m_2=1,2,3,4$. In both scans, the normalized Casimir matrices at $\widetilde m_2=1,2$ show visible deviations from the identity, indicating departures from an irreducible fuzzy-sphere configuration. At $\widetilde m_2=3,4$, by contrast, they are nearly equal to the identity, consistently with an almost exact fuzzy-sphere configuration at the level of this diagnostic. This change suggests a transition from a deformed configuration to a fuzzy sphere configuration between the sampled masses $\widetilde m_2=2$ and $3$. The change occurs between the same sampled masses for both $\widetilde m_1=0.2$ and $0.8$, suggesting that the critical value of \(\tilde m_2\) does not or only weakly depend on $\widetilde m_1$. Keeping $\widetilde m_1$ small separates this behavior from the special zero-mode ratios discussed above. The departures at smaller $\widetilde m_2$ are consistent with stronger bosonic matrix fluctuations about the fuzzy-sphere background as the mass deformation is reduced. A quantitative explanation of these fluctuations and the apparent weak dependence of the transition location on $\widetilde m_1$ will require further investigation.

\begin{table}[t]
\centering
\small
\setlength{\tabcolsep}{4pt}
\begin{tabular}{cccccccc}
\hline
$\widetilde m_2$ & $M$
& $\bar w_{\{11\}}$
& $\bar w_{\{22\}}$
& $\bar w_{\{33\}}$
& $\Delta_{SO(3)}$
& $\operatorname{Re}C_M$
& $\operatorname{Re}C_M/C^{\mathrm{FS}}$ \\
\hline
12 & 4  & $11.26-0.00i$ & $11.26-0.00i$ & $11.26-0.00i$
   & $0.000$ & $33.777$ & $0.938$ \\
12 & 6  & $11.67+0.00i$ & $11.67+0.00i$ & $11.67+0.00i$
   & $0.000$ & $35.004$ & $0.972$ \\
12 & 8  & $11.81+0.00i$ & $11.81+0.00i$ & $11.81+0.00i$
   & $0.000$ & $35.442$ & $0.984$ \\
12 & 10 & $11.88-0.00i$ & $11.88-0.00i$ & $11.88+0.00i$
   & $0.000$ & $35.644$ & $0.990$ \\
\hline
10 & 4  & $7.82+0.00i$ & $7.83-0.00i$ & $7.83+0.00i$
   & $0.000$ & $23.478$ & $0.939$ \\
10 & 6  & $8.10+0.00i$ & $8.11-0.00i$ & $8.11+0.00i$
   & $0.000$ & $24.318$ & $0.973$ \\
10 & 8  & $8.21+0.00i$ & $8.21+0.00i$ & $8.21-0.00i$
   & $0.000$ & $24.619$ & $0.985$ \\
10 & 10 & $8.25-0.00i$ & $8.25-0.00i$ & $8.25-0.00i$
   & $0.000$ & $24.754$ & $0.990$ \\
\hline
8 & 4  & $5.02+0.00i$ & $5.02+0.00i$ & $5.02+0.00i$
  & $0.000$ & $15.068$ & $0.942$ \\
8 & 6  & $5.18+0.00i$ & $5.20+0.01i$ & $5.19+0.00i$
  & $0.002$ & $15.569$ & $0.973$ \\
8 & 8  & $5.26+0.00i$ & $5.26-0.00i$ & $5.26-0.00i$
  & $0.000$ & $15.771$ & $0.986$ \\
8 & 10 & $5.29-0.00i$ & $5.29-0.00i$ & $5.29-0.00i$
  & $0.000$ & $15.857$ & $0.991$ \\
\hline
6 & 4  & $2.86-0.00i$ & $2.86+0.00i$ & $2.86+0.00i$
  & $0.001$ & $8.578$ & $0.953$ \\
6 & 6  & $2.89+0.00i$ & $2.92-0.00i$ & $2.90+0.00i$
  & $0.005$ & $8.716$ & $0.968$ \\
6 & 8  & $2.97-0.00i$ & $2.97-0.00i$ & $2.97-0.00i$
  & $0.000$ & $8.903$ & $0.989$ \\
6 & 10 & $2.98+0.00i$ & $2.98+0.00i$ & $2.98+0.00i$
  & $0.000$ & $8.943$ & $0.994$ \\
\hline
5 & 4  & $2.03+0.00i$ & $2.03-0.00i$ & $2.03+0.00i$
  & $0.001$ & $6.092$ & $0.975$ \\
5 & 6  & $1.99+0.04i$ & $2.04+0.03i$ & $2.00+0.04i$
  & $0.011$ & $6.019$ & $0.963$ \\
5 & 8  & $2.08-0.00i$ & $2.08-0.00i$ & $2.08-0.00i$
  & $0.000$ & $6.228$ & $0.996$ \\
5 & 10 & $2.08+0.00i$ & $2.08-0.00i$ & $2.08-0.00i$
  & $0.000$ & $6.243$ & $0.999$ \\
\hline
4 & 4  & $0.01-0.00i$ & $0.01-0.00i$ & $0.01-0.00i$
  & $0.060$ & $0.016$ & $0.004$ \\
4 & 6  & $0.02+0.01i$ & $0.04-0.01i$ & $0.02+0.00i$
  & $0.578$ & $0.084$ & $0.021$ \\
4 & 8  & $0.45-0.01i$ & $-0.09-0.02i$ & $0.02+0.01i$
  & $0.897$ & $0.384$ & $0.096$ \\
4 & 10 & $0.05-0.00i$ & $0.05-0.00i$ & $0.04-0.01i$
  & $0.113$ & $0.144$ & $0.036$ \\
\hline
3 & 4  & $0.88-0.00i$ & $0.89+0.00i$ & $0.89-0.00i$
  & $0.006$ & $2.659$ & $1.182$ \\
3 & 6  & $0.75-0.01i$ & $0.81-0.01i$ & $0.77-0.01i$
  & $0.031$ & $2.323$ & $1.033$ \\
3 & 8  & $0.81+0.00i$ & $0.82+0.00i$ & $0.81-0.00i$
  & $0.003$ & $2.445$ & $1.087$ \\
3 & 10 & $0.87-0.01i$ & $0.84-0.00i$ & $0.86-0.01i$
  & $0.011$ & $2.570$ & $1.142$ \\
\hline
2 & 4  & $0.19-0.01i$ & $0.15+0.01i$ & $0.22-0.04i$
  & $0.170$ & $0.556$ & $0.556$ \\
2 & 6  & $0.05+0.00i$ & $0.09-0.00i$ & $0.07-0.00i$
  & $0.244$ & $0.219$ & $0.219$ \\
2 & 8  & $0.07+0.01i$ & $0.06-0.00i$ & $0.07+0.02i$
  & $0.352$ & $0.208$ & $0.208$ \\
2 & 10 & $0.04-0.00i$ & $0.07-0.01i$ & $0.07+0.02i$
  & $0.353$ & $0.178$ & $0.178$ \\
\hline
\end{tabular}
\caption{ Values of $\bar w_{\{11\}}$, $\bar w_{\{22\}}$, $\bar w_{\{33\}}$, $\Delta_{SO(3)}$, $\operatorname{Re}C_M$, and $\operatorname{Re}C_M/C^{\mathrm{FS}}$ for $M=4,6,8,10$ at fixed $\widetilde m_1=2$ and $\widetilde m_2=2,3,4,5,6,8,10,12$. The corresponding values of $X$ are $75$, $81$, $145$, and $218$, in ascending order of $M$. The classical reference is $C^{\mathrm{FS}}=\widetilde m_2^2/4$. The second moments are shown to two decimal places, while $\Delta_{SO(3)}$, $\operatorname{Re}C_M$, and the ratio are shown to three decimal places. }
\label{tab:three_matrix_M_dependence}
\end{table}
We next investigate the $M$ dependence of the optimization results, focusing on the departures from the fuzzy-sphere configuration observed at small masses. Since the RMA describes the large-$N$ limit and $M$ is a regularization parameter, comparing different values of $M$ allows us to examine whether these features persist as the regularization is refined. We fix $\widetilde m_1=2$ and perform optimizations at $\widetilde m_2=2,3,4,5,6,8,10,12$ for $M=4,6,8,10$. The corresponding values of $X$ are $75$, $81$, $145$, and $218$, respectively. Table~\ref{tab:three_matrix_M_dependence} lists the diagonal quadratic moments, $\Delta_{SO(3)}$, $\operatorname{Re}C_M$, and $\operatorname{Re}C_M/C^{\mathrm{FS}}$.

A common feature across all values of $M$ is that the quadratic moments remain approximately $SO(3)$ symmetric at the sampled masses $\widetilde m_2\geq5$, with $\Delta_{SO(3)}\ll0.1$.
In this regime, $\operatorname{Re}C_M/C^{\mathrm{FS}}$ approaches unity as $M$ increases. At $\widetilde m_2=2$, however, we find $\Delta_{SO(3)}>0.1$ for all values of $M$ considered.
The quadratic size decreases as $M$ increases, while the moments remain anisotropic.
Thus the departure persists as the regularization is refined, consistently with the fermion-kernel obstruction at the equal-mass reference sphere for every $M$ considered here.
At $\widetilde m_2=4$, the degree of anisotropy depends nonmonotonically on $M$. We find $\Delta_{SO(3)}<0.1$ for $M=4$, $\Delta_{SO(3)}\gg0.1$ for $M=6,8$, and $\Delta_{SO(3)}\gtrsim0.1$ for $M=10$.
Nevertheless, $\operatorname{Re}C_M/C^{\mathrm{FS}}<0.1$ in every case, indicating a strong suppression of the fuzzy-sphere size even when the quadratic moments remain approximately isotropic.
In contrast, at $\widetilde m_2=3$, $\Delta_{SO(3)}\ll0.1$ and the quadratic size remains close to the classical prediction for all values of $M$.
This contrast is consistent with the zero-mode condition discussed above, which is met at $\widetilde m_2=2,4$ but not at $\widetilde m_2=3$.
Another notable feature is that, at larger masses $\widetilde m_2 \geq 6$, $\operatorname{Re}C_M/C^{\mathrm{FS}}$ approaches unity monotonically as $M$ increases. This behavior is consistent with an increasingly accurate reproduction of the classical fuzzy-sphere size.

There is one subtlety concerning the stability of the configurations found by the optimization. A configuration may be approximately isotropic and reproduce the classical fuzzy-sphere size even when the corresponding classical saddle is metastable rather than preferred. For positive masses with $\widetilde m_1<\widetilde m_2<2\widetilde m_1$, the fuzzy sphere is a radial local minimum but has higher classical action than the trivial configuration, whereas for $\widetilde m_2>2\widetilde m_1$ it has lower action. As a simple check, we compare $(\widetilde m_1,\widetilde m_2)=(2,3)$ in the former region with $(2,5)$ in the latter, using each optimized $M=10$, $X=218$ configuration as the initial condition for optimization with $X=230$. After 100 function evaluations, $\operatorname{Re}C_M$ decreases from $2.569863$ to $2.451184$ at $(2,3)$, a change of $4.62\%$, compared with $6.242948$ to $6.242234$ at $(2,5)$, a change of only $0.0114\%$. The greater sensitivity at $(2,3)$ is consistent with its classical metastable interpretation. As an example, this provides a simple check of the stability of the optimized configurations as additional loop equations are included.

\FloatBarrier
\section{Four-Matrix Model}\label{sec:four_matrix}
We next consider the mass-deformed four-dimensional Type-II supersymmetric Yang--Mills matrix model of Ref.~\cite{Martina:2025kwc}. The dynamical variables are four Hermitian bosonic matrices $A_I$, $I=1,\ldots,4$, and four fermionic matrices $\Psi_\alpha$, $\alpha=1,\ldots,4$. The action is
\begin{align}
    S_{4\mathrm M}[A,\Psi]
    =
    N\operatorname{Tr}\Bigg[
    &-\frac14[A_I,A_J]^2
    -\frac{i}{2}\Psi_\alpha(C\Gamma^I)_{\alpha\beta}[A_I,\Psi_\beta]
    +\frac{2i}{3}(m_1+m_2)\epsilon_{ijk}A_iA_jA_k
    \nonumber\\
    &+
    \left(
    \frac{2}{9}m_1^2+\frac13m_1m_2
    \right)A_iA_i
    +\frac13m_1m_2A_4^2
    -\frac13m_1
    \Psi_\alpha(C\Gamma^{123})_{\alpha\beta}\Psi_\beta
    \Bigg],
\label{eq:4matrix_action}
\end{align}
where $I,J=1,\ldots,4$ while $i,j,k=1,2,3$.

Using the same collective-index notation and moment conventions as in Sec.~\ref{subsec:three_matrix_setup}, the bosonic loop equations for varying the three bosonic matrices \(A_i\), $i=1,2,3$ are
\begin{align}
&\sum_{k=1}^{n}
\delta_{z_k,i}\,
w\{z_1\cdots z_{k-1}\}
w\{z_{k+1}\cdots z_n\}
-
w\{z_1\cdots z_n[J,[J,i]]\}
\nonumber\\
&-
\frac{i}{2}(C\Gamma^i)_{\alpha\beta}
w\{z_1\cdots z_n\{\alpha,\beta\}\}
-
\left(
\frac49m_1^2+\frac23m_1m_2
\right)
w\{z_1\cdots z_n i\}
\nonumber\\
&-
i(m_1+m_2)\epsilon_{ijk}
w\{z_1\cdots z_n[j,k]\}
=0 .
\label{eq:4matrix_bosonic_loop_123}
\end{align}

For the fourth bosonic matrix \(A_4\), the corresponding loop equation is
\begin{align}
&\sum_{k=1}^{n}
\delta_{z_k,4}\,
w\{z_1\cdots z_{k-1}\}
w\{z_{k+1}\cdots z_n\}
-
w\{z_1\cdots z_n[J,[J,4]]\}
\nonumber\\
&-
\frac{i}{2}(C\Gamma^4)_{\alpha\beta}
w\{z_1\cdots z_n\{\alpha,\beta\}\}
-
\frac23m_1m_2
w\{z_1\cdots z_n4\}
=0 .
\label{eq:4matrix_bosonic_loop_4}
\end{align}

We apply the same numerical RMA procedure used for the three-matrix model. The regularized master field is represented by four complex $M\times M$ matrices, with $\hat A_1^{(M)}$ chosen diagonal and complex and $\hat A_{2,3,4}^{(M)}$ taken to be generic complex matrices. After accounting for the residual gauge redundancy, the number of independent real degrees of freedom is \(6M^2+2\). We use both fuzzy-sphere and random initial configurations for the four matrices. The fuzzy-sphere initial configurations are constructed from the generators of the $M$-dimensional irreducible representation of $SU(2)$, with $\hat A_i^{(M)}=(\tfrac23m_1+m_2)L_i$ for $i=1,2,3$ and $\hat A_4^{(M)}=0$. For random initial configurations, the real and imaginary parts of the independent matrix entries are sampled from a Gaussian distribution with zero mean and standard deviation $0.1$, keeping $\hat A_1^{(M)}$ diagonal. The results below are obtained using fuzzy-sphere initial configurations. Random initial configurations do not lead to fuzzy spheres in the runs studied here.

We first examine the dependence of the quadratic size of the first three matrix directions on the mass deformation. Following the normalization used for the three-matrix model, we define
\begin{equation}
C_M\equiv\sum_{i=1}^{3}\bar w^{(M)}_{\{ii\}}=\frac{1}{M^2}\operatorname{Tr}\!\left[\sum_{i=1}^{3}(\hat A_i^{(M)})^2\right].
\label{eq:four_matrix_CM}
\end{equation}
As shown in Appendix~\ref{app:four_matrix_saddles}, the stable fuzzy-sphere branch for positive masses is
\begin{equation}
A_i=\left(\frac23m_1+m_2\right)L_i,\qquad A_4=0.
\end{equation}
As in the three-matrix model, we define the rescaled masses by $m_a=\widetilde m_a/\sqrt M$, with $\widetilde m_a=\mathcal O(M^0)$ held fixed as $M$ varies. This scaling is motivated by requiring the action evaluated at a generic fuzzy-sphere saddle in the original rank-$N$ theory to be of order $N^2$, and gives finite normalized quadratic moments for the regularized configurations.
Its normalized large-$N$ Casimir is
\begin{equation}
C^{\mathrm{FS}}=\frac14\left(\frac23\widetilde m_1+\widetilde m_2\right)^2.
\label{eq:four_matrix_classical_C}
\end{equation}

\begin{figure}[p]
\centering
\begin{subfigure}[t]{0.85\textwidth}
\centering
\includegraphics[width=\linewidth]{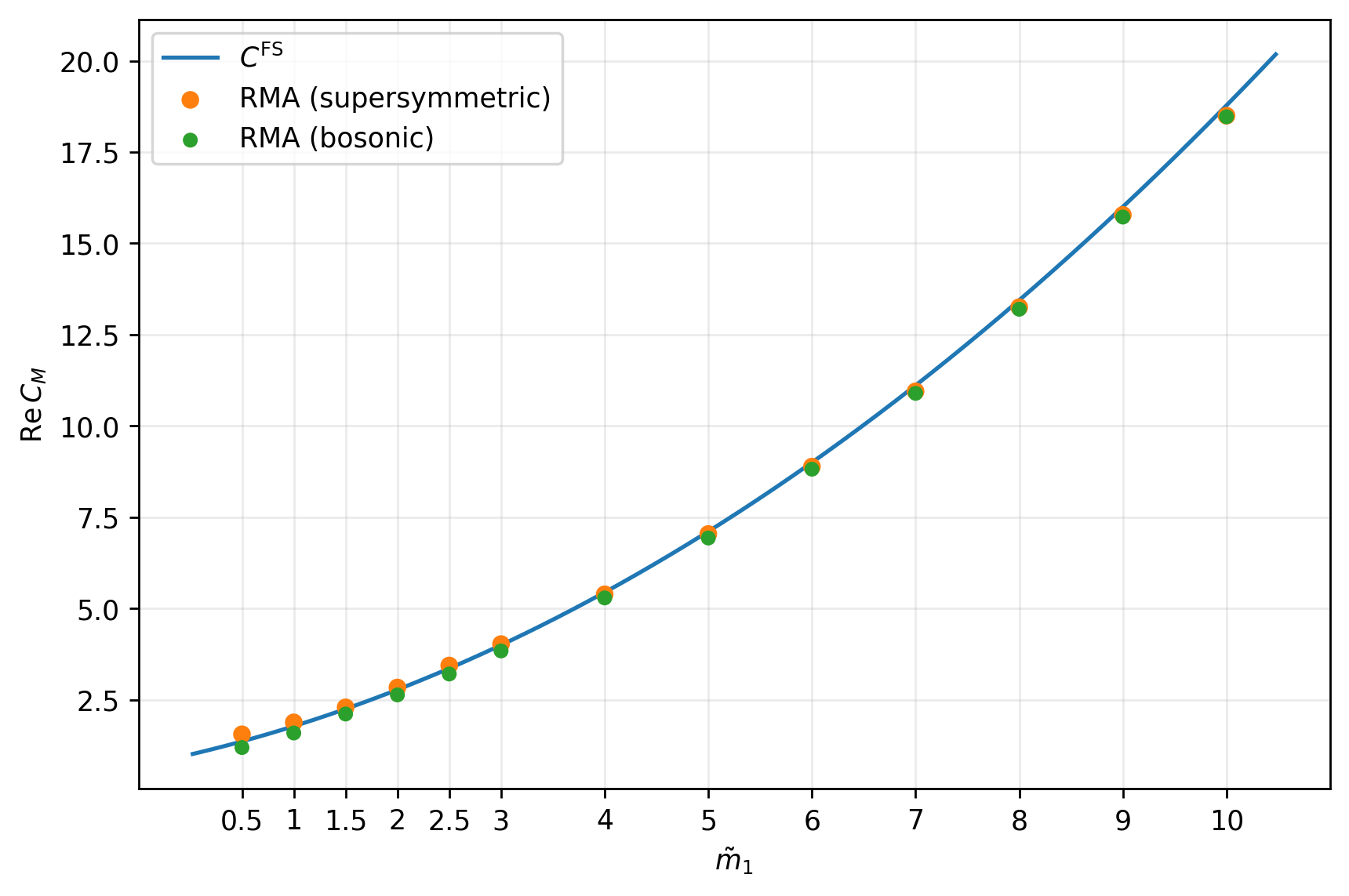}
\caption{$\widetilde m_2=2$ fixed.}
\end{subfigure}

\medskip
\begin{subfigure}[t]{0.85\textwidth}
\centering
\includegraphics[width=\linewidth]{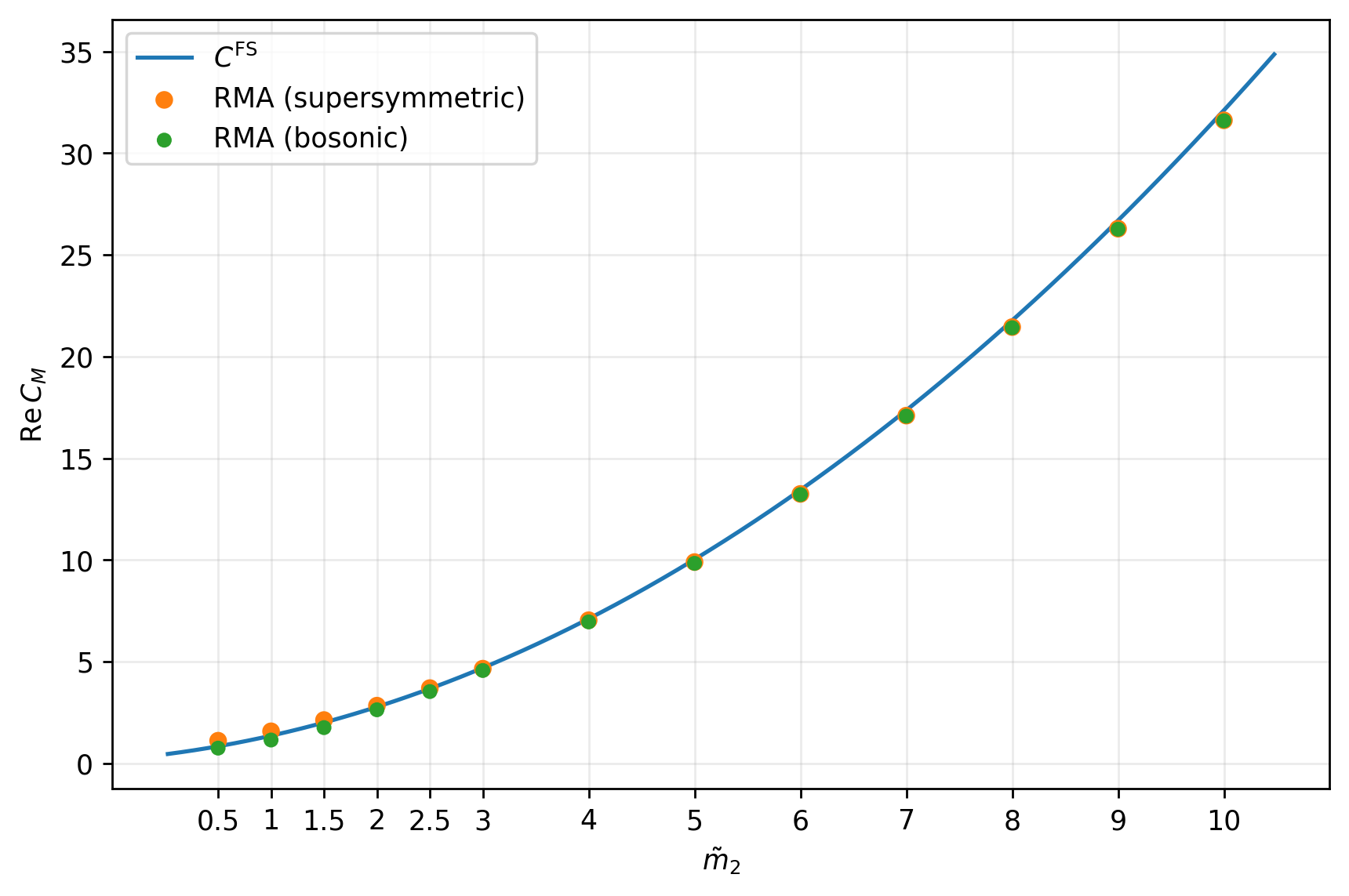}
\caption{$\widetilde m_1=2$ fixed.}
\end{subfigure}
\caption{Comparison of $\operatorname{Re}C_{M}$ with $C^{\mathrm{FS}}$ for the four-matrix model at $M=8$ and $X=200$. The upper and lower panels vary $\widetilde m_1$ and $\widetilde m_2$, respectively, while holding the other rescaled mass at $2$. Orange and green dots represent $\operatorname{Re}C_M$ for the supersymmetric and bosonic models, respectively, while the solid blue line represents $C^{\mathrm{FS}}=\tfrac14(\tfrac23\widetilde m_1+\widetilde m_2)^2$.}
\label{fig:four_matrix_CM_scan}
\end{figure}

Figure~\ref{fig:four_matrix_CM_scan} compares the supersymmetric and bosonic RMA results at $M=8$ and $X=200$. The upper panel varies $\widetilde m_1$ at fixed $\widetilde m_2=2$, while the lower panel varies $\widetilde m_2$ at fixed $\widetilde m_1=2$. In both scans, the results closely follow the classical curve at larger masses, consistently with a stable fuzzy-sphere configuration in this regime. At smaller masses, the supersymmetric results lie above the classical reference, and the difference from the bosonic results becomes more visible. Unlike the three-matrix scans, these plots do not exhibit a pronounced suppression of $C_M$ at equal masses or at $\widetilde m_2=2\widetilde m_1$.

The four-matrix results show no analogous indication of the instability associated with zero modes of the Dirac operator discussed for the three-matrix model. The structure of the four-matrix Dirac operator is different from that of the three-matrix model,
\begin{equation}
\mathcal D\Psi
=-i\sum_{I=1}^{4}\Gamma^I[\hat A_I^{(M)},\Psi]
-\frac23m_1\Gamma^{123}\Psi.
\end{equation}
On the exact fuzzy-sphere background $\hat A_i^{(M)}=aL_i$, $\hat A_4^{(M)}=0$, with $a=\tfrac23m_1+m_2$, we multiply the Dirac operator by $\Gamma^{123}$. Since $(\Gamma^{123})^2=-\mathbf1$, this multiplication preserves its zero modes and turns the mass term into the scalar $\tfrac23m_1$. The adjoint action $[L_i,\cdot]$ on $M\times M$ matrices decomposes into angular-momentum sectors labeled by $l=0,\ldots,M-1$, each containing $2l+1$ independent matrices. The matrices $-i\Gamma^{123}\Gamma^i$ satisfy the Pauli algebra and therefore act as twice the spin-$1/2$ generators. Coupling this spin to the matrix angular momentum $l$ gives total angular momentum $J=l+\tfrac12$ and, for $l\geq1$, $J=l-\tfrac12$. The corresponding eigenvalues of the spin-dependent term are $a[J(J+1)-l(l+1)-\tfrac34]$, giving $al$ and $-a(l+1)$, respectively. Thus the eigenvalues of $\Gamma^{123}\mathcal D$ are
\begin{equation}
\begin{aligned}
&al+\frac23m_1,
&&l=0,\ldots,M-1,\\
&-a(l+1)+\frac23m_1,
&&l=1,\ldots,M-1.
\end{aligned}
\end{equation}
For positive masses, $a>\tfrac23m_1>0$, so neither branch vanishes. Thus the Dirac operator on this fuzzy-sphere branch has no zero modes, unlike the three-matrix model.
The supersymmetric results remain close to the bosonic results through the mass values where the three-matrix scans showed strong departures associated with singular fermion propagators. This comparison concerns the fuzzy sphere configurations and mass ranges studied here and does not exclude zero modes on other four-matrix backgrounds.

\begin{table}[t]
\centering
\small
\setlength{\tabcolsep}{3pt}
\begin{tabular}{c|cccccc}
\hline
$\widetilde m_2$ & $\bar w^{(8)}_{\{11\}}$ & $\bar w^{(8)}_{\{22\}}$ & $\bar w^{(8)}_{\{33\}}$ & $\Delta_{SO(3)}$ & $\operatorname{Re}C_M$ & $\operatorname{Re}C_M/C^{\mathrm{FS}}$ \\
\hline
10 & $10.5394+0.0005i$ & $10.5389+0.0007i$ & $10.5388+0.0011i$ & $6.23\times10^{-5}$ & $31.6171$ & $0.9846$ \\
9 & $8.7626-0.0003i$ & $8.7618+0.0001i$ & $8.7639-0.0004i$ & $1.29\times10^{-4}$ & $26.2883$ & $0.9848$ \\
8 & $7.1492+0.0010i$ & $7.1492+0.0005i$ & $7.1513+0.0020i$ & $2.38\times10^{-4}$ & $21.4497$ & $0.9849$ \\
7 & $5.7018-0.0017i$ & $5.6959-0.0010i$ & $5.6996+0.0008i$ & $5.30\times10^{-4}$ & $17.0973$ & $0.9848$ \\
6 & $4.4160-0.0006i$ & $4.4163-0.0005i$ & $4.4156+0.0001i$ & $2.21\times10^{-4}$ & $13.2480$ & $0.9854$ \\
5 & $3.3005+0.0016i$ & $3.2966+0.0016i$ & $3.2959+0.0026i$ & $7.70\times10^{-4}$ & $9.8930$ & $0.9866$ \\
4 & $2.3518+0.0008i$ & $2.3412+0.0016i$ & $2.3461-0.0028i$ & $2.17\times10^{-3}$ & $7.0391$ & $0.9899$ \\
3 & $1.5551-0.0010i$ & $1.5443-0.0060i$ & $1.5574-0.0030i$ & $5.25\times10^{-3}$ & $4.6568$ & $0.9920$ \\
2.5 & $1.2390+0.0014i$ & $1.2288+0.0022i$ & $1.2332+0.0028i$ & $4.44\times10^{-3}$ & $3.7010$ & $1.0074$ \\
2 & $0.9485-0.0019i$ & $0.9332-0.0066i$ & $0.9563-0.0029i$ & $1.05\times10^{-2}$ & $2.8379$ & $1.0216$ \\
1.5 & $0.7298-0.0039i$ & $0.6926+0.0053i$ & $0.7083+0.0093i$ & $2.38\times10^{-2}$ & $2.1307$ & $1.0617$ \\
1 & $0.5406+0.0018i$ & $0.4989-0.0014i$ & $0.5372+0.0008i$ & $3.81\times10^{-2}$ & $1.5768$ & $1.1584$ \\
0.8 & $0.4845-0.0014i$ & $0.4520+0.0001i$ & $0.4706-0.0014i$ & $2.97\times10^{-2}$ & $1.4072$ & $1.2368$ \\
0.5 & $0.4092+0.0034i$ & $0.3298+0.0263i$ & $0.3756+0.0160i$ & $9.29\times10^{-2}$ & $1.1145$ & $1.3264$ \\
0.2 & $0.3267+0.0250i$ & $0.3083+0.0012i$ & $0.3674-0.0107i$ & $9.96\times10^{-2}$ & $1.0024$ & $1.7054$ \\
\hline
\end{tabular}
\caption{Quadratic moments, their deviation from $SO(3)$ isotropy, $\operatorname{Re}C_M$, and $\operatorname{Re}C_M/C^{\mathrm{FS}}$ for the supersymmetric four-matrix RMA solutions at fixed $\widetilde m_1=2$, $M=8$, and $X=200$, as $\widetilde m_2$ is varied. The definition of $\Delta_{SO(3)}$ is that of Eq.~\eqref{eq:SO3_deviation}, restricted to the first three directions and evaluated using the unrounded data. The moments, $\operatorname{Re}C_M$, and the ratio are rounded to four decimal places, with imaginary parts that round to zero omitted. The classical large-$N$ reference is $C^{\mathrm{FS}}=\tfrac14(\tfrac23\widetilde m_1+\widetilde m_2)^2$.}
\label{tab:four_matrix_mass_scan_M8}
\end{table}

Table~\ref{tab:four_matrix_mass_scan_M8} gives the quadratic moments, $\Delta_{SO(3)}$, $\operatorname{Re}C_M$, and $\operatorname{Re}C_M/C^{\mathrm{FS}}$ for the supersymmetric scan at fixed $\widetilde m_1=2$. The last two columns measure the overall quadratic size of the first three directions and its deviation from the classical large-$N$ reference. At the sampled masses $\widetilde m_2\geq5$, the three diagonal moments are nearly equal, $\Delta_{SO(3)}<8\times10^{-4}$, and the ratio remains close to unity, between approximately $0.9846$ and $0.9866$. Together, these observations support an approximately isotropic fuzzy-sphere configuration at larger masses.

As $\widetilde m_2$ decreases, $\operatorname{Re}C_M$ decreases from $31.6171$ at $\widetilde m_2=10$ to $1.0024$ at $\widetilde m_2=0.2$. However, its ratio to the classical reference increases above unity at the sampled masses $\widetilde m_2\leq2.5$, indicating a quadratic size larger than the classical prediction. At $\widetilde m_2=0.5$ and $0.2$, the ratios reach approximately $1.33$ and $1.71$, respectively, while $\Delta_{SO(3)}$ reaches approximately $0.093$ and $0.100$. Thus the small-mass configurations show both an enhanced size relative to the classical large-$N$ reference and increasing anisotropy of the quadratic moments.

\begin{figure}[p]
\centering
\includegraphics[width=\textwidth,height=0.48\textheight,keepaspectratio]{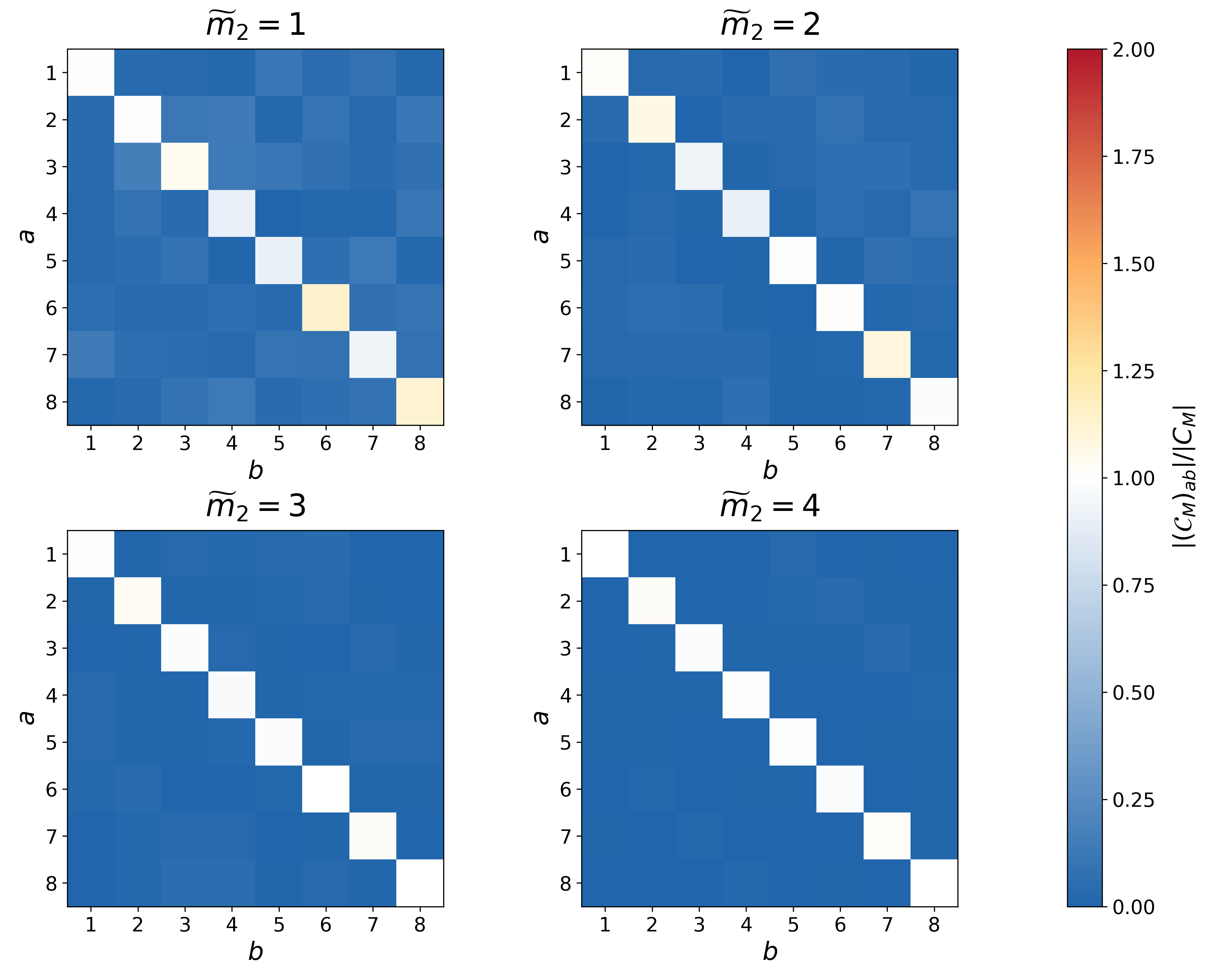}

\medskip
\includegraphics[width=\textwidth,height=0.48\textheight,keepaspectratio]{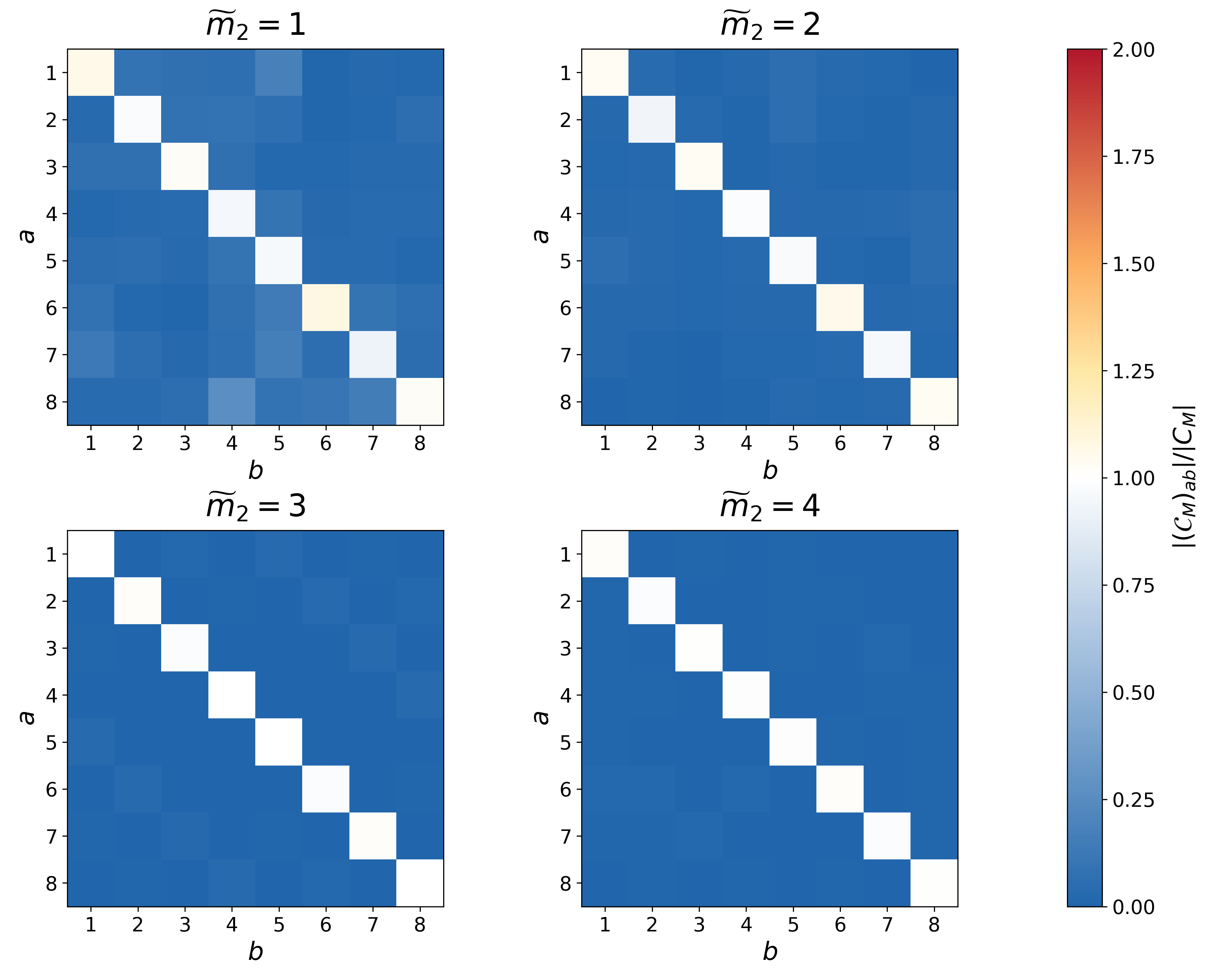}
\caption{Heatmaps of the Casimir matrix $|\mathcal C_{M,ab}/C_M|$ for the four-matrix model at $M=8$ and $X=200$. The upper and lower grids have fixed $\widetilde m_1=0.2$ and $0.8$, respectively. Within each grid, the upper left, upper right, lower left, and lower right panels correspond to $\widetilde m_2=1,2,3,4$, respectively.}
\label{fig:four_matrix_small_mass_heatmaps}
\end{figure}

To investigate the loss of fuzzy-sphere structure at smaller masses, Figure~\ref{fig:four_matrix_small_mass_heatmaps} shows $|({\cal C}_{M})_{ab}|/|C_{M}|$, where $\mathcal C_M=M^{-1}\sum_{i=1}^{3}(\hat A_i^{(M)})^2$, at fixed $\widetilde m_1=0.2$ and $0.8$, with $\widetilde m_2=1,2,3,4$. At larger $\widetilde m_2$, the diagonal entries approach unity and the off-diagonal entries are suppressed, consistently with an irreducible fuzzy sphere. At smaller $\widetilde m_2$, off-diagonal contributions and deviations of the diagonal entries become more visible.

\begin{figure}[p]
\centering
\includegraphics[width=\textwidth,height=0.48\textheight,keepaspectratio]{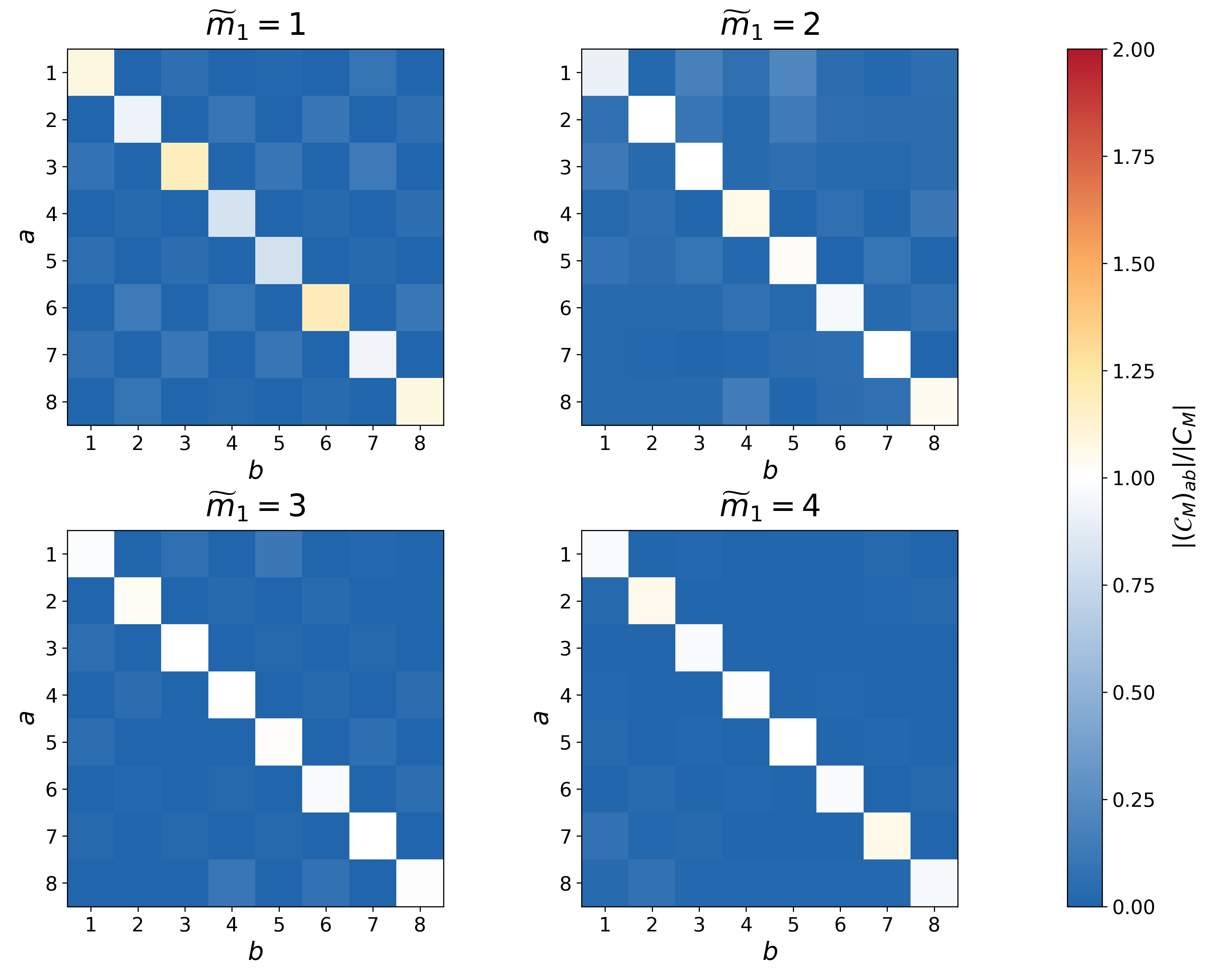}

\par\vspace{4pt}
\includegraphics[width=\textwidth,height=0.48\textheight,keepaspectratio]{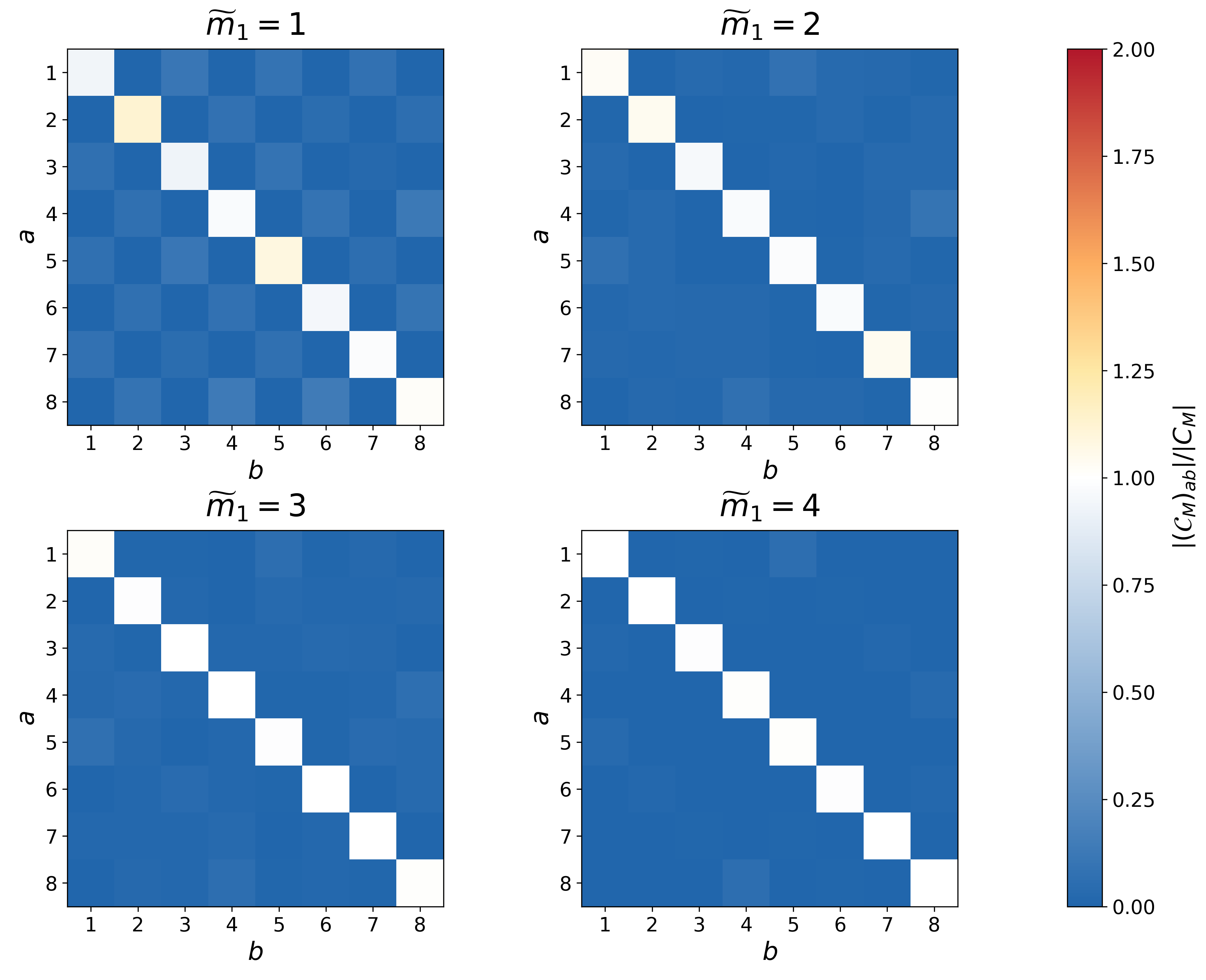}
\caption{Heatmaps of the normalized Casimir matrix $|\mathcal C_{M,ab}/C_M|$ for the four-matrix model at $M=8$ and $X=200$. The upper and lower grids have fixed $\widetilde m_2=0.2$ and $0.8$, respectively. Within each grid, the upper left, upper right, lower left, and lower right panels correspond to $\widetilde m_1=1,2,3,4$, respectively.}
\label{fig:four_matrix_small_mass_heatmaps_m2fixed}
\end{figure}

This behavior is qualitatively similar to the phase-transition-like behavior observed in the three-matrix model. Figure~\ref{fig:four_matrix_small_mass_heatmaps_m2fixed} shows a similar change when $\widetilde m_2$ is fixed at $0.2$ or $0.8$ and $\widetilde m_1=1,2,3,4$ is varied. However, when the fixed mass is increased to $2$ or $5$, all four heatmaps over the same range of the varied mass are close to the identity, consistently with an almost exact fuzzy-sphere configuration. This suggests that a sufficiently large value of either mass maintains the fuzzy-sphere structure, while departures occur when both masses are small. Increasing either mass increases the classical radius parameter $\tfrac23\widetilde m_1+\widetilde m_2$ and the action evaluated at this saddle becomes more negative relative to the trivial configuration, so the fuzzy-sphere saddle is increasingly favored by the classical action.

\begin{figure}[t]
\centering
\includegraphics[width=\textwidth]{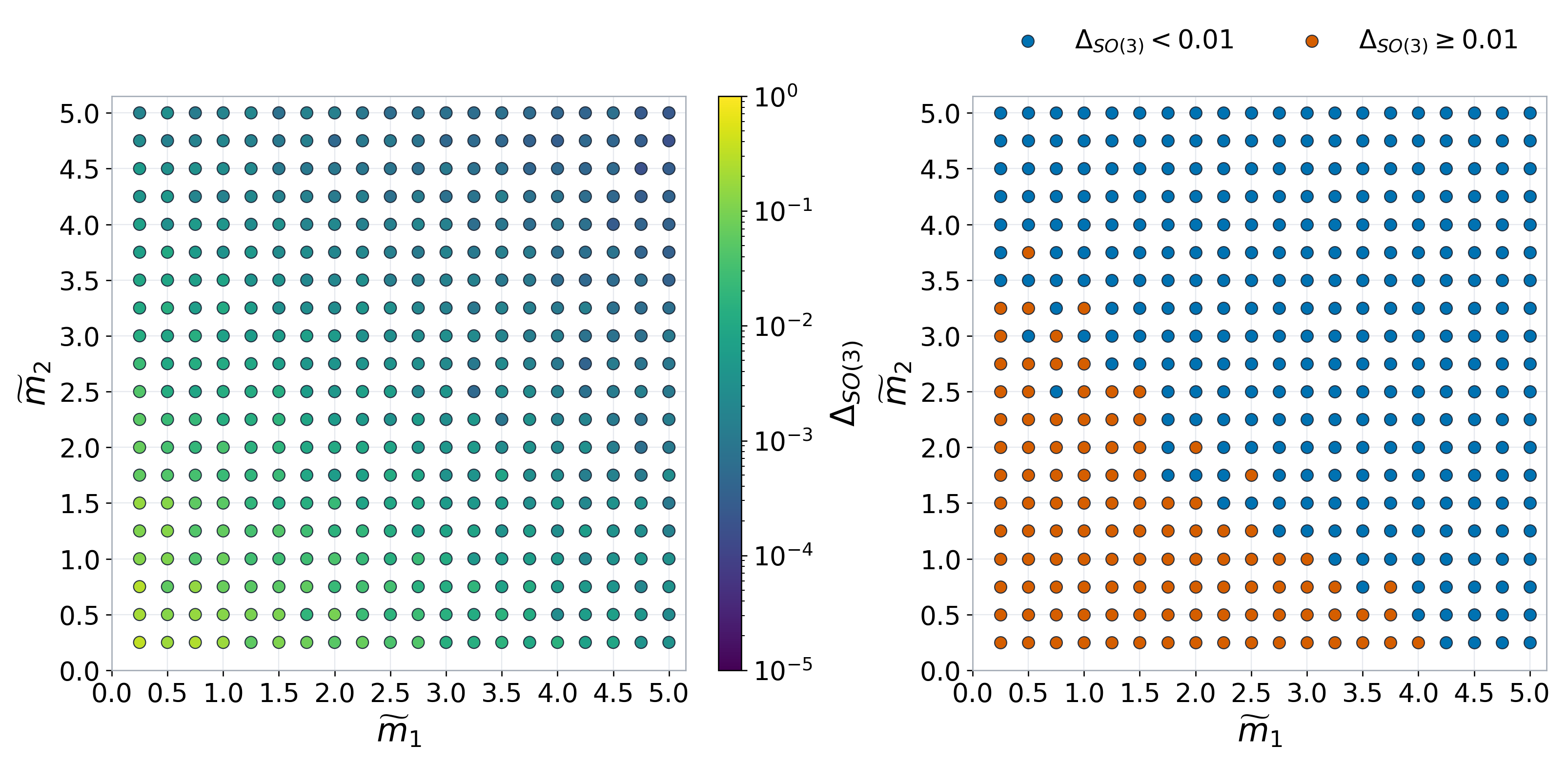}
\caption{Mass dependence of $\Delta_{SO(3)}$ in the supersymmetric
four-matrix model at $M=8$ and $X=200$.
Left: $\Delta_{SO(3)}$ on a logarithmic color scale.
Right: blue and orange indicate $\Delta_{SO(3)}<10^{-2}$
and $\Delta_{SO(3)}\geq10^{-2}$, respectively.}
\label{fig:four_matrix_mass_plane}
\end{figure}

Figure~\ref{fig:four_matrix_mass_plane} extends the one-dimensional scans to a phase diagram on the two-dimensional mass plane, with both rescaled masses sampled from $0.25$ to $5$ in steps of $0.25$.
The left panel shows that the quadratic-moment anisotropy generally decreases as either mass increases.
The right panel uses the cutoff $\Delta_{SO(3)}=10^{-2}$ to distinguish approximately isotropic and anisotropic configurations.
The anisotropic configurations predominantly occupy a triangular region in the lower-left corner, separated from the approximately isotropic region by a roughly straight, negatively sloped boundary. The relation between the critical values of the two masses follows
\begin{equation}
    \tilde{m}_{1,\textrm{cr}}+\tilde{m}_{2,\textrm{cr}}\approx3.5.
\end{equation}
Together with the Casimir-matrix diagnostics above, this pattern supports the breakdown of fuzzy-sphere structure when both mass deformations are small. 

Finally, we characterize the fourth matrix relative to the configuration in the first three directions. Since the optimized matrices are generally complex, we use the Hermitian positive-semidefinite matrix Laplacian
\begin{equation}
\mathcal H(X)\equiv\sum_{i=1}^{3}[(\hat A_i^{(M)})^\dagger,[\hat A_i^{(M)},X]].
\label{eq:four_matrix_extra_laplacian}
\end{equation}
For an exact fuzzy sphere, its eigenvalues are proportional to $\ell(\ell+1)$, with degeneracy $2\ell+1$ on the traceless matrix space. Let $Y_{\ell r}$, $r=1,\ldots,2\ell+1$, be orthonormal eigenmatrices, with $\operatorname{Tr}(Y_{\ell r}^\dagger Y_{\ell' r'})=\delta_{\ell\ell'}\delta_{rr'}$. We define the mode fractions of the traceless part $A_{4,0}\equiv\hat A_4^{(M)}-M^{-1}\operatorname{Tr}(\hat A_4^{(M)})\mathbf1_M$ by
\begin{equation}
P_\ell\equiv\frac{\displaystyle\sum_{r=1}^{2\ell+1}|\operatorname{Tr}(Y_{\ell r}^\dagger A_{4,0})|^2}{\|A_{4,0}\|_F^2},\qquad \|X\|_F^2\equiv\operatorname{Tr}(X^\dagger X).
\label{eq:four_matrix_extra_Pell}
\end{equation}
Thus $\sum_{\ell=1}^{M-1}P_\ell=1$. The identity component is negligible for all configurations tabulated below, so subtracting it does not affect the displayed percentages. Away from an exact fuzzy sphere, we order the eigenmatrices by increasing eigenvalue and assign the first three to $\ell=1$, the next five to $\ell=2$, and so on. In this case, $\ell$ labels groups of modes and need not be an exact angular momentum.

To compare the extent of the extra direction with that of the first three, we also define
\begin{equation}
\varsigma\equiv\frac{\|\hat A_4^{(M)}\|_F^2}{\displaystyle\sum_{i=1}^{3}\|\hat A_i^{(M)}\|_F^2}.
\label{eq:four_matrix_extra_varsigma}
\end{equation}
Table~\ref{tab:four_matrix_extra_direction} lists the mode percentages and $\varsigma$ for all 15 configurations in the fixed-$\widetilde m_1=2$ scan.

\begin{table}[t]
\centering
\small
\setlength{\tabcolsep}{3.5pt}
\begin{tabular}{c c c c c c c c c}
\hline
$\widetilde m_2$ & $100P_1$ & $100P_2$ & $100P_3$ & $100P_4$ & $100P_5$ & $100P_6$ & $100P_7$ & $\varsigma$ \\
\hline
10 & 0.000 & 0.001 & 0.006 & 20.679 & 22.067 & 27.597 & 29.650 & $1.102\times10^{-5}$ \\
9 & 0.001 & 0.001 & 0.009 & 23.858 & 15.986 & 36.424 & 23.720 & $1.552\times10^{-5}$ \\
8 & 0.002 & 0.002 & 0.025 & 22.465 & 16.883 & 39.910 & 20.712 & $3.850\times10^{-5}$ \\
7 & 0.002 & 0.009 & 0.018 & 18.386 & 27.391 & 28.404 & 25.790 & $1.125\times10^{-4}$ \\
6 & 0.001 & 0.004 & 0.022 & 5.492 & 50.915 & 20.898 & 22.668 & $5.332\times10^{-5}$ \\
5 & 0.002 & 0.011 & 0.070 & 13.919 & 38.686 & 14.195 & 33.117 & $1.031\times10^{-4}$ \\
4 & 0.082 & 0.137 & 0.347 & 32.515 & 21.563 & 20.768 & 24.588 & $3.239\times10^{-4}$ \\
3 & 0.103 & 0.176 & 0.829 & 30.029 & 36.041 & 22.864 & 9.957 & $6.053\times10^{-4}$ \\
2.5 & 0.012 & 0.112 & 0.234 & 19.491 & 16.861 & 37.715 & 25.575 & $8.482\times10^{-4}$ \\
2 & 0.040 & 0.084 & 0.771 & 5.483 & 43.107 & 5.598 & 44.918 & $9.297\times10^{-4}$ \\
1.5 & 0.175 & 0.275 & 2.020 & 14.965 & 34.188 & 4.864 & 43.513 & $1.846\times10^{-3}$ \\
1 & 0.226 & 0.072 & 0.680 & 8.707 & 71.423 & 6.160 & 12.731 & $3.769\times10^{-3}$ \\
0.8 & 0.229 & 0.154 & 2.799 & 15.925 & 46.898 & 4.158 & 29.836 & $5.370\times10^{-3}$ \\
0.5 & 0.663 & 1.242 & 2.465 & 5.121 & 46.086 & 8.195 & 36.228 & $6.463\times10^{-3}$ \\
0.2 & 3.968 & 2.200 & 6.269 & 20.450 & 34.560 & 16.623 & 15.930 & $9.737\times10^{-3}$ \\
\hline
\end{tabular}
\caption{Percentage contributions ($100P_\ell$) of the matrix-Laplacian modes and relative size $\varsigma$ of the fourth matrix, for all available supersymmetric configurations at fixed $\widetilde m_1=2$, $M=8$, and $X=200$. The modes are grouped in increasing order of eigenvalue, with $2\ell+1$ modes in each group. By definition, $\sum_{\ell=1}^{7}P_\ell=1$; displayed percentages may differ slightly from $100$ because of rounding.}
\label{tab:four_matrix_extra_direction}
\end{table}

Table~\ref{tab:four_matrix_extra_direction} shows that the fourth matrix has very little support in the lowest Laplacian modes at larger masses. For $\widetilde m_2\geq5$, the combined contribution of $\ell=1,2$ is below $0.02\%$, and that of $\ell=1,2,3$ is below $0.1\%$. Almost all of the norm therefore lies in the groups $\ell=4,\ldots,7$.

As the mass decreases, the low-mode contributions generally increase, although not monotonically at every sampled mass. The combined contribution of $\ell=1,2$ reaches approximately $1.90\%$ at $\widetilde m_2=0.5$ and $6.17\%$ at $\widetilde m_2=0.2$. This increase accompanies the growing anisotropy of the quadratic moments discussed above, suggesting that $P_1$ and $P_2$ provide complementary diagnostics of departures from the fuzzy-sphere configuration. At smaller masses, $\ell$ labels the ordered mode groups defined above rather than an exact angular momentum.

The relative extent of the fourth direction also grows: $\varsigma$ increases overall from approximately $1.10\times10^{-5}$ at $\widetilde m_2=10$ to $9.74\times10^{-3}$ at $\widetilde m_2=0.2$. Thus the small-mass configurations show both a larger low-mode contribution within $\hat A_4^{(M)}$ and a larger fourth direction relative to the first three. Nevertheless, its squared norm remains below $1\%$ of their combined squared norm throughout this scan.

\begin{figure}[p]
\centering
\includegraphics[width=\textwidth,height=0.31\textheight,keepaspectratio]{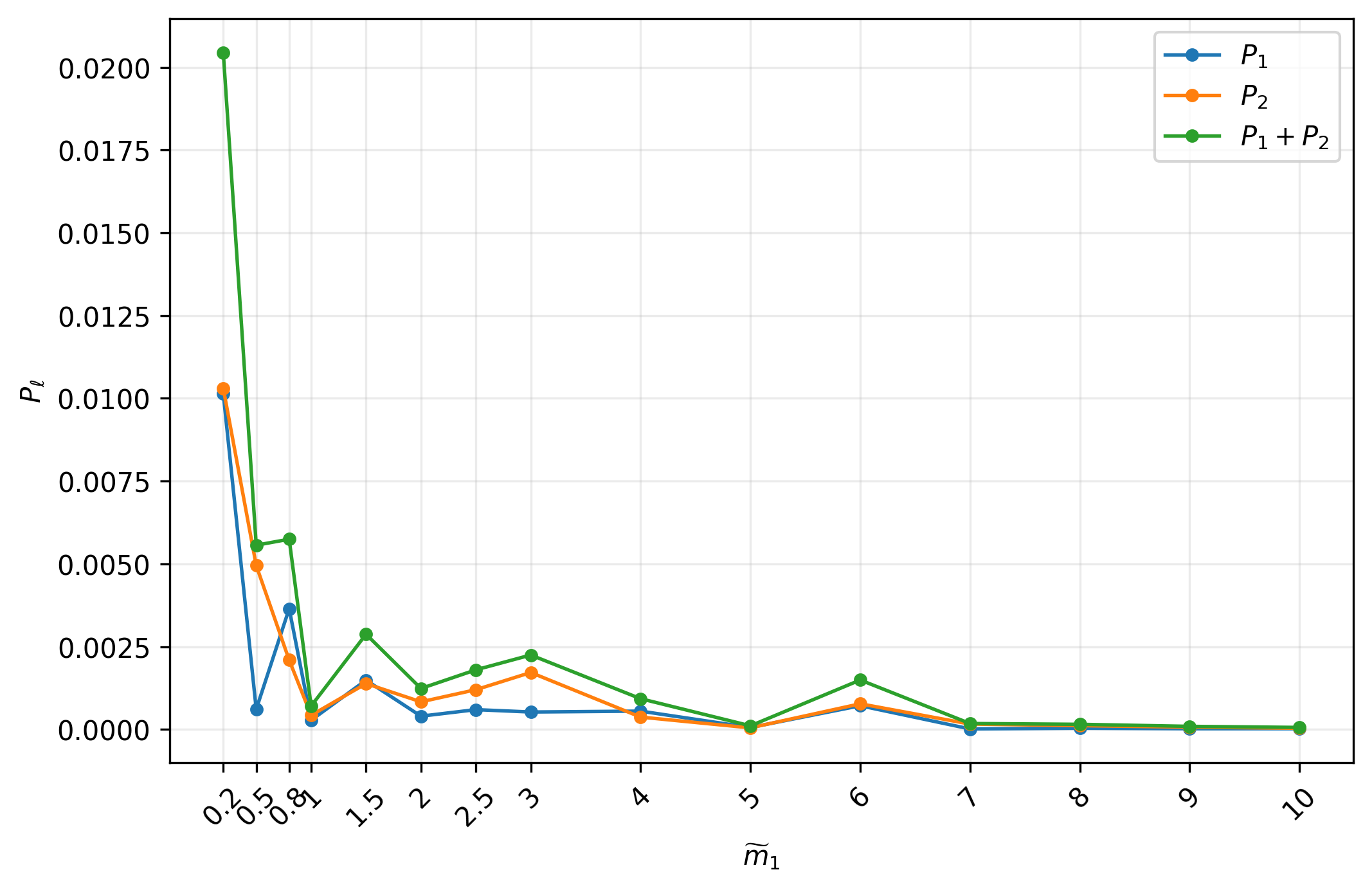}

\par\vspace{4pt}
\includegraphics[width=\textwidth,height=0.31\textheight,keepaspectratio]{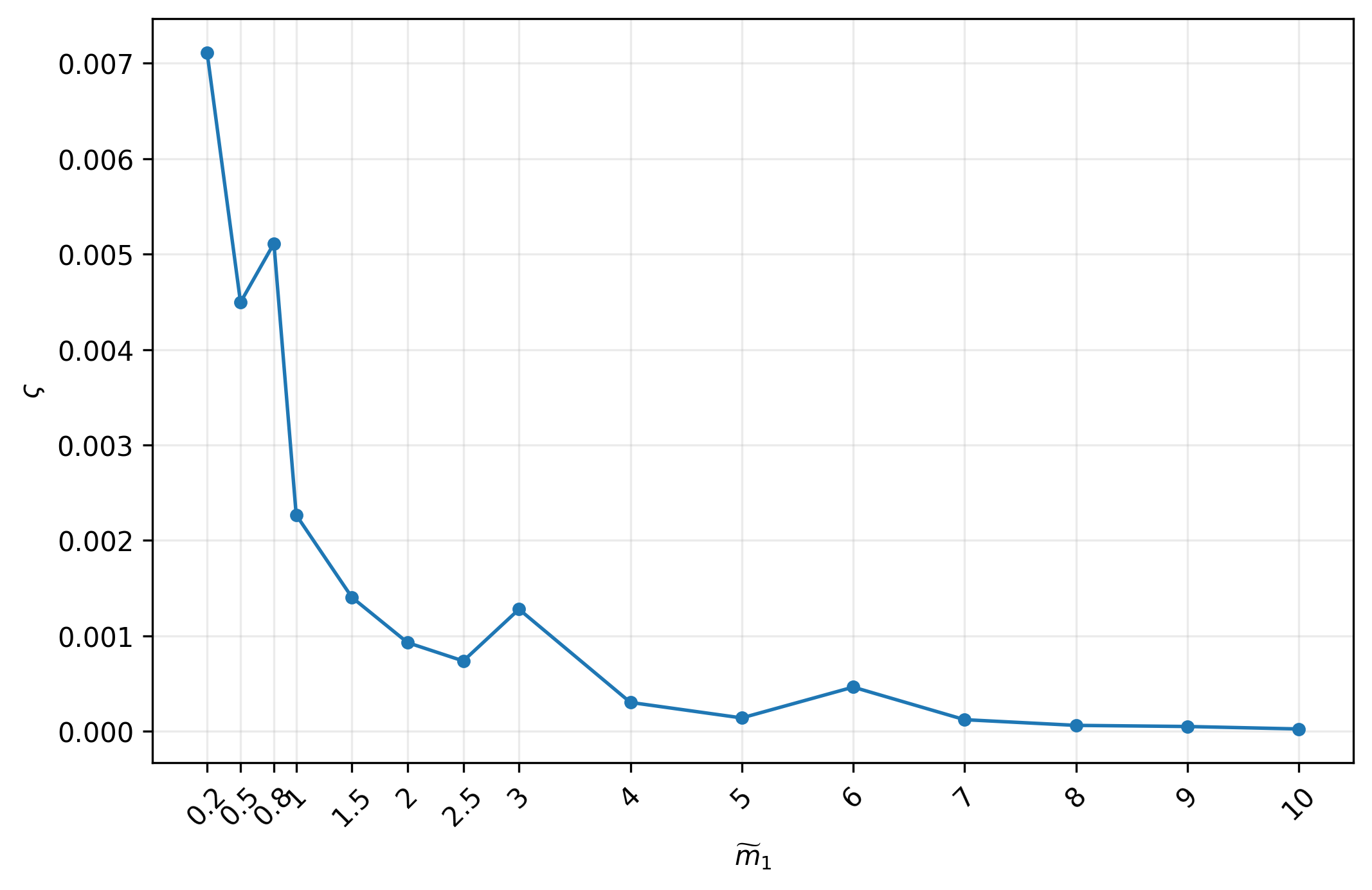}

\par\vspace{4pt}
\includegraphics[width=\textwidth,height=0.31\textheight,keepaspectratio]{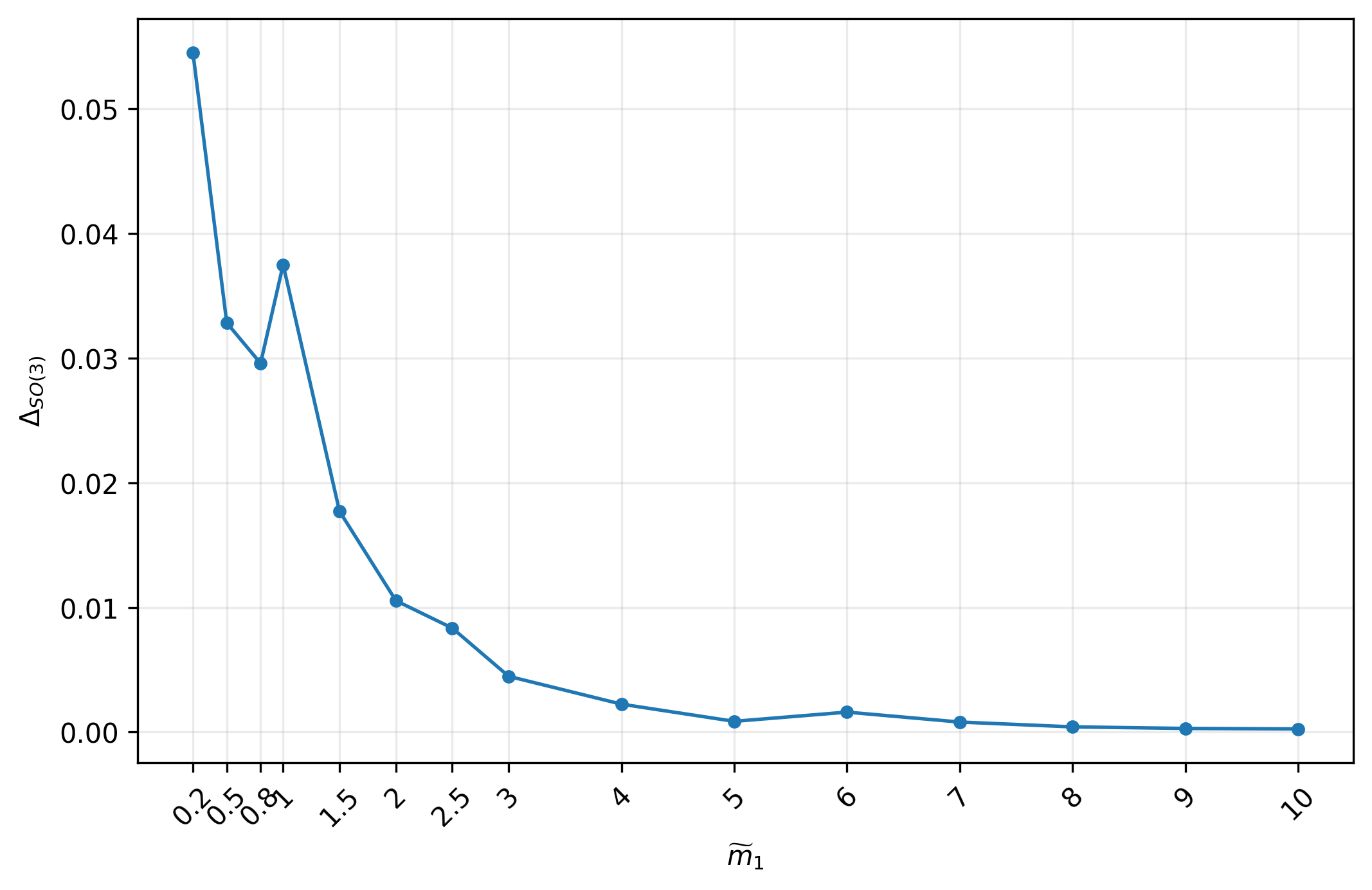}
\caption{Mass dependence of $P_1$, $P_2$, $P_1+P_2$, $\varsigma$, and $\Delta_{SO(3)}$ for  $M=8$ and $X=200$, varying $\widetilde m_1$ at fixed $\widetilde m_2=2$. The top, middle, and bottom panels show $P_1$, $P_2$, and $P_1+P_2$; $\varsigma$; and $\Delta_{SO(3)}$, respectively. All panels use the same sampled masses $\widetilde m_1=0.2,0.5,0.8,1,1.5,2,2.5,3,4,5,6,7,8,9,10$.}
\label{fig:four_matrix_order_parameters_m1}
\end{figure}

\begin{figure}[p]
\centering
\includegraphics[width=\textwidth,height=0.31\textheight,keepaspectratio]{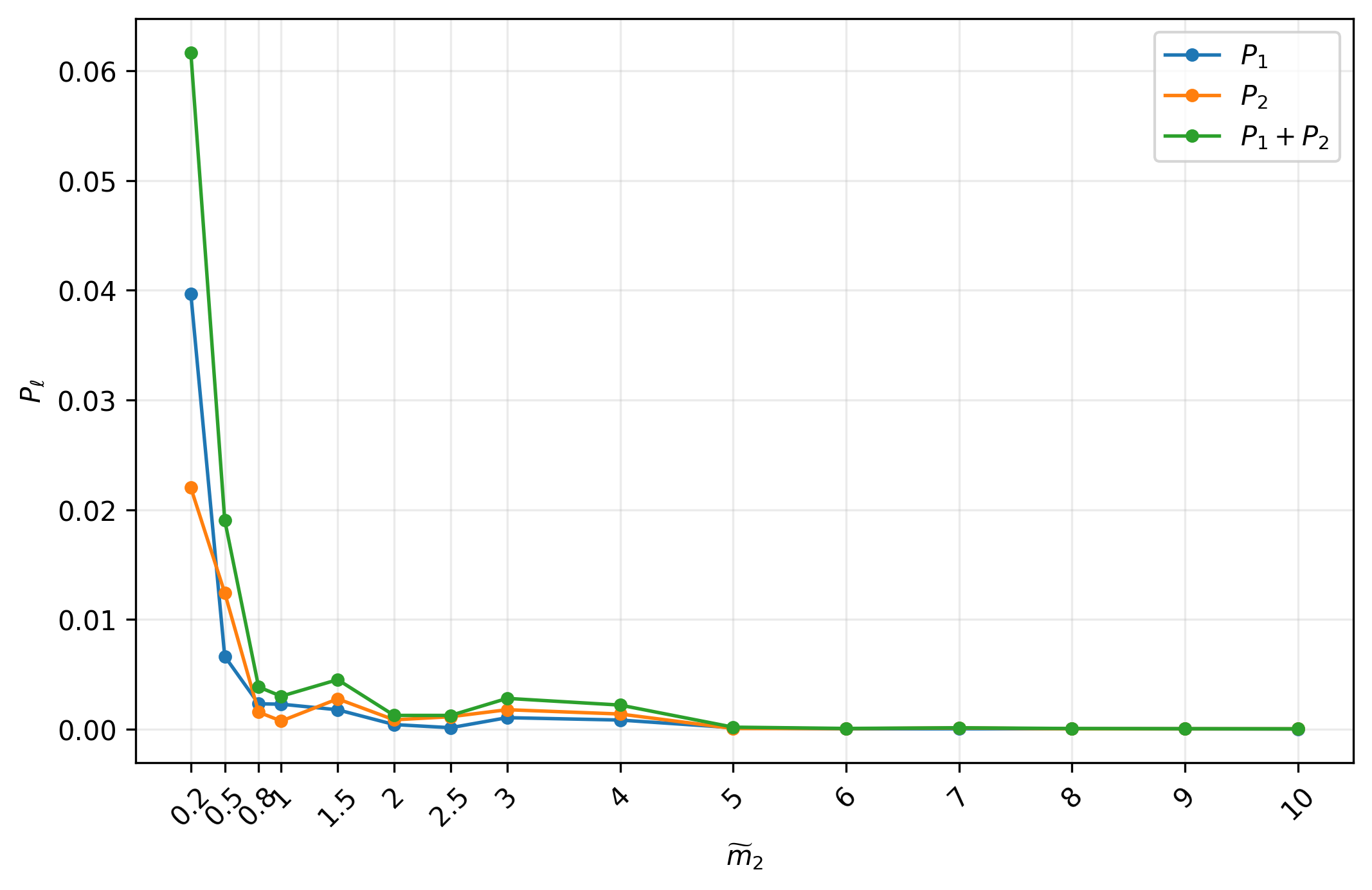}

\par\vspace{4pt}
\includegraphics[width=\textwidth,height=0.31\textheight,keepaspectratio]{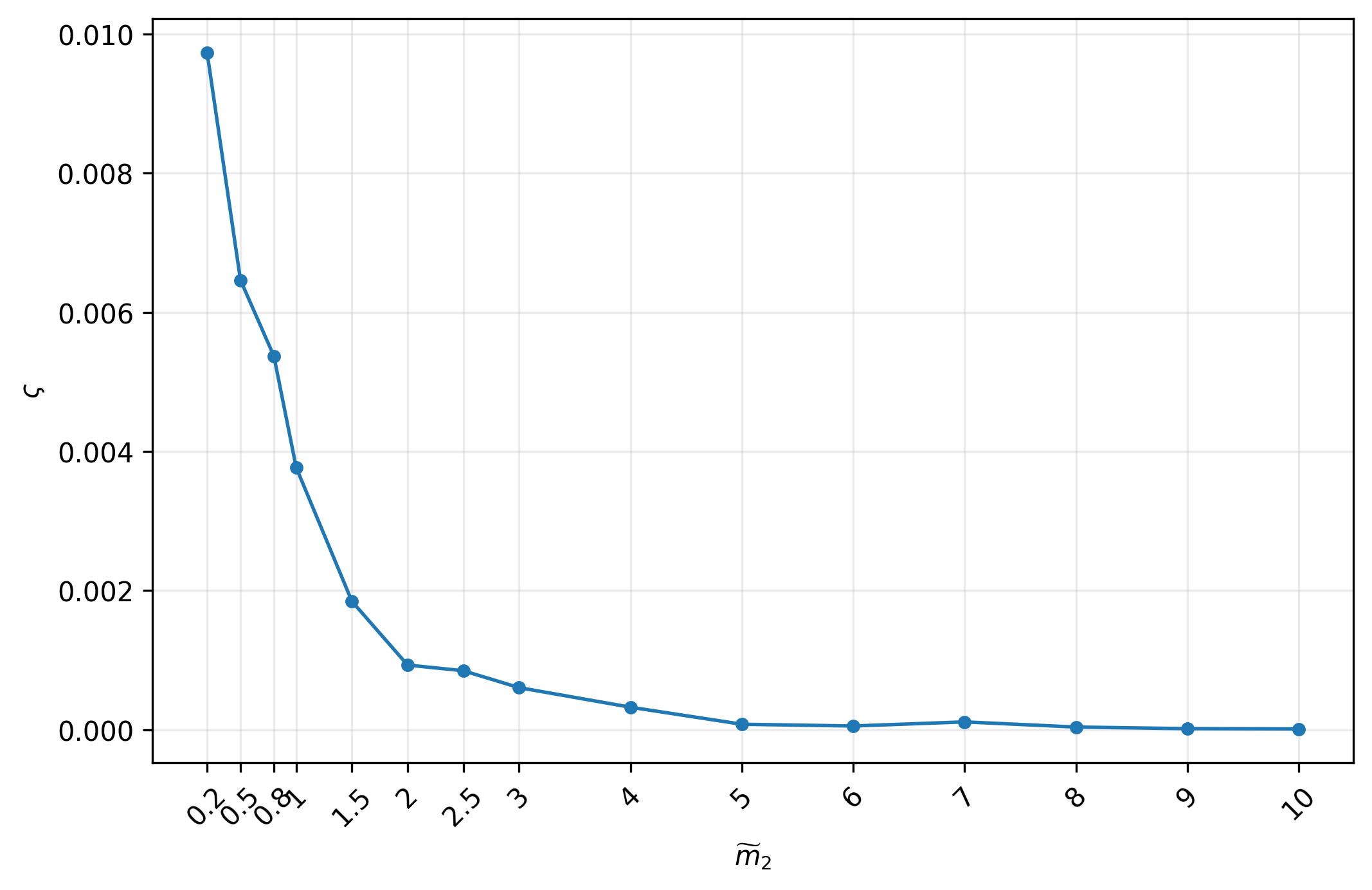}

\par\vspace{4pt}
\includegraphics[width=\textwidth,height=0.31\textheight,keepaspectratio]{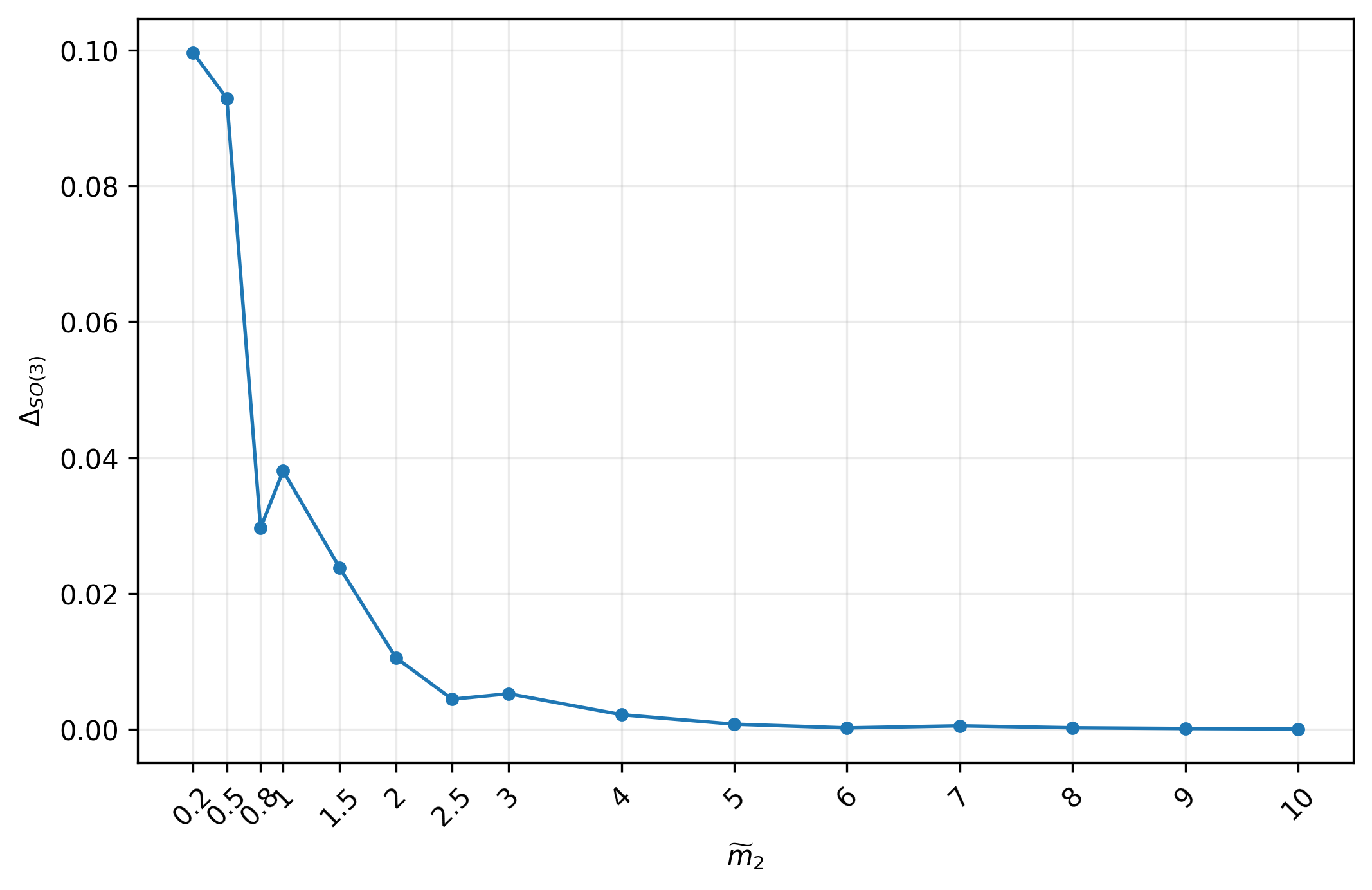}
\caption{Mass dependence of $P_1$, $P_2$, $P_1+P_2$, $\varsigma$, and $\Delta_{SO(3)}$ for $M=8$ and $X=200$, varying $\widetilde m_2$ at fixed $\widetilde m_1=2$. The top, middle, and bottom panels show $P_1$, $P_2$, and $P_1+P_2$; $\varsigma$; and $\Delta_{SO(3)}$, respectively. All panels use the same sampled masses $\widetilde m_2=0.2,0.5,0.8,1,1.5,2,2.5,3,4,5,6,7,8,9,10$.}
\label{fig:four_matrix_order_parameters_m2}
\end{figure}

These numerical results suggest that $P_1$, $P_2$, and $\varsigma$ may also serve as indicators of departures from the fuzzy-sphere configuration. We therefore plot these quantities in Figures~\ref{fig:four_matrix_order_parameters_m1} and~\ref{fig:four_matrix_order_parameters_m2}, together with $\Delta_{SO(3)}$ for comparison. The first figure varies $\widetilde m_1$ at fixed $\widetilde m_2=2$, while the second varies $\widetilde m_2$ at fixed $\widetilde m_1=2$. In each figure, the top panel shows $P_1$, $P_2$, and $P_1+P_2$, the middle panel shows $\varsigma$, and the bottom panel shows $\Delta_{SO(3)}$. The mode weights are plotted as fractions rather than the percentages listed in Table~\ref{tab:four_matrix_extra_direction}.

In both scans, $P_1$, $P_2$, and $\varsigma$ remain small at larger masses and generally increase as the varied mass decreases. The increase is most pronounced at the smallest sampled masses, where $\Delta_{SO(3)}$ also increases. At the smallest mass, the scan with fixed $\widetilde m_1=2$ gives larger values of $P_1+P_2$ and $\Delta_{SO(3)}$ than the scan with fixed $\widetilde m_2=2$. The two scans therefore show the same overall trend, although the values and their variation with mass differ. Neither scan shows a sufficiently clear onset to determine a critical mass.
The quantities in Table~\ref{tab:four_matrix_extra_direction} depend on the choice of gauge under similarity transformations, $\hat A_I^{(M)}\mapsto S\hat A_I^{(M)}S^{-1}$.
Nevertheless, fixing the residual gauge produces only slight changes in $P_1$, $P_2$, $P_1+P_2$, and $\varsigma$, leaving their qualitative mass dependence unchanged.

By definition, $\Delta_{SO(3)}$ measures the anisotropy of the quadratic moments. However, the loss of fuzzy-sphere structure does not necessarily imply that these moments become anisotropic, so the connection between $\Delta_{SO(3)}$ and a possible phase transition is indirect. The quantities $P_1$ and $P_2$ measure the contributions of the lowest Laplacian modes to the fourth matrix and may reveal changes that are not apparent in the quadratic moments. At small masses, these modes need not have definite angular momentum, as explained above.

The ratio $\varsigma$ measures the size of the fourth direction relative to the first three. Its increase at smaller masses shows that the fourth direction becomes less suppressed. Like $\Delta_{SO(3)}$, however, it does not directly measure the fuzzy-sphere structure of the first three matrices. The observed increases in $P_1$, $P_2$, $\varsigma$, and $\Delta_{SO(3)}$ suggest that these quantities may help identify the onset of a phase transition, but further calculations are needed to establish whether a transition occurs and to determine its location.

\FloatBarrier
\section{Ten-Matrix Model}\label{sec:ten_matrix}

The mass-deformed SYM-type ten-matrix model, also known as the polarised IKKT matrix model \cite{Bonelli:2002mb, Hartnoll:2024csr, Komatsu:2024bop, Komatsu:2024ydh}, has recently attracted renewed attention because of its connection to holographic duality.
Its action is given by
\begin{align}
S_{\text{p-IKKT}}[A,\Psi] & \equiv S_{\text{IKKT}}[A,\Psi]+S_{m}[A,\Psi],\\
S_{\text{IKKT}}[A,\Psi] & \equiv N\operatorname{Tr}\{-\frac{1}{4}[A_{\mu},A_{\nu}]^{2}+\frac{1}{2}\bar{\Psi}\Gamma^{\mu}[A_{\mu},\Psi]\},\\
S_{m}[A,\Psi] & \equiv N\operatorname{Tr}(\frac{m^{2}}{64}A^{2}_{I}+\frac{3m^{2}}{64}A^{2}_{i}+\frac{m}{3}i\epsilon_{ijk}A_{i}A_{j}A_{k}+\frac{im}{8}\Psi^{T}\Sigma\Psi).
\end{align}
Here $i,j,k=1,2,3$ and $I=4,\ldots,10$.
We adopt the gamma-matrix representation of
Ref.~\cite{Hartnoll:2024csr}, with the matrix indices
reassigned to match the convention $i=1,2,3$ and
$I=4,\ldots,10$ used in this paper.
We introduce nine $16\times16$ matrices
\begin{equation}
\begin{aligned}
\gamma^1&=
\begin{pmatrix}
0&I_8\\
I_8&0
\end{pmatrix},
\qquad
\gamma^2=
\begin{pmatrix}
I_8&0\\
0&-I_8
\end{pmatrix},\\
\gamma^I&=
\begin{pmatrix}
0&k^{I-3}\\
-k^{I-3}&0
\end{pmatrix}
\quad (I=4,\ldots,10),
\end{aligned}
\end{equation}
where the matrices $k^r$ ($r=1,\ldots,7$)
are constructed from the octonionic structure constants.
The Weyl blocks of the ten-dimensional gamma matrices
are then given by
\begin{equation}
\begin{aligned}
\sigma^1&=\bar\sigma^1=\gamma^1,\\
\sigma^2&=\bar\sigma^2=\gamma^2,\\
\sigma^3&=iI_{16},
\qquad \bar\sigma^3=-iI_{16},\\
\sigma^I&=\bar\sigma^I=\gamma^I
\quad (I=4,\ldots,10).
\end{aligned}
\end{equation}
The charge-conjugation matrix and fermionic
mass matrix are defined by
\begin{equation}
\mathcal C=-i\Gamma^3,
\qquad
\Sigma=i\gamma^1\gamma^2.
\end{equation}

Using these conventions, the bosonic loop equations in the large-$N$ limit are
\begin{align}
{\cal L}_{\mu;\mu_1\cdots\mu_n}
={}&
\sum_{k=1}^n\delta_{\mu_k,\mu}
w\{\mu_1\cdots\mu_{k-1}\}
w\{\mu_{k+1}\cdots\mu_n\}
\nonumber\\
&-w\{\mu_1\cdots\mu_n[\nu,[\nu,\mu]]\}
\nonumber\\
&+\frac12(\bar{\sigma}^\mu)_{\alpha\beta}
w\{\mu_1\cdots\mu_n\{\alpha,\beta\}\}
\nonumber\\
&-c_\mu w\{\mu_1\cdots\mu_n\mu\}
\nonumber\\
&-\frac{im}{2}\epsilon_{\mu jk}
w\{\mu_1\cdots\mu_n[j,k]\}
=0,
\label{eq:pIKKT_bosonic_loop}
\end{align}
where
\begin{equation}
c_\mu=
\begin{cases}
3m^2/32 & (\mu=1,2,3),\\
m^2/32 & (\mu=4,\ldots,10).
\end{cases}
\end{equation}
Here $\mu,\nu,\mu_1,\ldots,\mu_n=1,\ldots,10$ and $j,k=1,2,3$, with $\epsilon_{123}=1$.
The inserted commutators and anticommutators follow the notation introduced in Sec.~\ref{subsec:three_matrix_loops}.
As in the three-matrix model, we use only bosonic loop equations in the RMA, while the fermionic contributions are evaluated by Wick contraction.

Numerical calculations for this model were carried out in \cite{Chou:2025rwy} using the Lefschetz-thimble method.
The results showed excellent agreement with the analytic predictions obtained by localization and also demonstrated quantitatively that several distinct saddles contribute to the matrix integral.
For matrix integrals with a complex weight, the Lefschetz-thimble method deforms the integration contour into the complexified configuration space so as to reduce the oscillations associated with the complex phase, after which the integral is evaluated numerically.
Provided that the contour is deformed appropriately, this deformation does not change the value of the matrix integral.
Thus, like conventional Monte Carlo methods, the Lefschetz-thimble method evaluates the matrix integral itself numerically and may in this sense be regarded as a first-principles computational method.

The Lefschetz-thimble method is, however, known to be computationally expensive.
Since the computational cost grows as the matrix size $N$ is increased, studying the behavior of the model in the large-$N$ limit requires substantial computational resources.\footnote{Like the mass-deformed three- and four-matrix models discussed above, the polarised IKKT matrix model is a fuzzy-sphere-type matrix model and has a clear physical interpretation even at finite $N$.
Analytically and systematically evaluating finite-$N$ effects is often difficult, whereas Monte Carlo and Lefschetz-thimble calculations automatically include all such effects.
It should therefore be emphasized that methods designed to treat large-$N$ theories efficiently are not necessarily superior in every context.
Given the generality and power of Monte Carlo methods and their extensions, it is more appropriate to regard RMA as a complementary method that compensates for some of their limitations.}
By contrast, RMA is specifically designed for large-$N$ reduced matrix models with complex weights and allows the optimization to be performed at relatively low computational cost.
This advantage becomes particularly important for models such as the ten-matrix model, in which the number of different matrices is itself large.

The ten-matrix model has already been studied using localization techniques, and phase transitions have been reported to occur depending on the mass parameter $m$ and the matrix size $N$ \cite{Hartnoll:2024csr, Hartnoll:2025ecj}.
For our action, the critical mass at fixed $N$ is expressed as
\begin{equation}
\begin{aligned}
\bar{m}_{N} &\equiv N^{\frac{3}{4}}m,\\
\bar{m}_{N,\text{cr}} &\sim 8.06.
\end{aligned}
\end{equation}
In other words, the phase transition occurs when $m$ becomes as small as $m\sim N^{-3/4}$.
Since we know in advance that the relevant scale is $m\sim N^{-3/4}$, we use the corresponding mass parameter $\bar{m}$ from the outset.
As in the previous cases, we implement this scaling within the RMA by replacing $N$ with $M$ and defining
\begin{equation}
\bar{m}\equiv M^{\frac{3}{4}}m.
\end{equation}
We then investigate whether a phase transition occurs near $\bar{m}\sim 8$.

However, there is a theoretical subtlety that should be noted here.
When using the original parameter $\bar{m}_{N}$, we must take $m\to0$ as $N\to\infty$ to keep $\bar{m}_{N}\equiv N^{\frac{3}{4}}m$ of order unity and induce the phase transition.
In contrast, our definition $\bar{m}\equiv M^{\frac{3}{4}}m$ uses finite $M$ in place of $N$, so $m$ remains finite when $\bar{m}\sim 8$.
Since the RMA always assumes the large-$N$ limit, it is not obvious whether phase-transition-like behavior can be captured at such a finite value of $m$, and this must be examined numerically.

Apart from the phase transition, it is also important, as in the previous cases, to compare the radius of the fuzzy sphere with its saddle-point value.
The fuzzy-sphere solution (a saddle point of the equations of motion) in the ten-matrix model is given by $A^{\text{FS}}_{i}=\frac{3m}{8}L_{i}$.
Combining this with the definition of $\bar{m}_{N}$ gives
\begin{equation}
\begin{aligned}
(A^{\text{FS}}_{1})^{2}+(A^{\text{FS}}_{2})^{2}+(A^{\text{FS}}_{3})^{2}
&=\frac{9m^{2}}{256}(N^{2}-1)\mathbf{1}_{N}\\
&=\frac{9\bar{m}^{2}_{N}}{256}(N^{\frac{1}{2}}-N^{-\frac{3}{2}})\mathbf{1}_{N}.
\end{aligned}
\end{equation}
Since only the leading term in $N$ matters in the large-$N$ limit, the value of $C_{\text{FS}}$ expressed in terms of $\bar{m}_{N}$ is given by
\begin{equation}
C_{\text{FS}}\equiv\lim_{N\to\infty}\frac{1}{N^{\frac{3}{2}}}\text{Tr}_{N}\{(A^{\text{FS}}_{1})^{2}+(A^{\text{FS}}_{2})^{2}+(A^{\text{FS}}_{3})^{2}\}=\frac{9\bar{m}^{2}_{N}}{256}.
\end{equation}
Similarly, we define the normalized second moments as
\begin{equation}
\bar{w}^{(M)}\{\mu\nu\}\equiv\frac{1}{M^{\frac{3}{2}}}\text{Tr}_{M}(\hat{A}^{(M)}_{\mu}\hat{A}^{(M)}_{\nu}).
\end{equation}
We use these quantities to analyze the numerical results obtained from the RMA.

Unlike the three-matrix and four-matrix models, the ten-matrix model considered here has only one mass parameter, $\bar{m}$.
We therefore perform the RMA for different values of $\bar{m}$ to investigate how the physical observables change.
As before, we generally expect quantum fluctuations to become less significant as $\bar{m}$ increases, leading to better agreement between $C_{M}$ and $C_{\text{FS}}$.
Conversely, as $\bar{m}$ decreases, we expect quantum fluctuations to produce an increasingly large discrepancy between $C_{M}$ and $C_{\text{FS}}$, eventually leading to a phase transition in which the fuzzy-sphere solution becomes unstable.

For the optimization, we use the fuzzy-sphere configuration as the initial condition.
For computational efficiency, we primarily use the \texttt{complex64} data type in our CUDA/CuPy calculations, which provides approximately seven significant decimal digits of precision.
We generally set the matrix size to $M=8$ and choose $X$ so that $F=0$ forms a square system (i.e., the number of independent loop equations equals the number of independent variables).
This is because the use of \texttt{complex64} makes the optimization prone to termination upon satisfying the \texttt{xtol} criterion, and the value of $X$ for which $F\sim10^{-16}$ can be achieved tends to be smaller than the value specified above.
When investigating the $M$ dependence, we use different values of $M$ but choose $X$ according to the same criterion.
In this section, we take $A_\mu$ to be traceless Hermitian matrices in $\mathfrak{su}(N)$ and, correspondingly, the regularized master fields $\hat{A}^{(M)}_\mu$ to be traceless complex matrices.
This is because, if $\hat{A}^{(M)}_\mu$ were allowed to be general complex matrices without the tracelessness condition, the Dirac operator would necessarily have zero modes in the massless limit.

\begin{figure}[t] 
\centering
\includegraphics[width=0.7\textwidth]{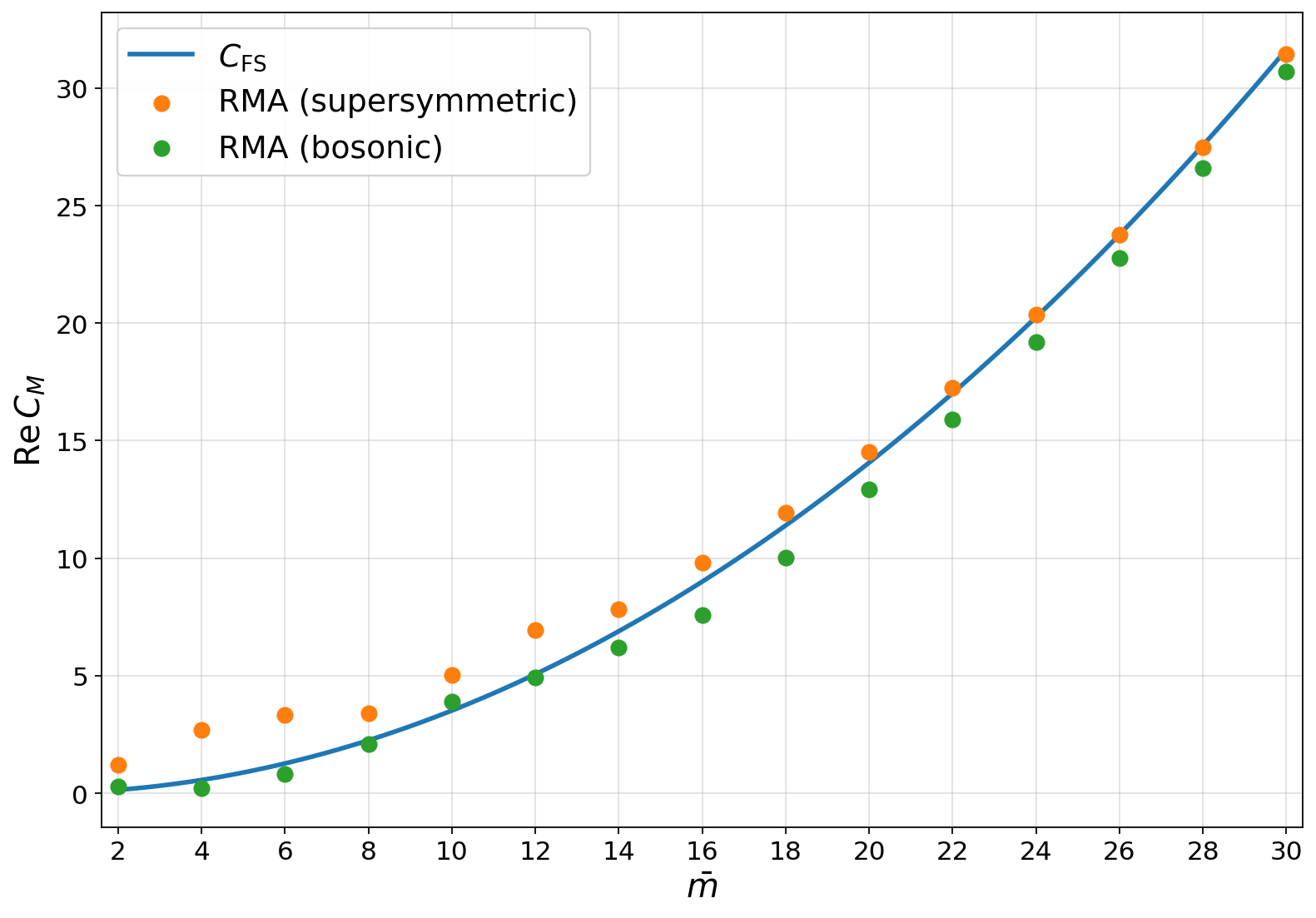}
\caption{
Comparison of $\text{Re}C_{M}$ and $C_{\text{FS}}$ for $M=8$, $X=601$, and $\bar{m}=2,4,\ldots,30$.
The solid blue line represents $C_{\text{FS}}$, while the orange and green dots represent the values of $\text{Re}C_{M}$ obtained by applying the RMA to the supersymmetric ten-matrix model with fermions and the bosonic ten-matrix model without fermions, respectively.
}
\label{fig:pIKKT_2nd_moment_M8_X601}
\end{figure}

In Figure \ref{fig:pIKKT_2nd_moment_M8_X601}, we first plot $\text{Re}C_{M}$ for $M=8$ and $\bar{m}=2,4,\ldots,30$ and compare it with $C_{\text{FS}}$.
We also include results for the bosonic model without fermions in this figure.
Overall, $\text{Re}C_{M}$ agrees reasonably well with $C_{\text{FS}}$, regardless of whether fermions are included.
More specifically, as the mass parameter $\bar{m}$ increases, $\text{Re}C_{M}$ approaches $C_{\text{FS}}$, namely the saddle-point value, whereas the discrepancy between them grows as $\bar{m}$ decreases.
Furthermore, the values of $\text{Re}C_{M}$ obtained with fermions (orange points) are larger than those obtained in the bosonic model (green points).
As in the three-matrix and four-matrix models, this can be interpreted as an indication that fermions tend to increase the radius of the fuzzy sphere.
On the other hand, compared with these two models, both the deviation of $\text{Re}C_M$ from $C_{\text{FS}}$ and the difference between the results with and without fermions are larger.
We suspect that this is because the ten-matrix model has more bosonic and fermionic matrices, resulting in larger bosonic matrix fluctuations and fermionic contributions.

\begin{table}[t]
\centering
\begin{tabular}{c c c c c}
\hline
$\bar{m}$ & $\bar{w}^{(M)}\{11\}$ & $\bar{w}^{(M)}\{22\}$ & $\bar{w}^{(M)}\{33\}$ & $\Delta_{SO(3)}$ \\
\hline
30 & $10.48-0.00i$ & $10.47-0.00i$ & $10.48+0.00i$ & $0.00099$ \\
20 & $4.84+0.00i$ & $4.84+0.01i$ & $4.84+0.03i$ & $0.00353$ \\
18 & $3.99+0.00i$ & $3.96-0.00i$ & $3.97+0.01i$ & $0.00520$ \\
16 & $3.27-0.03i$ & $3.26-0.07i$ & $3.26-0.05i$ & $0.00778$ \\
14 & $2.58-0.02i$ & $2.67-0.04i$ & $2.60-0.03i$ & $0.02190$ \\
12 & $2.30-0.04i$ & $2.30-0.09i$ & $2.34-0.03i$ & $0.02053$ \\
10 & $1.71+0.01i$ & $1.72+0.03i$ & $1.59+0.10i$ & $0.04765$ \\
 8 & $1.20-0.07i$ & $1.07-0.07i$ & $1.13-0.03i$ & $0.05733$ \\
 6 & $1.13-0.02i$ & $1.09-0.02i$ & $1.10+0.04i$ & $0.07744$ \\
 4 & $0.93-0.06i$ & $0.89+0.00i$ & $0.86+0.02i$ & $0.08184$ \\
 2 & $0.84+0.02i$ & $0.18+0.02i$ & $0.18-0.01i$ & $0.62122$ \\
\hline
\end{tabular}
\caption{Values of $\bar{w}^{(M)}\{11\}$, $\bar{w}^{(M)}\{22\}$, $\bar{w}^{(M)}\{33\}$, and $\Delta_{SO(3)}$ for $M=8$, $X=601$, and $\bar{m}=2,4,\ldots,20,30$.
The values of $\bar{w}^{(M)}\{11\}$, $\bar{w}^{(M)}\{22\}$, and $\bar{w}^{(M)}\{33\}$ are shown to two decimal places.}
\label{tab:pIKKT_SO3}
\end{table}

Although the graph shows numerical agreement between $\text{Re}C_{M}$ and $C_{\text{FS}}$, this does not necessarily imply that a stable fuzzy-sphere phase is realized for all values of $\bar{m}$.
In fact, in Figure \ref{fig:pIKKT_2nd_moment_M8_X601}, the RMA results exhibit smooth behavior beyond $\bar{m}\sim10$, below which somewhat irregular fluctuations appear.
The origin of this instability can be understood by examining the numerical results in greater detail.

Table \ref{tab:pIKKT_SO3} lists the values of $\bar{w}^{(M)}\{11\}$, $\bar{w}^{(M)}\{22\}$, $\bar{w}^{(M)}\{33\}$, and $\Delta_{SO(3)}$, which measures the degree of $SO(3)$ symmetry breaking in the second moments (see equation \eqref{eq:SO3_deviation} for its definition), for $M=8$ and $\bar{m}=2,4,\ldots,20,30$.\footnote{
Although we use the term SO(3) symmetry breaking here, we are only measuring whether the second moments exhibit SO(3) symmetry breaking in an optimization initialized with the fuzzy-sphere configuration.
If the fuzzy-sphere configuration is unstable, it is not a stable vacuum in the first place, so even a large value of $\Delta_{SO(3)}$ does not imply that the true vacuum breaks SO(3) symmetry, i.e., that spontaneous SO(3) symmetry breaking occurs.
}
Notably, even though all the optimizations start from the fuzzy-sphere configuration, the breaking of $SO(3)$ symmetry in the second moments becomes increasingly pronounced below a certain value of $\bar{m}$.
More specifically, the $SO(3)$ symmetry is preserved to very high precision at $\bar{m}=30$, and this remains the case down to approximately $\bar{m}\sim20$.
However, as $\bar{m}$ decreases to around $15$, $\Delta_{SO(3)}$ begins to increase gradually, albeit slightly, and then rises sharply at $\bar{m}=2$.
This appears to be the phase-transition-like behavior we anticipated.

\begin{figure}[t] 
\centering
\includegraphics[width=1.0\textwidth]{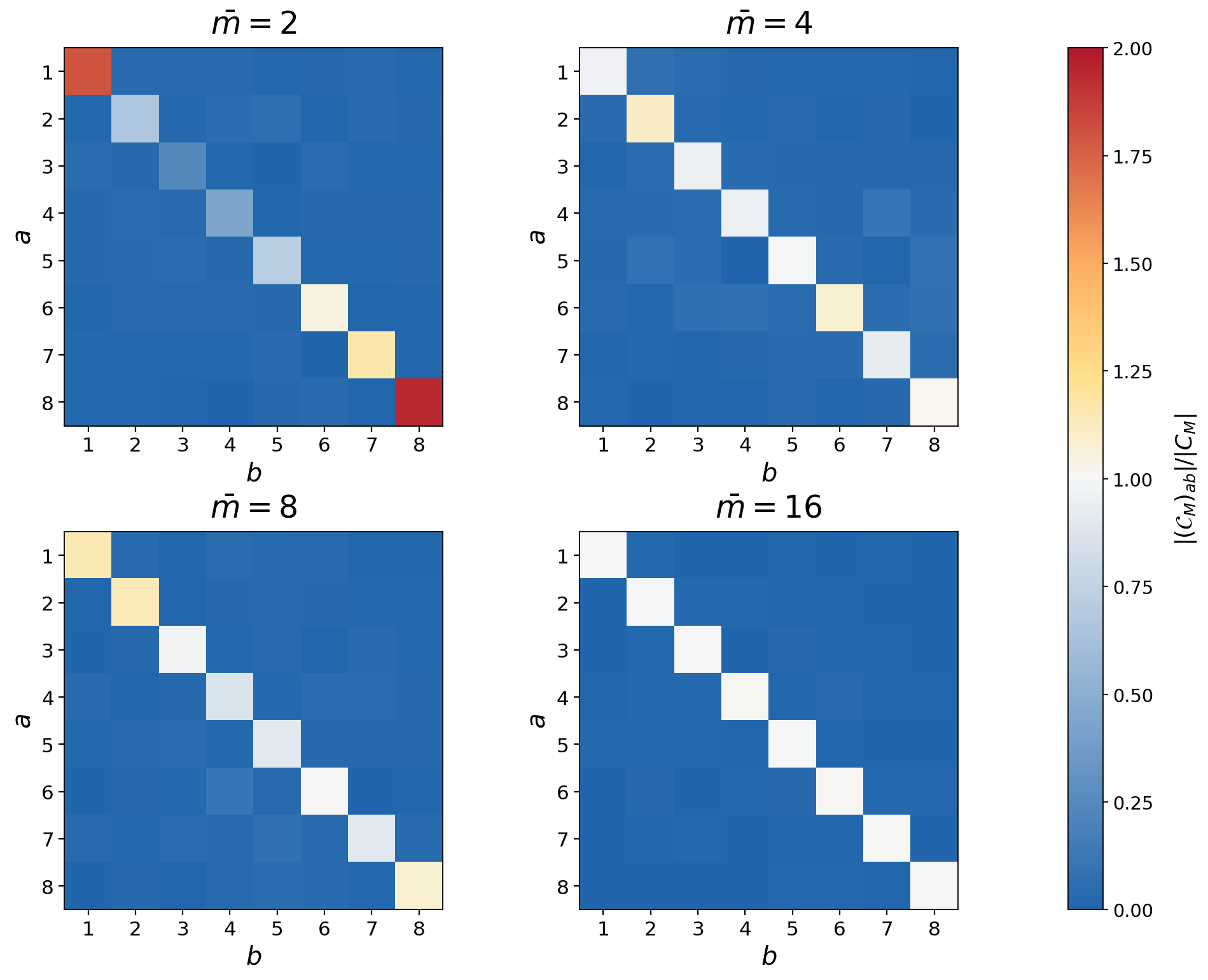}
\caption{
Heatmaps of $|({\cal C}_{M})_{ab}|/|C_{M}|$ for $M=8$, $X=601$, and $\bar{m}=2,4,8,16$.
The upper left, upper right, lower left, and lower right panels correspond to $\bar{m}=2$, $4$, $8$, and $16$, respectively.
}
\label{fig:pIKKT_heatmaps}
\end{figure}

To examine this from another perspective, we also calculate the Casimir matrix ${\cal C}_{M}$ for $M=8$.
Here, we compute $|({\cal C}_{M})_{ab}|/|C_{M}|$ for $\bar{m}=2,4,8,16$ and present the results as heatmaps in Figure \ref{fig:pIKKT_heatmaps}.
While the matrix with elements $|({\cal C}_{M})_{ab}|/|C_{M}|$ is nearly equal to the identity matrix at $\bar{m}=16$, the off-diagonal elements of ${\cal C}_{M}$ become slightly larger at $\bar{m}=4,8$, and some variations also appear among the diagonal elements.
At $\bar{m}=2$, the identity-matrix-like structure has clearly broken down.
This is a clear indication of the instability of the fuzzy-sphere configuration.

\begin{figure*}[t]
\centering
\begin{minipage}[t]{0.49\textwidth}
    \centering
    \includegraphics[width=\linewidth]{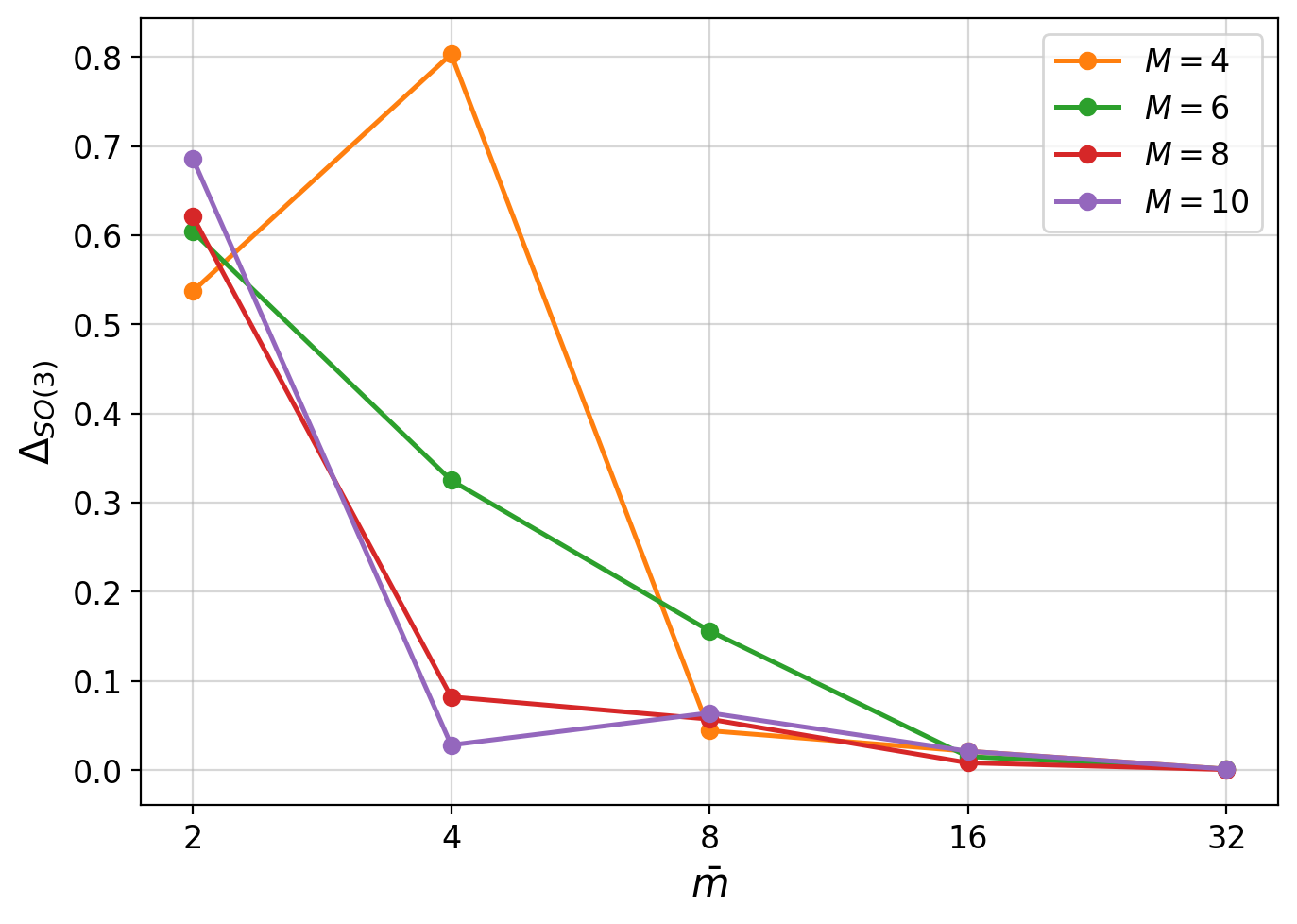}
\end{minipage}
\hfill
\begin{minipage}[t]{0.49\textwidth}
    \centering
    \includegraphics[width=\linewidth]{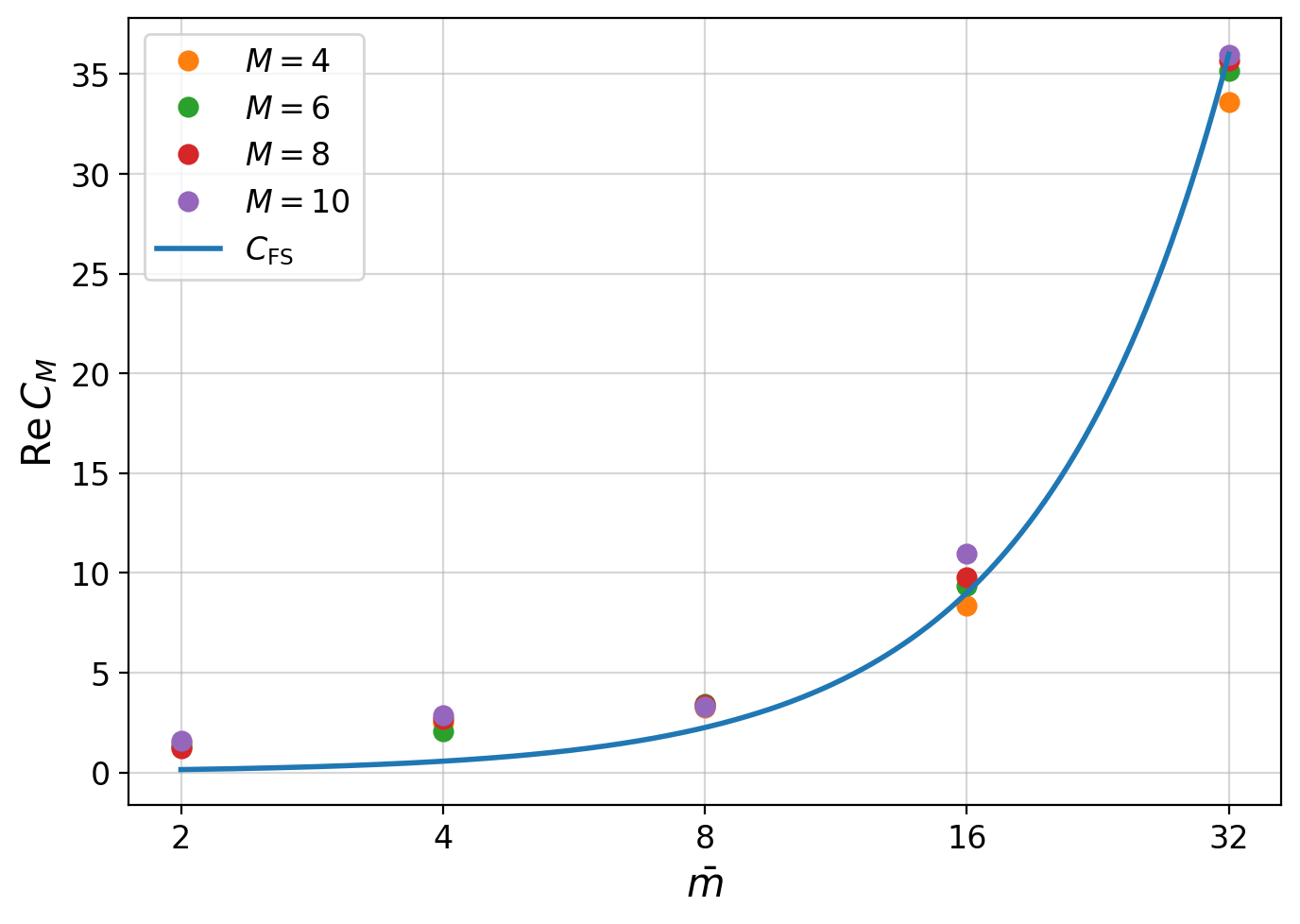}
\end{minipage}
\caption{Plots of $\Delta_{SO(3)}$ and $\operatorname{Re}C_{M}$ using the data presented in Table \ref{tab:pIKKT_M_dependence}.
The left and right panels show $\Delta_{SO(3)}$ and $\operatorname{Re}C_{M}$, respectively.
The vertical axes represent $\Delta_{SO(3)}$ and $\operatorname{Re}C_{M}$, while the horizontal axes use a logarithmic scale with $\bar{m}=2,4,8,16,32$.
The orange, green, red, and purple points correspond to $M=4,6,8,10$, respectively.}
\label{fig:pIKKT_M_dependence}
\end{figure*}

\begin{table*}[t]
\centering
\small
\setlength{\tabcolsep}{4pt}
\begin{tabular}{cccccccc}
\hline
$\bar{m}$ & $M$ & $\bar{w}^{(M)}\{11\}$ & $\bar{w}^{(M)}\{22\}$ & $\bar{w}^{(M)}\{33\}$ & $\Delta_{SO(3)}$ & $\operatorname{Re}C_{M}$ & $\operatorname{Re}C_{M}/C_{\text{FS}}$ \\
\hline
32 & 4 & $11.20-0.01i$ & $11.22-0.01i$ & $11.19-0.02i$ & $0.001$ & $33.611$ & $0.934$ \\
32 & 6 & $11.72-0.01i$ & $11.71-0.01i$ & $11.72-0.01i$ & $0.001$ & $35.144$ & $0.976$ \\
32 & 8 & $11.90+0.01i$ & $11.89+0.01i$ & $11.89+0.01i$ & $0.000$ & $35.684$ & $0.991$ \\
32 & 10 & $12.01+0.00i$ & $11.98+0.00i$ & $11.98+0.00i$ & $0.001$ & $35.972$ & $0.999$ \\
\hline
16 & 4 & $2.74+0.03i$ & $2.81+0.03i$ & $2.83+0.03i$ & $0.021$ & $8.375$ & $0.931$ \\
16 & 6 & $3.09+0.00i$ & $3.10-0.02i$ & $3.16+0.01i$ & $0.015$ & $9.348$ & $1.039$ \\
16 & 8 & $3.27-0.03i$ & $3.26-0.07i$ & $3.26-0.05i$ & $0.008$ & $9.795$ & $1.088$ \\
16 & 10 & $3.74-0.02i$ & $3.58-0.03i$ & $3.62-0.03i$ & $0.021$ & $10.944$ & $1.216$ \\
\hline
8 & 4 & $1.11+0.00i$ & $1.13+0.00i$ & $1.02+0.00i$ & $0.044$ & $3.260$ & $1.449$ \\
8 & 6 & $0.96+0.12i$ & $1.29+0.00i$ & $1.18-0.03i$ & $0.156$ & $3.440$ & $1.529$ \\
8 & 8 & $1.20-0.07i$ & $1.07-0.07i$ & $1.13-0.03i$ & $0.057$ & $3.405$ & $1.513$ \\
8 & 10 & $1.17+0.02i$ & $1.07-0.01i$ & $1.05-0.01i$ & $0.064$ & $3.294$ & $1.464$ \\
\hline
4 & 4 & $0.05+0.00i$ & $0.46+0.11i$ & $1.98-0.09i$ & $0.803$ & $2.480$ & $4.409$ \\
4 & 6 & $0.55+0.09i$ & $0.89-0.15i$ & $0.61+0.04i$ & $0.325$ & $2.059$ & $3.661$ \\
4 & 8 & $0.93-0.06i$ & $0.89+0.00i$ & $0.86+0.02i$ & $0.082$ & $2.676$ & $4.757$ \\
4 & 10 & $0.95-0.02i$ & $0.96+0.02i$ & $0.94+0.00i$ & $0.028$ & $2.850$ & $5.067$ \\
\hline
2 & 4 & $0.77+0.34i$ & $0.49+0.03i$ & $0.24+0.16i$ & $0.537$ & $1.494$ & $10.623$ \\
2 & 6 & $0.91-0.04i$ & $0.20+0.01i$ & $0.22+0.00i$ & $0.604$ & $1.327$ & $9.439$ \\
2 & 8 & $0.84+0.02i$ & $0.18+0.02i$ & $0.18-0.01i$ & $0.621$ & $1.194$ & $8.493$ \\
2 & 10 & $1.25-0.01i$ & $0.18+0.00i$ & $0.18+0.00i$ & $0.686$ & $1.612$ & $11.463$ \\
\hline
\end{tabular}
\caption{Values of $\bar{w}^{(M)}\{11\}$, $\bar{w}^{(M)}\{22\}$, $\bar{w}^{(M)}\{33\}$, $\Delta_{SO(3)}$, $\operatorname{Re}C_{M}$, and $\operatorname{Re}C_{M}/C_{\text{FS}}$ for $M=4,6,8,10$ and $\bar{m}=2,4,8,16,32$.
The corresponding values of $X$ are $169$, $349$, $601$, and $925$, in ascending order of $M$.
The second moments are shown to two decimal places, while $\Delta_{SO(3)}$ is shown to three decimal places.}
\label{tab:pIKKT_M_dependence}
\end{table*}

Next, we investigate the $M$ dependence of the optimization results.
In \cite{Hartnoll:2024csr,Hartnoll:2025ecj}, a phase transition was identified near $\bar{m}_{N,\text{cr}}\sim 8$, where $\bar{m}_{N}\equiv N^{\frac{3}{4}}m$, by performing calculations at different values of $N$.
However, the theory we investigate here using the RMA is already in the large-$N$ limit, and $M$ is merely a regularization parameter.
It is therefore necessary to perform calculations at different values of $M$ to determine whether $\bar{m}\sim 8$ is indeed the correct critical value for $\bar{m}\equiv M^{\frac{3}{4}}m$.
Here, we perform optimizations for different mass parameters $\bar{m}=2,4,8,16,32$ and matrix sizes $M=4,6,8,10$.
The corresponding values of $X$ are $169$, $349$, $601$, and $925$, respectively.
Table \ref{tab:pIKKT_M_dependence} lists the values of $\bar{w}^{(M)}\{11\}$, $\bar{w}^{(M)}\{22\}$, $\bar{w}^{(M)}\{33\}$, $\Delta_{SO(3)}$, $\operatorname{Re}C_{M}$, and $\operatorname{Re}C_{M}/C_{\text{FS}}$, while Figure \ref{fig:pIKKT_M_dependence} shows plots of $\Delta_{SO(3)}$ and $\operatorname{Re}C_{M}$.

A common feature across all values of $M$ is that the fuzzy-sphere phase becomes increasingly unstable as $\bar{m}$ decreases, with $\Delta_{SO(3)}$ eventually reaching values of order $0.1$--$1.0$.
However, there is a slight $M$ dependence: at fixed $\bar{m}$, the $SO(3)$ symmetry of the second moments tends to break more readily for smaller $M$.
For example, at $\bar{m}=4$, we find $\Delta_{SO(3)}<0.1$ for $M=8,10$, whereas $\Delta_{SO(3)}\gg0.1$ for $M=4,6$.
In other words, while the $SO(3)$ symmetry of the second moments is approximately preserved in the former cases, it is strongly broken in the latter cases, indicating that the fuzzy sphere has effectively collapsed.
On the other hand, at $\bar{m}=2$, the $SO(3)$ symmetry is broken for all values of $M$ considered, confirming phase-transition-like behavior in every case examined here.

Another notable feature is that $\operatorname{Re}C_{M}$, which characterizes the radius of the fuzzy sphere, depends on $M$.
For $\bar{m}=16,32$, the fuzzy-sphere phase is expected to be stable for all values of $M$ considered, yet $\operatorname{Re}C_{M}/C_{\text{FS}}$ increases with $M$.
In particular, at $\bar{m}=16$ and $M=10$, we find $\operatorname{Re}C_{M}/C_{\text{FS}}\approxeq1.216$, which is significantly larger than unity.
This suggests that increasing $X$ allows the optimization to capture more of the effects that enlarge the radius of the fuzzy sphere.
Nevertheless, at $\bar{m}=32$, $\operatorname{Re}C_{M}/C_{\text{FS}}$ is close to unity, so the qualitative behavior remains unchanged: quantum effects become smaller as $\bar{m}$ increases, reducing the discrepancy between $C_{M}$ and $C_{\text{FS}}$.
Since we generally expect the numerical results to be more reliable at larger $M$, we suspect that the smaller values of $\operatorname{Re}C_{M}$ obtained at smaller $M$ for $\bar{m}=16,32$ are due to undesirable finite-$M$ effects.
We note that the fuzzy-sphere phase becomes unstable for $\bar{m}\lesssim 8$, so the $M$ dependence of $\operatorname{Re}C_{M}$ in this regime is not considered particularly important.

The destabilization of the fuzzy-sphere configuration at small mass parameters is also observed in the three-matrix and four-matrix models, but the important result here is that the RMA approximately reproduces the critical value $\bar{m}_{N,\text{cr}}\sim8.06$ found in previous calculations.
As noted above, the RMA uses $\bar{m}$ rather than $\bar{m}_{N}$, so it is far from obvious that the fuzzy-sphere configuration should become unstable near $\bar{m}\sim8$.

However, several important caveats should be noted: the observed phase-transition-like behavior exhibits some dependence on $M$; although we use $\Delta_{SO(3)}$ as an order parameter, it may not always serve as the most reliable indicator; for $M=6$, $\Delta_{SO(3)}$ increases continuously rather than discontinuously; and when the optimization starts from a random configuration, the fuzzy-sphere phase is not reached even as $\bar{m}$ increases, and no phase-transition-like behavior is observed.
Taken together, although the numerical calculations using the RMA clearly capture signatures of a phase transition, we should conclude that the results are not yet sufficiently precise to determine the critical value of $\bar{m}$.
As we discuss below, quantities constructed from the matrices in the extra-dimensional directions may also serve as order parameters for the transition.

Finally, we examine the seven matrices $\hat A_\mu^{(M)}$
($\mu=4,\ldots,10$) in the directions other than those forming the
fuzzy sphere. We first consider the matrix Laplacian
\begin{equation}
  \mathcal H(X) \equiv
  \sum_{i=1}^{3}
  [ (\hat A_i^{(M)})^\dagger, [\hat A_i^{(M)},X] ].
  \label{eq:pIKKT_extra_laplacian}
\end{equation}
We use this form because the optimized matrices are generally complex:
$\mathcal H$ is Hermitian and positive semidefinite, so its eigenmatrices
can be chosen orthonormal. For an exact fuzzy sphere, the eigenvalues
are proportional to $\ell(\ell+1)$, with degeneracy $2\ell+1$ for
$\ell=1,\ldots,M-1$ on the space of traceless matrices.
Let $Y_{\ell r}$ ($r=1,\ldots,2\ell+1$) denote these eigenmatrices,
normalized by $\operatorname{Tr}(Y_{\ell r}^\dagger Y_{\ell' r'})
=\delta_{\ell\ell'}\delta_{rr'}$. We define
\begin{equation}
  P_\ell \equiv
  \frac{
    \displaystyle\sum_{\mu=4}^{10}\sum_{r=1}^{2\ell+1}
      \left|\operatorname{Tr}(Y_{\ell r}^\dagger\hat A_\mu^{(M)})\right|^2
  }{
    \displaystyle\sum_{\mu=4}^{10}\|\hat A_\mu^{(M)}\|_F^2
  },\qquad
  \|X\|_F^2\equiv\operatorname{Tr}(X^\dagger X).
  \label{eq:pIKKT_extra_Pell}
\end{equation}
If the first three matrices depart substantially from the $SU(2)$
algebra, the eigenvalues no longer have the expected degeneracies.
We nevertheless arrange the eigenmatrices in increasing order of their
eigenvalues, assigning the first three to $\ell=1$, the next five to
$\ell=2$, and so on. Thus, at small $\bar m$, $\ell$ is only a label
for this ordering and should not be interpreted as an exact angular
momentum.

To compare the magnitude of the seven matrices with that of the first
three, we also define
\begin{equation}
  \varsigma \equiv
  \frac{
    \displaystyle\sum_{\mu=4}^{10}\|\hat A_\mu^{(M)}\|_F^2
  }{
    \displaystyle\sum_{i=1}^{3}\|\hat A_i^{(M)}\|_F^2
  }.
  \label{eq:pIKKT_extra_varsigma}
\end{equation}

\begin{table}[t]
\centering
\small
\setlength{\tabcolsep}{3.5pt}
\begin{tabular}{c c c c c c c c c}
\hline
$\bar m$ & $100P_1$ & $100P_2$ & $100P_3$ &
$100P_4$ & $100P_5$ & $100P_6$ & $100P_7$ & $\varsigma$ \\
\hline
30 & 0.002 & 0.010 & 14.119 & 8.579 & 29.927 & 16.914 & 30.450 & 0.000542 \\
20 & 0.012 & 0.266 & 12.460 & 9.317 & 24.968 & 25.541 & 27.435 & 0.005812 \\
18 & 0.013 & 0.162 & 12.036 & 12.042 & 24.046 & 23.053 & 28.647 & 0.006867 \\
16 & 0.056 & 0.284 & 9.948 & 12.242 & 13.641 & 18.324 & 45.506 & 0.017809 \\
14 & 0.151 & 0.828 & 12.842 & 10.429 & 15.365 & 20.536 & 39.848 & 0.029498 \\
12 & 0.207 & 1.285 & 18.946 & 9.816 & 16.588 & 28.334 & 24.825 & 0.093755 \\
10 & 0.430 & 0.764 & 11.905 & 7.560 & 25.514 & 16.549 & 37.277 & 0.148393 \\
8 & 1.448 & 1.847 & 12.691 & 19.240 & 14.553 & 21.079 & 29.141 & 0.581147 \\
6 & 0.302 & 2.976 & 12.560 & 8.541 & 18.916 & 17.582 & 39.122 & 0.425867 \\
4 & 1.170 & 8.555 & 13.260 & 14.530 & 20.325 & 16.863 & 25.298 & 1.221398 \\
2 & 1.587 & 5.792 & 5.893 & 18.433 & 22.073 & 22.222 & 24.001 & 0.248234 \\
\hline
\end{tabular}
\caption{
Percentage contributions ($100P_\ell$) of the matrix-Laplacian modes
and the relative size $\varsigma$ of the seven extra-dimensional matrices
for $M=8$, $X=601$, and $\bar m=2,4,\ldots,20,30$.
By definition, $\sum_{\ell=1}^{7}P_\ell=1$.
At small $\bar m$, the index $\ell$ is assigned by grouping the
matrix-Laplacian eigenmodes in increasing order of eigenvalue, with
$2\ell+1$ eigenmodes in each group, and need not represent an exact
angular momentum.
}
\label{tab:pIKKT_extra_directions}
\end{table}

Our previous calculations have shown that, for $M=8$, the fuzzy-sphere configuration is very stable at $\bar m\gg 10$, begins to become unstable around $\bar m\sim 15$, and is completely unstable at $\bar m=2$.
With this in mind, let us examine Table~\ref{tab:pIKKT_extra_directions}.

The first observation is that the low-angular-momentum modes $\ell=1,2$ make only very small contributions for $\bar m\gtrsim 10$.
In particular, at $\bar m=30$, the $\ell=1$ and $\ell=2$ modes contribute only $0.002\%$ and $0.010\%$, respectively, and are therefore practically absent.
As $\bar m$ decreases, the contributions from these modes gradually increase, but for $\bar m\gtrsim 10$, their combined contribution remains of order $1\%$ or less.

However, this does not necessarily mean that the contributions from higher-angular-momentum modes increase with $\bar m$.
In fact, the $\ell=3$ mode contributes about $14\%$ at $\bar m=30$, which is the third-largest value among the 11 mass parameters considered.
Conversely, the highest-angular-momentum mode, $\ell=7$, contributes about $30\%$ at $\bar m=30$, which is not particularly large compared with its contributions at other masses.

Around $\bar m\sim 15$, where the fuzzy sphere begins to become unstable, and at $\bar m=8$, which is close to the theoretical critical value, we observe a substantial increase in the contributions from the low-angular-momentum modes.
The same tendency is observed at the smaller masses $\bar m=6,4,2$.
At these masses, $\hat A_1^{(M)}$, $\hat A_2^{(M)}$, and $\hat A_3^{(M)}$ are not necessarily close to the $SU(2)$ generators, so the term "low-angular-momentum modes" is not strictly accurate.
Nevertheless, these results suggest that $P_1$ and $P_2$ are related to the destabilization of the fuzzy sphere.
We also observe that the relative size $\varsigma$ of the extra-dimensional matrices generally tends to increase as the mass decreases.

\clearpage

\begin{figure}[p]
\centering
\centering

\includegraphics[
    width=\textwidth,
    height=0.3\textheight,
    keepaspectratio
]{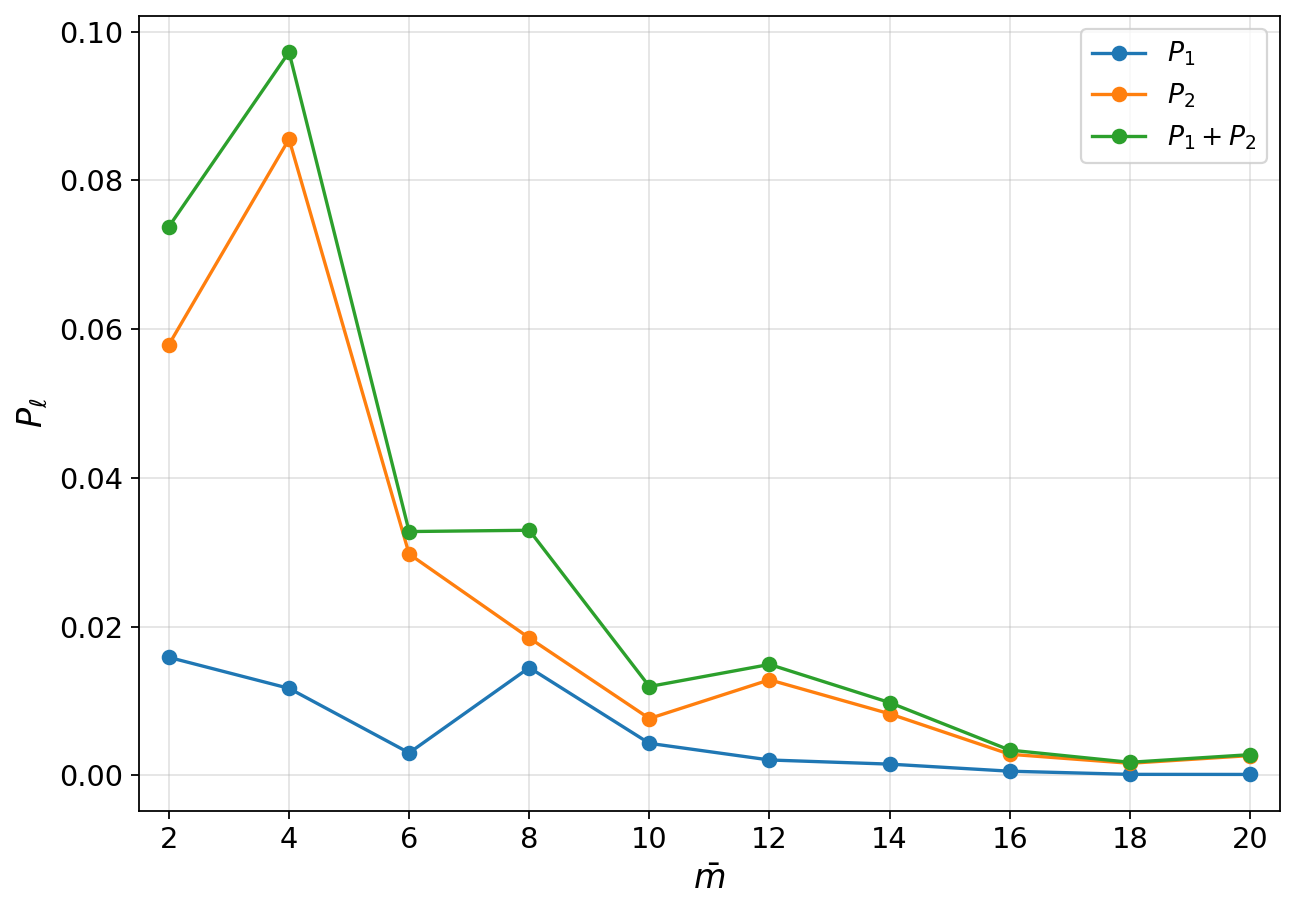}

\vspace{1.5em}

\includegraphics[
    width=\textwidth,
    height=0.3\textheight,
    keepaspectratio
]{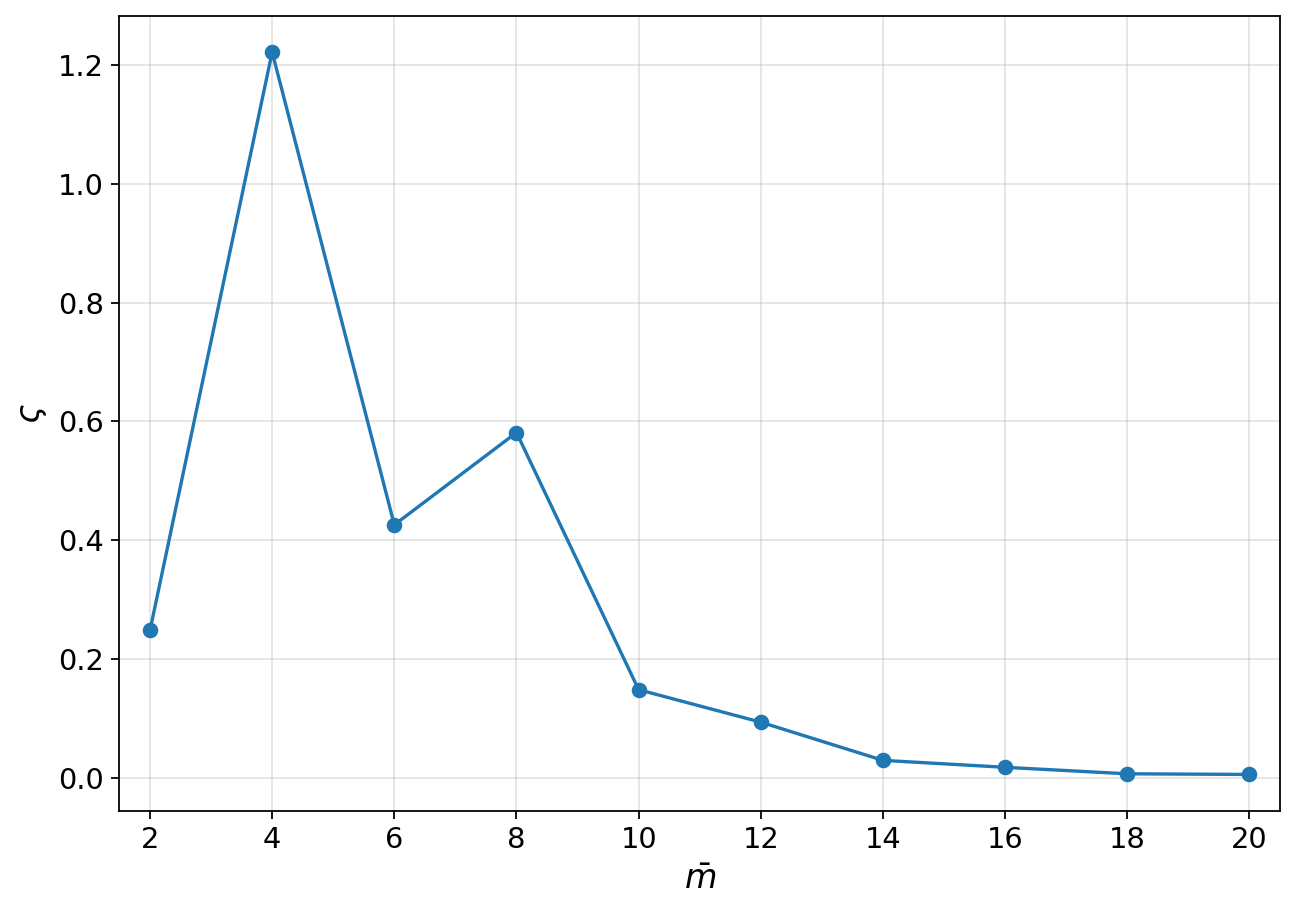}

\vspace{1.5em}

\includegraphics[
    width=\textwidth,
    height=0.3\textheight,
    keepaspectratio
]{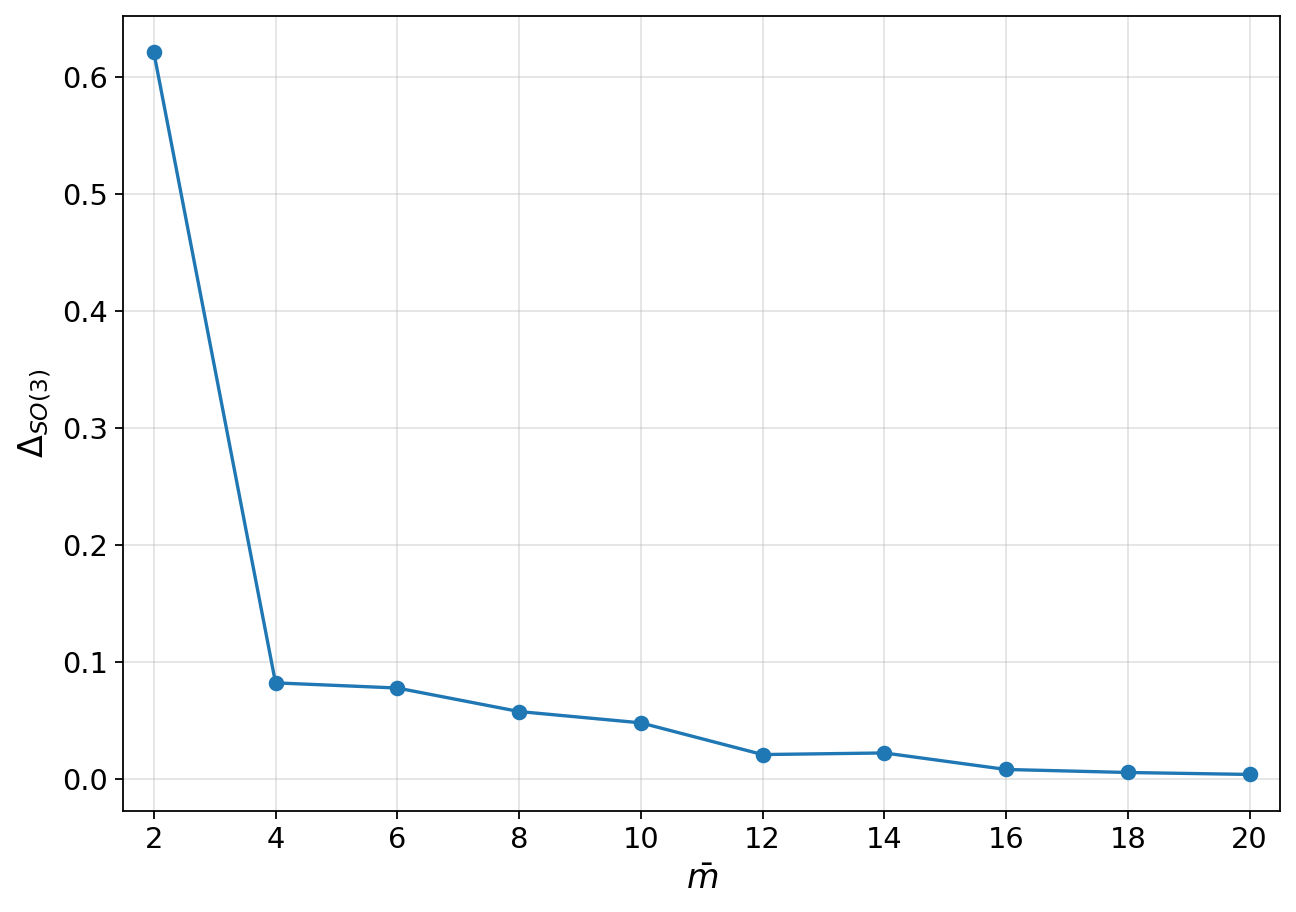}

\caption{
Mass dependence of $P_1$, $P_2$, $P_1+P_2$, $\varsigma$, and $\Delta_{SO(3)}$ for $M=8$ and $X=601$.
The top, middle, and bottom panels show $P_1$, $P_2$, and $P_1+P_2$; $\varsigma$; and $\Delta_{SO(3)}$, respectively.
The horizontal axis represents $\bar m=2,4,\ldots,20$ in all three panels.
}
\label{fig:pIKKT_order_parameters}
\end{figure}

\clearpage

These numerical results suggest that $P_{1}$, $P_{2}$, and $\varsigma$ may also serve as indicators of the onset of a phase transition.
We therefore plot these quantities in Figure \ref{fig:pIKKT_order_parameters}, together with $\Delta_{SO(3)}$ for comparison.
The horizontal axis in all three panels represents the normalized mass $\bar{m}=2,4,\ldots,20$, while the vertical axes show $P_{1}$, $P_{2}$, and $P_{1}+P_{2}$ (top), $\varsigma$ (middle), and $\Delta_{SO(3)}$ (bottom).
In all three panels, the values begin to increase around $\bar{m}\sim15$, with a more pronounced increase near $\bar{m}\sim10$.
Notably, the increase near $\bar{m}\sim10$ is more pronounced for the low-angular-momentum mode weights $P_{1}$, $P_{2}$, and $P_{1}+P_{2}$, as well as for the size of the extra-dimensional directions $\varsigma$.

By definition, $\Delta_{SO(3)}$ measures the breaking of $SO(3)$ symmetry, but the destabilization of the fuzzy-sphere phase does not necessarily imply that $SO(3)$ symmetry is broken.
In this sense, the connection between $\Delta_{SO(3)}$ and the phase transition is indirect.
In contrast, the low-angular-momentum modes are more closely related to the $SO(3)$ structure and may therefore provide more sensitive indicators of the phase transition.
As for $\varsigma$, which measures the size of the extra-dimensional directions, it may also serve as an indicator of the phase transition, although its connection to the instability of the fuzzy sphere is likewise indirect, as in the case of $\Delta_{SO(3)}$.
We should also note that, in the ten-matrix model considered here, the relative importance of matrix fluctuations increases as $\bar{m}$ decreases.

Finally, we comment on the massless limit.
When the mass parameter of the ten-matrix model is set to zero, its classical action coincides exactly with that of the original IKKT matrix model.
One might therefore expect that gradually decreasing $m$ in the RMA would eventually allow us to perform numerical calculations for the IKKT matrix model.
However, this strategy does not work.
Here we explain why the present framework fails in the massless limit.

One serious obstacle is that many of the loop equations become degenerate when $m$ is set to zero.
This degeneracy causes the following problem:
For example, when $m=0.1$, we typically choose $X$ so that the system $F=0$ is square.
In this case, a configuration $\hat{A}_{\mu}^{(M)}$ satisfying $F=0$ to machine precision generically exists and is expected to provide a good approximation to the low-order moments.
If such a configuration is found, it should also be reasonably close to the solution at $m=0.01$ for the same $X$, that is, the same set of loop equations.
Importantly, the system $F=0$ generically remains square at $m=0.01$.
Thus, by using the configuration $\hat{A}_{\mu}^{(M)}$ obtained at $m=0.1$ as the initial condition for the optimization, we can efficiently obtain a reliable solution at $m=0.01$.

However, when $m$ is set to zero while keeping $X$ fixed, the system $F=0$ suddenly becomes highly underdetermined.
Consequently, even if we perform the optimization using the configuration $\hat{A}_{\mu}^{(M)}$ obtained at $m=0.01$ as the initial condition, the resulting solution cannot be regarded as a reliable representation of the true vacuum.
This is because a highly underdetermined system $F=0$ is more likely to lead to spurious convergence of $\hat{A}_{\mu}^{(M)}$.

In fact, the situation is even worse.
When we actually decrease $m$ using our current code, we find that the objective function $F$ eventually fails to become sufficiently small.
This means that the optimization fails to find a solution to the loop equations.
At present, the reason for this behavior remains unclear.
We examine this issue further in the following discussion.

\section{Discussion and Summary} \label{sec:discussion}

In this paper, we first theoretically demonstrated that the RMA can be applied to matrix models containing fermions.
In our approach, the RMA is implemented by integrating out all the fermionic matrices to obtain a bosonic effective theory, and the key points justifying its applicability are that the additional contribution $\log\text{Pf}({\cal M}[A])$ is of order $N^{2}$ and that quantities obtained through fermionic Wick contractions can still be expressed in terms of normalized traces.
We then performed numerical calculations for mass-deformed supersymmetric Yang--Mills-type matrix models.
For all three cases, namely the three-matrix model, the four-matrix model, and the ten-matrix model (known as the polarised IKKT matrix model), we confirmed that the fuzzy-sphere phase can be reproduced using relatively small regularized master fields with $M\sim10$, and that the RMA can also capture phase-transition-like behavior in which the fuzzy-sphere phase becomes unstable.
Among the matrix models studied here, the three-matrix and ten-matrix models have negative or complex $\text{Pf}({\cal M}[A])$ respectively, making them difficult to study using standard Monte Carlo methods or matrix bootstrap methods.
\footnote{
However, complex Langevin methods and generalized Lefschetz thimble methods can also handle complex weights in some cases.
In \cite{Chou:2025rwy}, numerical calculations for the ten-matrix model, namely the polarised IKKT matrix model, were performed at $N=2$.
}
In contrast, the RMA can be applied even when the matrix-integral weight is complex, and the success of our numerical calculations can be regarded as a demonstration of this advantage of the RMA.

Throughout the numerical calculations in this paper, we take the regularized master fields $\hat{A}^{(M)}_{\mu}$ to be complex $M\times M$ matrices.
Despite this choice, in the parameter region where the fuzzy-sphere phase is stable, the three matrices $\hat{A}^{(M)}_{1}$, $\hat{A}^{(M)}_{2}$, and $\hat{A}^{(M)}_{3}$ corresponding to the fuzzy sphere are remarkably close to the generators of a unitary representation of the $SU(2)$ algebra, exhibiting almost no non-Hermiticity.
This means that non-Hermitian matrix fluctuations are somehow almost entirely absent from the master field, which is a nontrivial result.
On the other hand, the matrices in the extra-dimensional directions are not necessarily Hermitian.
The physical significance of these behaviors is not yet fully understood, and further theoretical investigation is needed.

Turning to the individual models, the three-matrix model has two mass parameters, $m_{1}$ and $m_{2}$.
At the classical level, the fuzzy-sphere configuration is a solution to the equations of motion unless both mass parameters vanish, with the fermionic matrices set to zero.
In contrast, the RMA incorporates quantum effects, including fermionic contributions, and reveals that the fuzzy-sphere configuration becomes unstable in three cases: when $m_1<m_2$ and both are small, when $m_1\geq m_2$, and when $2m_1=m_2$.
The first case corresponds to a phase transition driven by enhanced quantum fluctuations of the bosonic matrices, whereas the latter two are attributed to zero modes of the Dirac operator.
In particular, the latter two cases provide examples in which fermionic effects play an essential role, as demonstrated quantitatively within the RMA by a sharp increase in the fermion propagator.

Zero modes of the Dirac operator also play an important role in numerical calculations using Monte Carlo methods: if the Dirac operator has zero modes, the corresponding Pfaffian vanishes, and such configurations do not contribute to the matrix integral.
Thus, both in the RMA and in Monte Carlo methods, we expect configurations in which the Dirac operator has zero or near-zero modes to be generally disfavored.
Our results for the three-matrix model demonstrate precisely such a phenomenon, in which the fuzzy sphere, which is stable in the bosonic theory, becomes unstable upon the inclusion of fermions due to these zero modes.

We conjecture that a similar phenomenon may occur in the original IKKT matrix model without mass deformation.
When we perform the RMA for the IKKT matrix model with all the fermionic matrices omitted, the regularized master fields $\hat{A}^{(M)}_{\mu}$ are close to zero for all $\mu$.
However, for the configuration in which $A_{\mu}=0$ exactly for all $\mu$, the Dirac operator itself vanishes, making this configuration highly singular.
We therefore expect that configurations in which all the master fields are close to zero, $\hat{A}_{1},\ldots,\hat{A}_{10}\sim0$, are disfavored in the original IKKT matrix model with fermions.
In this regard, it has already been shown that the Dirac operator necessarily has zero modes when $A_{\mu}=0$ for at least eight of the ten values of $\mu$ \cite{Krauth:1998xh, Nishimura:2000ds}.\footnote{
We learned this fact from Prof. Tsuchiya during a discussion at the workshop \textit{From Quantum Gravity to Cosmology}, held at Kavli IPMU in Kashiwa, Japan.
}
This may provide a possible explanation for why the emergent spacetime arising from the IKKT matrix model has at least three dimensions in the framework of the RMA.

The ten-matrix model, also known as the polarised IKKT matrix model, has only one mass parameter and is therefore simpler in this respect than the three-matrix and four-matrix models.
The main achievement of our numerical calculations for this model is that we approximately reproduced the critical mass $\bar{m}_{\text{cr}}\sim8$ previously obtained in \cite{Hartnoll:2024csr, Hartnoll:2025ecj} within the framework of the RMA.
We also quantitatively demonstrated that both matrix fluctuations in the extra-dimensional directions and low-angular-momentum modes become more pronounced in the small-mass region.
Furthermore, calculations of the fuzzy-sphere radius $C_{M}$ revealed that the effects of matrix fluctuations are generally larger in the ten-matrix model than in the three-matrix and four-matrix models, and that fermionic effects are also more significant.
We suspect that this is because the ten-matrix model contains larger numbers of both bosonic and fermionic matrices, resulting in correspondingly larger contributions from their fluctuations.

Among the possible future applications of the RMA, the most important is the massless IKKT matrix model.
However, as shown in \cite{Komatsu:2024ydh, Chou:2025rwy}, taking the mass parameters of the polarised IKKT matrix model to be very small does not necessarily make it equivalent to the massless IKKT matrix model.
In our present implementation, an additional difficulty arises: when the mass parameters are set exactly to zero, the objective function no longer approaches zero.
This means that the optimization for the massless IKKT matrix model fails.
The same phenomenon was also observed in the three-matrix and four-matrix models.
Thus, we must conclude that the current RMA offers no particular advantage in this respect.

The fact that the objective function does not become sufficiently small suggests that there is no regularized master field $\hat A^{(M)}_{\mu}$ that simultaneously satisfies a finite set of loop equations ${\cal L}_{\mu;\mu_{1}...\mu_{n}}(\hat A^{(M)}_{\mu})=0$.
A similar phenomenon was observed, for example, in the Euclidean-type one-matrix model studied in \cite{Maeta:2026miu}.
For the Euclidean-type one-matrix model with a quartic interaction, if the interaction term is assigned a large negative coefficient (more precisely, if the coupling constant $g$ falls below the critical value $g_{c}=-1/12$), the theory can no longer be defined even in the large-$N$ limit.
Correspondingly, it was found that the objective function in the RMA does not become sufficiently small in this region.
This means that for $g<g_{c}$ there is no regularized master field $\hat{\phi}^{(M)}$ satisfying the loop equations, which is consistent with the breakdown of the theory in this region.

The situation is different, however, for the massless SYM-type matrix models.
The partition functions of these matrix models have been shown to be finite (at finite $N$ and with an appropriate gauge fixing), suggesting that, unlike the Euclidean-type one-matrix model with a negative coupling constant, they are well-defined theories \cite{Krauth:1998yu, Krauth:1998xh, Moore:1998et, Austing:2001pk}.
Indeed, numerical calculations using Monte Carlo methods have been performed for the massless SYM-type four-matrix model \cite{Ambjorn:2000bf}, where no anomalous behavior analogous to that observed in the RMA has been found.
Thus, the failure of the RMA optimization for the massless SYM-type matrix models is likely not due to a problem with the theories themselves, but rather to a limitation of the current optimization scheme.

Our conjecture is that this problem may originate from the shape of the eigenvalue distributions of the massless SYM-type matrix models.
Since these matrix models have no mass terms and also receive contributions from fermions, their eigenvalue distributions are expected to have much longer tails than those of ordinary bosonic matrix models or mass-deformed SYM-type matrix models.
On the other hand, as shown in \cite{Maeta:2026miu}, $\hat A^{(M)}_{\mu}$ in the RMA reproduces the endpoints of the eigenvalue distributions well.
If the endpoints of the eigenvalue distributions are located at infinity, however, they cannot be reproduced on a computer.
In other words, we conjecture that the loop equations appear to have no solution because $\hat A^{(M)}_{\mu}$ on a computer can only have finite entries.
It will therefore be important to first examine whether this conjecture is correct and then modify the RMA so that numerical calculations can also be successfully performed for massless SYM-type matrix models.

Nevertheless, except for the massless models, the RMA can be applied to a remarkably wide variety of large-$N$ reduced matrix models.
Other advantages of the RMA are that it is free from the sign problem and that it provides access to the $N\to\infty$ limit at low computational cost, thereby complementing the shortcomings of conventional Monte Carlo methods and matrix bootstrap methods.
An important direction for future work is to perform detailed calculations for more physically important models, such as large-$N$ reduced lattice gauge theories and the (polarised) IKKT matrix model, using these methods in a complementary manner.

\acknowledgments

The authors would like to thank Robert Brandenberger for support and helpful discussions. 
RM would like to thank Takehiro Azuma, Takeshi Morita, Jun Nishimura and Asato Tsuchiya for helpful discussions.
RM is supported by JST SPRING, Grant Number JPMJSP2132.


 \bibliographystyle{JHEP}
 \bibliography{biblio}


\appendix

\section{Classical Saddle Structure}
\label{app:classical-saddles}

We restrict the classical bosonic action to $A_i=rL_i$, where $L_i$ generate the $N$-dimensional irreducible representation of $SU(2)$, and set the fermions to zero. The stability statements below refer to variations of the radius parameter $r$ within this ansatz.

\subsection{Three-matrix model}
\label{app:three_matrix_saddles}

The bosonic action of the massive three-matrix model is
\begin{equation}
S_B
=
N\,\operatorname{Tr}\left[
-\frac14[A_i,A_j]^2
+m_1m_2 A_iA_i
+\frac{2i}{3}(m_1+m_2)\epsilon_{ijk}A_iA_jA_k
\right].
\end{equation}
At a classical saddle the fermionic matrices may be set to zero, so the bosonic equation of motion is
\begin{equation}
[A_j,[A_j,A_i]]
+
2m_1m_2A_i
+
2i(m_1+m_2)\epsilon_{ijk}A_jA_k
=0.
\end{equation}
Using the fuzzy-sphere ansatz
\begin{equation}
A_i=rL_i,
\qquad
[L_i,L_j]=i\epsilon_{ijk}L_k,
\end{equation}
the equation of motion reduces to
\begin{equation}
r(r-m_1)(r-m_2)=0.
\end{equation}
Thus there are three radial stationary points,
\begin{equation}
    r=0,\qquad r=m_1,\qquad r=m_2,
\end{equation}
corresponding to the trivial configuration and two nontrivial fuzzy-sphere solutions.

Restricting the action to this ansatz gives
\begin{equation}
S_B(r)
=
N\,\operatorname{Tr} L^2
\left[
\frac12 r^4
-\frac23(m_1+m_2)r^3
+m_1m_2r^2
\right].
\end{equation}
At the two nontrivial stationary points,
\begin{align}
\left.\frac{d^2S_B}{dr^2}\right|_{r=m_1}
&=
2N\,m_1(m_1-m_2)\operatorname{Tr} L^2,
\\
\left.\frac{d^2S_B}{dr^2}\right|_{r=m_2}
&=
2N\,m_2(m_2-m_1)\operatorname{Tr} L^2.
\end{align}
Hence, for positive masses and $m_2>m_1$, the $r=m_1$ saddle is unstable while $r=m_2$ is the stable nontrivial minimum.
If $m_1>m_2$, the roles are reversed.
The stable nontrivial saddle therefore has
\begin{equation}
r=\max(m_1,m_2).
\end{equation}

To determine whether the trivial or the stable nontrivial saddle is classically preferred, consider $m_2>m_1$.
The trivial configuration has
\begin{equation}
S_B(0)=0,
\end{equation}
while the stable fuzzy-sphere saddle has
\begin{equation}
S_B(m_2)
=
-\frac{N}{6}
m_2^3(m_2-2m_1)\operatorname{Tr} L^2.
\end{equation}
Therefore,
\begin{equation}
    \begin{array}{ll}
    m_1<m_2<2m_1
    &\Longrightarrow
    S_B(0)<S_B(m_2),
    \\[1mm]
    m_2=2m_1
    &\Longrightarrow
    S_B(0)=S_B(m_2),
    \\[1mm]
    m_2>2m_1
    &\Longrightarrow
    S_B(m_2)<S_B(0).
    \end{array}
\end{equation}
Thus $m_1=m_2$ determines which of the two nontrivial stationary points is locally stable, while $m_2=2m_1$ marks the crossing at which the stable fuzzy-sphere minimum becomes deeper than the trivial one.
The case $m_1>m_2$ follows by interchanging $m_1$ and $m_2$.

The bosonic action evaluated at the fuzzy-sphere saddles also motivates the mass normalization used in Section~\ref{sec:three_matrix}. Using $\operatorname{Tr}L^2=N(N^2-1)/4$, its value on either nontrivial branch is
\begin{equation}
S_B(m_a)
=-\frac{N^2(N^2-1)}{24}
m_a^3(m_a-2m_b),
\qquad a,b\in\{1,2\},\quad b\neq a.
\end{equation}
At fixed mass ratios, the action scales as $N^4m^4$. We therefore take $m_a=\widetilde m_a/\sqrt N$, holding $\widetilde m_a$ fixed, so that
\begin{equation}
S_B(m_a)
=-\frac{N^2-1}{24}
\widetilde m_a^3(\widetilde m_a-2\widetilde m_b).
\end{equation}
The action evaluated at either saddle is then generically of order $N^2$; at $m_a=2m_b$, it vanishes because the contributions cancel. For the regularized configurations, we adopt the corresponding prescription $m_a=\widetilde m_a/\sqrt M$, with $\widetilde m_a=\mathcal O(M^0)$ held fixed as $M$ varies.

\subsection{Four-matrix model}
\label{app:four_matrix_saddles}

For the four-matrix action in Eq.~\eqref{eq:4matrix_action}, we take $A_i=rL_i$ for $i=1,2,3$ and $A_4=0$. The equation of motion for $A_4$ is then satisfied, and the remaining equations reduce to
\begin{equation}
r\left(r-\frac{m_1}{3}\right)\left(r-\frac{2m_1}{3}-m_2\right)=0.
\end{equation}
Besides $r=0$, there are therefore two nontrivial stationary branches,
\begin{equation}
r_-=\frac{m_1}{3},\qquad r_+=\frac{2m_1}{3}+m_2.
\end{equation}
Restricting the bosonic action to this ansatz gives
\begin{equation}
S_B(r)=N\operatorname{Tr}L^2\left[\frac12r^4-\frac23(m_1+m_2)r^3+\left(\frac29m_1^2+\frac13m_1m_2\right)r^2\right].
\end{equation}
For positive masses, $r_+>r_->0$, and the radial curvatures are
\begin{equation}
\left.\frac{d^2S_B}{dr^2}\right|_{r=r_\pm}=\pm2N\operatorname{Tr}L^2\,r_\pm(r_+-r_-).
\end{equation}
Thus the smaller-radius branch is radially unstable, whereas the larger-radius branch is a radial local minimum throughout the positive-mass region. The action evaluated on the latter branch is
\begin{equation}
S_B(r_+)=-\frac{N}{6}\operatorname{Tr}L^2\,m_2r_+^3<0=S_B(0).
\end{equation}
Consequently, the branch $r_+=\tfrac23m_1+m_2$ always has lower classical action than the trivial configuration for positive masses. Unlike the three-matrix model, the two nontrivial branches do not coalesce in this region, and there is no equality of the actions at the larger-radius sphere and the trivial configuration. The reference used in Section~\ref{sec:four_matrix} is this radially stable branch.

The same mass scaling as in the three-matrix model gives an action of order $N^2$. Substituting $m_a=\widetilde m_a/\sqrt N$ and $\operatorname{Tr}L^2=N(N^2-1)/4$ gives
\begin{equation}
S_B(r_+)=-\frac{N^2-1}{24}\widetilde m_2\left(\frac23\widetilde m_1+\widetilde m_2\right)^3.
\end{equation}
Holding the rescaled masses fixed therefore gives the required classical large-$N$ scaling. For the regularized configurations, we use $m_a=\widetilde m_a/\sqrt M$, with $\widetilde m_a=\mathcal O(M^0)$ held fixed as $M$ varies. The corresponding classical large-$N$ reference is
\begin{equation}
C^{\mathrm{FS}}=\lim_{N\to\infty}\frac{1}{N^2}\operatorname{Tr}\sum_{i=1}^{3}(r_+L_i)^2=\frac14\left(\frac23\widetilde m_1+\widetilde m_2\right)^2.
\end{equation}

\end{document}